\documentclass[11pt,reqno]{amsart}
\usepackage{amsmath}
\usepackage{amssymb}
\usepackage{verbatim}
\usepackage{subfigure}
\usepackage{multirow}
\usepackage{epsfig}
\usepackage{./wavelet}

\newcommand{\dtcwt}{\operatorname{DT-\mathbb{C}WT}}
\newcommand{\tpctf}{\operatorname{TP-\mathbb{C}TF}}

\newcommand{\ctf}{\operatorname{\mathbb{C}TF}}

\newcommand{\tcwt}{\operatorname{\mathbb{C}WT}}
\newcommand{\fcwt}{\operatorname{F\mathbb{C}WT}}
\newcommand{\ncwt}{\operatorname{N\mathbb{C}WT}}
\newcommand{\sgn}{\operatorname{sgn}}

\numberwithin{equation}{section}
\numberwithin{figure}{section}

\begin{document}

\title[Image Denoising by Directional Complex Tight Framelets]{Image Denoising Using Tensor Product Complex Tight Framelets with Increasing Directionality}

\author{Bin Han and Zhenpeng Zhao}

\thanks{Research supported in part by NSERC Canada under Grant
RGP 228051. \hfill  \today}

\address{Department of Mathematical and Statistical Sciences,
University of Alberta, Edmonton,\quad Alberta, Canada T6G 2G1.
\quad {\tt bhan@ualberta.ca}\quad {\tt zzhao7@ualberta.ca}\quad
{\tt http://www.ualberta.ca/$\sim$bhan}
}

\makeatletter \@addtoreset{equation}{section} \makeatother
\begin{abstract}
Tensor product real-valued wavelets have been employed in many applications such as image processing with impressive performance. Though edge singularities are ubiquitous and play a fundamental role in image processing and many other two-dimensional problems, tensor product real-valued wavelets are known to be only sub-optimal since they can only capture edges well along the coordinate axis directions (that is, the horizontal and vertical directions in dimension two).
Among several approaches in the literature to enhance the performance of tensor product real-valued wavelets, the dual tree complex wavelet transform ($\dtcwt$), proposed by Kingsbury \cite{K1999} and further developed by Selesnick et al. \cite{SBK}, is one of the most popular and successful enhancements of the classical tensor product real-valued wavelets by employing a correlated pair of orthogonal wavelet filter banks.
The two-dimensional $\dtcwt$ is obtained via tensor product and  offers improved directionality with $6$ directions.
In this paper we shall further enhance the performance of $\dtcwt$ for the problem of image denoising. Using framelet-based approach and the notion of discrete affine systems, we shall propose a family of tensor product complex tight framelets $\tpctf_n$ for all integers $n\ge 3$ with increasing directionality, where $n$ refers to the number of filters in the underlying one-dimensional complex tight framelet filter bank. For dimension two, such tensor product complex tight framelet $\tpctf_n$ offers $\frac{1}{2}(n-1)(n-3)+4$ directions when $n$ is odd, and $\frac{1}{2}(n-4)(n+2)+6$ directions when $n$ is even.
In particular, we shall show that $\tpctf_4$, which is different to $\dtcwt$ in both nature and design, provides an alternative to $\dtcwt$. Indeed, we shall see that $\tpctf_4$
behaves quite similar to $\dtcwt$ by offering $6$ directions in dimension two, employing the tensor product structure, and enjoying slightly less redundancy than $\dtcwt$. When $\tpctf_4$ is applied to image denoising, its performance is comparable to $\dtcwt$. Moreover, better results on image denoising can be obtained by using other $\tpctf_n$, for example, $n=6$, which has $14$ directions in dimension two. Moreover, $\tpctf_n$ allows us to further improve $\dtcwt$ by using $\tpctf_n$ as the first stage filter bank in $\dtcwt$.
Experiments on image denoising using $\tpctf_n$ and detailed comparison with $\dtcwt$ will be provided in this paper.
\end{abstract}


 \maketitle

\pagenumbering{arabic}

\section{Introduction and Motivations}

In this paper we shall take a framelet-based approach to enhance the performance of the classical tensor product real-valued wavelets by providing a family of tensor product complex tight framelet filter banks with increasing directionality. On the other hand, we provide alternatives and improvements to the well-known dual tree complex wavelet transform ($\dtcwt$) which has been proposed by Kingsbury in \cite{K1999,K2001} and further developed by Selesnick et al. in \cite{SBK}. We shall apply the constructed tensor product complex tight framelets for the problem of image denoising and we shall compare their performance with $\dtcwt$ in the area of image denoising.

To explain our motivations, let us first recall some definitions. By $\dlp{2}$ we denote the space of all complex-valued sequences $u=\{u(k)\}_{k\in \dZ}: \dZ\rightarrow \C$ such that $\|u\|_{\dlp{2}}:=(\sum_{k\in \dZ} |u(k)|^2)^{1/2}<\infty$. The Fourier series (or symbol) of a sequence $u\in \dlp{2}$ is defined to be $\wh{u}(\xi):=\sum_{k\in \dZ} u(k) e^{-ik\cdot \xi}, \xi\in \dR$, which is a $2\pi\dZ$-periodic measurable function in $\dTLp{2}$ such that $\|\wh{u}\|_{\dTLp{2}}^2:=\frac{1}{(2\pi)^d} \int_{[-\pi,\pi)^d} |\wh{u}(\xi)|^2 d\xi=\|u\|_{\dlp{2}}^2=\sum_{k\in \dZ} |u(k)|^2<\infty$.
If $u\in \dlp{1}$, that is, $\|u\|_{\dlp{1}}:=\sum_{k\in \dZ} |u(k)|<\infty$, then
$u\in \dlp{2}$ and $\wh{u}\in C(\dT)$ is a continuous function.

For filters $a, b_1, \ldots, b_s\in \dlp{1}$, $\{a; b_1, \ldots, b_s\}$ is called a $d$-dimensional (dyadic) tight framelet filter bank if
\begin{align}
&|\wh{a}(\xi)|^2+\sum_{\ell=1}^s |\wh{b_\ell}(\xi)|^2=1,
\label{tffb:pr:1}\\
&\wh{a}(\xi)\ol{\wh{a}(\xi+\pi \omega)}+\sum_{\ell=1}^s \wh{b_\ell}(\xi)\ol{\wh{b_\ell}(\xi+\pi \omega)}=0, \qquad \forall\; \omega\in \Omega\bs \{0\} \label{tffb:pr:0}
\end{align}
for all $\xi\in \dR$, where $\Omega:=[0,1]^d\cap \dZ$. The filter $a$ is called a low-pass filter since we often require $\wh{a}(0)=1$, and all the filters $b_1, \ldots, b_s$ are called high-pass filters since we often have $\wh{b_1}(0)=\cdots=\wh{b_s}(0)=0$. Note that if $\wh{a}(0)=1$ in a tight framelet filter bank $\{a; b_1, \ldots, b_s\}$, then it follows directly from \eqref{tffb:pr:1} that $\wh{b_1}(0)=\cdots=\wh{b_s}(0)=0$.
When $s=2^d-1$, a $d$-dimensional (dyadic) tight framelet filter bank $\{a; b_1, \ldots, b_{2^d-1} \}$ is called a $d$-dimensional (dyadic) orthogonal wavelet filter bank. Let $\{a; b_1, \ldots, b_s\}$ be a $d$-dimensional tight framelet filter bank. Under the mild condition $|1-\wh{a}(\xi)|\le C |\xi|^\tau, \xi\in [-\pi, \pi]^d$ for some positive numbers $C$ and $\tau$ (all our tight framelet filter banks constructed in this paper satisfy this condition with $\tau=1$), the function $\wh{\phi}(\xi):=\prod_{j=1}^\infty \wh{a}(2^{-j}\xi)$ is a well-defined function in $\dLp{2}$ and $\{\phi; \psi^1, \ldots, \psi^s\}$ is a tight framelet in $\dLp{2}$, that is,
the affine system
\[
\AS_0(\phi; \psi^1, \ldots, \psi^s):=\{\phi(\cdot-k) \setsp k\in \dZ\} \cup \{ 2^{dj/2} \psi^\ell(2^j\cdot-k) \setsp k\in \dZ, j\in \N \cup\{0\}, \ell=1, \ldots, s\}
\]
is a (normalized) tight frame for $\dLp{2}$ satisfying
\[
\|f\|_{\dLp{2}}^2=\sum_{k\in \dZ} |\la f, \phi(\cdot-k)\ra|^2+\sum_{j=0}^\infty \sum_{\ell=1}^s \sum_{k\in \dZ} |\la f, 2^{dj/2} \psi^\ell(2^j\cdot-k)\ra|^2, \qquad \forall\; f\in \dLp{2},
\]
where the functions $\psi^1, \ldots, \psi^s$ are defined by $\wh{\psi^\ell}(\xi):=\wh{b_\ell}(\xi/2) \wh{\phi}(\xi/2)$, $\ell=1, \ldots, s$.
Throughout this paper, the word framelet is the synonym for frame wavelet.
For more details on tight framelets and their applications, see \cite{CRSS,CHS,Daub:book,DGM,DHRS,Han:frame,Han:acha:2012,Han:acha:2013,HGS,RonShen:twf,S:acha,Shen} and many references therein. Due to this connection between a tight framelet filter bank and a tight framelet in the function space $\dLp{2}$, we shall concentrate on tight framelet filter banks instead of tight framelets in $\dLp{2}$ in this paper. In fact, to understand the properties and performance of a discrete framelet or wavelet transform, it is more important to study its underlying discrete affine systems $\DAS_J(\{a; b^1, \ldots, b^s\})$ than its associated functional affine system $\AS_0(\phi; \psi^1, \ldots, \psi^s)$ in $\dLp{2}$.
See \cite{Han:MMNP:2013} for more details on discrete affine systems. 

The simplest way to obtain a $d$-dimensional tight framelet filter bank is to use tensor product of one-dimensional tight framelet filter banks. For simplicity of presentation, in this paper we only discuss tensor product for dimension two. For two one-dimensional filters $u, v\in \lp{1}$, their tensor product filter $u\otimes v$ in dimension two is simply defined to be
$[u\otimes v](j,k)=u(j) v(k)$, $j,k\in \Z$.
Let $\{a; b_1, \ldots, b_s\}$ be a one-dimensional tight framelet filter bank. Then its tensor product tight framelet filter bank in dimension two is given by $\{a; b_1, \ldots, b_s\}\otimes\{a; b_1, \ldots, b_s\}$. More explicitly,
\[
\{a\otimes a\} \cup\{ a\otimes b_1, \ldots, a\otimes b_s\}\cup\{b_1\otimes a, \ldots, b_s\otimes a\}\cup\{b_\ell\otimes b_m \setsp \ell,m=1, \ldots, s\},
\]
where $a\otimes a$ is the low-pass filter and all other filters above are high-pass filters.
It is well known in the literature (\cite{CDDY,DV,FA,Han:acha:2012,K1999,K2001,KSZ,S:ieee,SBK})
that tensor product real-valued wavelets or framelets lack directionality. To see this point well, let us look at the simplest example of the tensor product Haar orthogonal wavelet filter bank in dimension two. The one-dimensional Haar orthogonal wavelet filter bank $\{a; b\}$ is given by $a=\{\frac{1}{2}, \frac{1}{2}\}_{[0,1]}$ and $b=\{\frac{1}{2}, -\frac{1}{2}\}_{[0,1]}$.
Then $\{a\otimes a; a\otimes b, b\otimes a, b\otimes b\}$ is a two-dimensional real-valued orthogonal wavelet filter bank, where
\be \label{haar:2d}
a\otimes a=\left[\begin{matrix} \tfrac{1}{4} &\tfrac{1}{4} \smallskip \\
\tfrac{1}{4} &\tfrac{1}{4}\end{matrix}\right]_{[0,1]^2}, \quad
a\otimes b=\left[\begin{matrix} -\tfrac{1}{4} &-\tfrac{1}{4}\smallskip \\
\tfrac{1}{4} &\tfrac{1}{4}\end{matrix}\right]_{[0,1]^2}, \quad
b\otimes a=\left[\begin{matrix} \tfrac{1}{4} &-\tfrac{1}{4} \smallskip \\
\tfrac{1}{4} &-\tfrac{1}{4}\end{matrix}\right]_{[0,1]^2}, \quad
b\otimes b=\left[\begin{matrix} -\tfrac{1}{4} &\tfrac{1}{4}\smallskip \\
\tfrac{1}{4} &-\tfrac{1}{4}\end{matrix}\right]_{[0,1]^2}.
\ee
Note that $a\otimes b$ has horizontal direction, $b\otimes a$ has vertical direction, but $b\otimes b$ does not exhibit any directionality (instead, $b\otimes b$ is sort of a saddle point). On the other hand, it is widely known (\cite{CDDY,DV,FA,K1999,KSZ,SBK} and many references therein) that edge singularities are ubiquitous and play a fundamental role in many two-dimensional problems such as image processing. To enhance the performance of tensor product real-valued wavelets by improving directionality, several approaches have been proposed in the literature. For example, curvelet transform \cite{CDDY} and shearlet transform \cite{KSZ} for dimension two on $\R^2$, steerable filter banks \cite{FA} and contourlets \cite{DV} in the discrete domain $\Z^2$, symmetric complex orthogonal wavelet filter banks \cite{Han:cw,LM}, and dual tree complex wavelet transform ($\dtcwt$) in \cite{K1999,K2001,S:ieee,SBK,SS:bs,SS:bslocal}, and etc.

Among all these approaches, $\dtcwt$ is probably one of the most popular and successful approaches to improve the performance of classical tensor product real-valued wavelet transform. The success of $\dtcwt$ largely lies in three major advantages of $\dtcwt$: (i) $\dtcwt$ offers $6$ directions (roughly along $\pm 15^\circ$, $\pm 45^\circ$, $\pm 75^\circ$), in comparison with only two directions (that is, horizontal and vertical directions) of classical tensor product real-valued wavelets. (ii) $\dtcwt$ is nearly shift-invariant without high redundancy, comparing with the shift-invariant undecimated wavelet transform.
(iii) $\dtcwt$ for high dimensions employs tensor product of one-dimensional $\dtcwt$, which is computationally efficient and simple for high dimensional problems. This also makes the implementation of $\dtcwt$ essentially the same as the classical tensor product real-valued wavelets (but with some degree of redundancy). Since in this paper we shall adopt the filter bank approach and preserve the tensor product structure, we shall only discuss $\dtcwt$ in this paper. Our goal of this paper is to provide alternatives and further improvements of $\dtcwt$ for the problem of image denoising.

To understand the key features and advantages of $\dtcwt$, let us first look at the possible shortcomings of tensor product real-valued filters.
For a one-dimensional filter $u: \Z \rightarrow \C$, it is straightforward to see that $u$ is a real-valued filter (that is, $u: \Z \rightarrow \R$) if and only if $\ol{\wh{u}(\xi)}=\wh{u}(-\xi)$. Therefore, for a real-valued filter $u$, we always have $|\wh{u}(-\xi)|=|\wh{u}(\xi)|$ and the magnitude of its frequency spectrum is symmetric about the origin. If both $u$ and $v$ are one-dimensional real-valued high-pass filters satisfying $\wh{u}(0)=\wh{v}(0)=0$, since the magnitudes of the frequency spectrums of $u$ and $v$ are symmetric about the origin, it is easy to see that the frequency spectrum of the two-dimensional real-valued tensor product filter $u\otimes v$ concentrates equally in the four quadrants (more precisely, the four corners) of the basic frequency square $[-\pi, \pi]^2$. Consequently, the filter $u\otimes v$ lacks directionality and behaves like a saddle point, just as the tensor product filter $b\otimes b$ in \eqref{haar:2d} in the two-dimensional tensor product Haar orthogonal wavelet filter bank. To achieve directionality while preserving the tensor product structure, as argued in \cite{K1999,K2001,S:ieee,SBK} and many other papers, it is natural to consider complex-valued high-pass filters $u$ and $v$ so that the frequency spectrums of $u$ and $v$ largely lie on either $[0, \pi)$ or $(-\pi, 0]$. The $\dtcwt$ achieves this goal by using a pair of correlated real-valued orthogonal wavelet filter banks which are linked to each other via an interesting half-shift condition and the Hilbert transform (see \cite{K1999,K2001,S:ieee,SBK}).
To understand better $\dtcwt$, we shall study the key ingredients of $\dtcwt$ in Section 2 using the framework of discrete affine systems, which have been introduced in \cite{Han:MMNP:2013}.

As demonstrated in many interesting works by the research groups of Kingsbury in \cite{K1999,K2001} and Selesnick in \cite{S:ieee,SBK,SS:bs,SS:bslocal}, $\dtcwt$ has impressive performance over the classical tensor product real-valued wavelets, for example, in image denoising in \cite{K1999,SS:bs,SS:bslocal} and many references therein. However, using dyadic orthogonal wavelet filter banks and Hilbert transform, to our best knowledge, it is not easy to generalize the $\dtcwt$ to have more directions. In this paper, we shall adopt a framelet-based approach and use discrete affine systems to provide alternatives and improvements to $\dtcwt$. This framelet-based approach allows us to achieve improved directionality while avoids the use of the Hilbert transform.

The structure of the paper is as follows. In order to understand the performance of discrete wavelet transform and $\dtcwt$, we shall first recall from \cite{Han:MMNP:2013} the notion of discrete affine systems associated with a multilevel discrete wavelet or framelet transform.
Then we shall discuss and analyze the main features of the $\dtcwt$ under the framework of discrete affine systems. The notion of discrete affine systems also plays a key role in our understanding and construction of tensor product complex tight framelet filter banks in Section~4. For application of $\dtcwt$ to image denoising, we shall demonstrate in Section~3 that $\dtcwt$, employing a pair of frequency-based (that is, bandlimited) correlated orthogonal wavelet filter banks, performs equally well as the original $\dtcwt$ employing a pair of finitely supported correlated
orthogonal wavelet filter banks proposed and used in \cite{K1999,K2001,SBK,SS:bslocal}.
In Section 4, we shall introduce and construct a family of tensor product complex tight framelets $\tpctf_n$ with $n\ge 3$, where $n$ refers to the number of filters in the underlying one-dimensional complex tight framelet filter banks.
Such tensor product complex tight framelet $\tpctf_n$ offers $\frac{1}{2}(n-1)(n-3)+4$ directions when $n$ is odd, and $\frac{1}{2}(n-4)(n+2)+6$ directions when $n$ is even.
In Section~4, we shall show that $\tpctf_4$, which is different to $\dtcwt$ in both nature and design, provides an alternative to $\dtcwt$. Indeed, we shall see that $\tpctf_4$
behaves quite similar to $\dtcwt$: $\tpctf_4$ offers $6$ directions in dimension two, employs the tensor product structure, and has slightly less redundancy than $\dtcwt$ by using only one low-pass filter in $\tpctf_4$ instead of four low-pass filters in $\dtcwt$ for dimension two. When $\tpctf_4$ is applied to the problem of image denoising, its performance is comparable to $\dtcwt$. Moreover, we shall demonstrate in Section~4 that better results on image denoising, in terms of PSNR, can be obtained by using other $\tpctf_n$, for example, $n=6$, which has $14$ directions in dimension two.
Experiments on image denoising using $\tpctf_n$ and detailed comparison with $\dtcwt$ will be provided in Section~4.
Finally, we shall discuss in Section~5 the choice of the initial filter banks for the first level of $\dtcwt$ for further improvements.
We shall show that $\tpctf_n$ allows us to further improve $\dtcwt$ by using $\tpctf_n$ as the first stage filter bank in $\dtcwt$.  We shall also make some remarks on $\tpctf_n$ in Section~5 for possible further improvements.

\section{Understand $\dtcwt$ Using Discrete Affine Systems}

Though wavelets and framelets have been extensively studied in the continuum domain,
to understand the performance of the classical discrete wavelet transform and $\dtcwt$, it is of fundamental importance to study discrete wavelet or framelet transform directly.
For this purpose, in this section we shall first recall the notion of discrete affine systems associated with a discrete wavelet transform or any discrete linear transform. Then we shall investigate the key features of $\dtcwt$ under the framework of discrete affine systems.

Though $\dtcwt$ has been extensively studied and discussed in \cite{K1999,K2001,S:ieee,SBK,SS:bs,SS:bslocal} and many other references in the setting of filter banks and functions in $\Lp{2}$ (that is, refinable functions and wavelet functions in $\Lp{2}$), our discussion on $\dtcwt$
in this section through the notion of discrete affine systems provides complimentary and probably more direct understanding on $\dtcwt$. Our investigation on $\dtcwt$ in this section by using discrete affine systems also allows us to see and understand better the advantages and possible places for further improvements of $\dtcwt$.

Let us first recall the multilevel discrete framelet transform using a $d$-dimensional tight framelet filter bank $\{a; b_1, \ldots, b_s\}$.
For a positive integer $J$ and an input signal $v\in \dlp{2}$,
a $J$-level discrete framelet decomposition computes the framelet/wavelet coefficients $v_j, w_{\ell,j}$ through the following recursive formulas:
\[
v_{j}:=2^{-d/2} \tz_a v_{j-1}, \quad w_{\ell,j}:=2^{-d/2} \tz_{b_\ell} v_{j-1}, \qquad \ell=1, \ldots,s, \; j=1, \ldots, J,
\]
where $v_0=v$ is the input signal and the transition operator $\tz_a: \dlp{2}\rightarrow \dlp{2}$ is defined to be
\[
[\tz_a v](n)=2^d \sum_{k\in \dZ} v(k) \ol{a(k-2n)}, \qquad n\in \dZ.
\]
A $J$-level discrete framelet reconstruction is used to recursively reconstruct the original signal as follows:
\[
\mathring{v}_{j-1}:=2^{-d/2} \sd_a \mathring{v}_{j}+2^{-d/2} \sum_{\ell=1}^s \sd_{b_\ell} \mathring{w}_{\ell,j}, \qquad j=J, \ldots, 1,
\]
where the subdivision operator $\sd_a: \dlp{2}\rightarrow \dlp{2}$ is defined to be
\[
[\sd_a v](n)=2^d \sum_{k\in \dZ} v(k) a(n-2k), \qquad n\in \dZ.
\]
Using convolution, upsampling and downsampling, we see that $\tz_a v=2^d (v*a^\star) \ds 2 I_d$ and $\sd_a v=2^d (v\us 2I_d)*a$, where the adjoint filter $a^\star$ is defined by
\be \label{astar}
a^\star(k):=\ol{a(-k)}, \qquad \forall\; k\in \dZ \qquad \mbox{or equivalently}\quad
\wh{a^\star}(\xi):=\ol{\wh{a}(\xi)}.
\ee
Following \cite{Han:MMNP:2013}, we define multilevel filters $\ta_j$ and $\tb_{\ell,j}$ with $j\in \N$ and $\ell=1, \ldots, s$ by
\be \label{ajbj}
\wh{a_j}(\xi):=2^{dj/2}\wh{a}(\xi) \wh{a}(2\xi) \cdots \wh{a}(2^{j-2}\xi) \wh{a}(2^{j-1}\xi)
\quad \mbox{and}\quad \wh{b_{\ell,j}}(\xi):=2^{dj/2} \wh{a}(\xi)\wh{a}(2\xi)\cdots \wh{a}(2^{j-2}\xi) \wh{b_\ell}(2^{j-1}\xi).
\ee
Note that $a_1=2^{d/2} a$ and $b_{\ell,1}=2^{d/2} b_\ell$. Now we define
\be \label{ajkbjk}
a_{j;k}:=a_j(\cdot-2^j k), \quad b_{\ell,j; k}:=b_{\ell,j}(\cdot-2^j k), \qquad k\in \dZ, j\in \N.
\ee
Note that $\dlp{2}$ is a Hilbert space equipped with the inner product $\la v,w\ra=\sum_{k\in \dZ} v(k) \ol{w(k)}$. As shown in \cite{Han:MMNP:2013}, we have
\[
v_j(k)=\la v, a_{j;k}\ra \quad \mbox{and}\quad w_{\ell,j}(k)=\la v, b_{\ell, j;k}\ra, \qquad k\in \dZ, \; j\in \N,\; \ell=1, \ldots, s.
\]
Consequently, a $J$-level discrete framelet transform is exactly to compute the following representation
\be \label{dfrt:expr}
v=\sum_{k\in \dZ} \la v, a_{J;k}\ra a_{J;k}+\sum_{j=1}^J \sum_{\ell=1}^s \sum_{k\in \dZ} \la v, b_{\ell, j; k}\ra b_{\ell, j; k}, \qquad \forall\; v\in \dlp{2}
\ee
with the series converging unconditionally in $\dlp{2}$.
Moreover, we have the following cascade structure on which the fast discrete framelet transform is based:
\[
\sum_{k\in \dZ} \la v, a_{j-1; k}\ra a_{j-1;k}=
\sum_{k\in \dZ} \la v, a_{j; k}\ra a_{j;k}+\sum_{\ell=1}^s \sum_{k\in \dZ} \la v, b_{\ell,j;k}\ra b_{\ell,j; k}, \qquad \forall\; v\in \dlp{2}, j\in \N.
\]
Following \cite{Han:MMNP:2013}, for every positive integer $J\in \N$, we define a $J$-level discrete affine system associated with the filter bank $\{a; b_1, \ldots, b_s\}$ by
\be \label{DAS}
\DAS_J(\{a; b_1, \ldots, b_s\}):=\{a_{J;k} \setsp k\in \dZ\}\cup \cup_{j=1}^J \{b_{\ell,j;k} \setsp k\in \dZ, \ell=1, \ldots, s\}.
\ee
It is not difficult to directly verify (\cite{Han:MMNP:2013}) that $\{a; b_1, \ldots, b_s\}$ is a tight framelet filter bank if and only if $\DAS_J(\{a; b_1, \ldots, b_s\})$ is a (normalized) tight frame for $\dlp{2}$ for every integer $J\in \N$, that is,
\[
\|v\|_{\dlp{2}}^2=\sum_{u\in \DAS_J(\{a; b_1, \ldots, b_s\})} |\la v, u\ra|^2, \qquad \forall\; v\in \dlp{2},
\]
which directly leads to the discrete representation in \eqref{dfrt:expr}.
Similarly, $\{a; b_1, \ldots, b_{2^d-1}\}$ is an orthogonal wavelet filter bank if and only if $\DAS_J(\{a; b_1, \ldots, b_{2^d-1}\})$ is an orthonormal basis for $\dlp{2}$ for every $J\in \N$. Therefore, the performance of a multilevel discrete framelet transform completely depends on its underlying discrete affine systems.

To discuss the key features of $\dtcwt$ using discrete affine systems, in the following let us first recall one-dimensional orthogonal wavelet filter bank $\{a; b\}$, which is simply a tight framelet filter bank with only one high-pass filter. More explicitly, for filters $a,b\in \lp{1}$, $\{a; b\}$ is an orthogonal wavelet filter bank if
\be \label{owfb}
\left [ \begin{matrix} \wh{a}(\xi) &\wh{b}(\xi)\\ \wh{a}(\xi+\pi) &\wh{b}(\xi+\pi)\end{matrix}\right]
\left[ \begin{matrix} \ol{\wh{a}(\xi)} &\ol{\wh{a}(\xi+\pi)}\\
\ol{\wh{b}(\xi)} &\ol{\wh{b}(\xi+\pi)}\end{matrix} \right]=
\left[ \begin{matrix} 1 &0\\ 0 &1\end{matrix}\right].
\ee
We often call $a$ a low-pass filter and $b$ a high-pass filter.
It is easy to see that the low-pass filter $a$ in an orthogonal wavelet filter bank $\{a; b\}$ must be an orthogonal filter satisfying $|\wh{a}(\xi)|^2+|\wh{a}(\xi+\pi)|^2=1$, and the high-pass filter $b$ is almost uniquely obtained from the orthogonal low-pass filter $a$ through the relation:
\be \label{owfb:b}
\wh{b}(\xi)=e^{-i\xi} \ol{\wh{a}(\xi+\pi)}.
\ee
In this paper, we always assume that the high-pass filter $b$ in a one-dimensional orthogonal wavelet filter bank $\{a; b\}$ is always obtained from an orthogonal low-pass filter $a$ through the relation in \eqref{owfb:b}.

In the following, we discuss some key features of $\dtcwt$ using discrete affine systems. As discussed in \cite{K1999,K2001,S:ieee,SBK}, a $\dtcwt$ employs three sets of real-valued orthogonal wavelet filter banks: $\{a^0; b^0\}$ and a correlated pair $\{a^1; b^1\}$ and $\{a^2, b^2\}$.
The initial filter bank $\{a^0; b^0\}$ can be any real-valued orthogonal wavelet filter bank and is used for the first level/stage in the dual tree complex wavelet transform.
The pair of correlated real-valued orthogonal wavelet filter banks $\{a^1; b^1\}$ and $\{a^2; b^2\}$ are linked to each other through the half-shift condition (see \cite{K1999,K2001,S:ieee,SBK}):
\be \label{halfshift}
\wh{a^2}(\xi)\approx e^{i\theta(\xi)} \wh{a^1}(\xi) \qquad \mbox{with} \qquad
\theta(\xi):=-\xi/2+ \pi \lfloor \tfrac{\xi+\pi}{2\pi}\rfloor, \qquad \xi\in \R,
\ee
where $\lfloor \cdot \rfloor$ is the floor function such that $\lfloor x\rfloor=n$ for $n\le x<n+1$ with $n$ being an integer. Note that $e^{i\theta(\xi)}$ is $2\pi$-periodic and the phase function $\theta(\xi)=-\xi/2$ for $\xi\in [-\pi,\pi)$, which corresponds to (approximate) half-shift in the discrete time domain $\Z$, that is, the half-shift condition is equivalent to saying that $a^2\approx a^1(\cdot-1/2)$, which should be interpreted properly since both filters $a^1$ and $a^2$ are defined only on $\Z$.
The pair of correlated real-valued orthogonal wavelet filter banks $\{a^1; b^1\}$ and $\{a^2; b^2\}$ is used for all other levels/stages except the first level in the dual tree complex wavelet transform. The half-shift condition in \eqref{halfshift} induces relations between the high-pass filters $b^1$ and $b^2$. Indeed,
\be \label{rel:b12}
\wh{b^2}(\xi)=e^{-i\xi}\ol{\wh{a^2}(\xi+\pi)}\approx e^{-i\xi} \ol{\wh{a^1}(\xi+\pi)} e^{-i\theta(\xi+\pi)}=\wh{b^1}(\xi) e^{i(\xi+\pi)/2} e^{-i\pi \lfloor \frac{\xi+2\pi}{2\pi}\rfloor}=-i e^{i\xi/2} \wh{b^1}(\xi) e^{-i\pi \lfloor \frac{\xi}{2\pi}\rfloor}.
\ee
Since $e^{-i\pi \lfloor \frac{\xi}{2\pi}\rfloor}=-1$ for $\xi\in [-\pi,0)$ and
$e^{-i\pi \lfloor \frac{\xi}{2\pi}\rfloor}=1$ for $\xi\in [0, \pi)$, on the basic frequency interval $[-\pi,\pi)$, we have
\be \label{rel:b12:basic}
\wh{b^2}(\xi)\approx -i \sgn (\xi) e^{i\xi/2} \wh{b^1}(\xi), \qquad \xi\in [-\pi, \pi),
\ee
where
\[
\sgn(\xi)=\begin{cases}
-1 &\text{if $\xi<0$,}\\
1 &\text{if $\xi \ge0$.}
\end{cases}
\]
In other words, the high-pass filters $b^1$ and $b^2$ are linked through a sort of Hilbert transform in \eqref{rel:b12:basic}, which plays a critical role to produce directionality in high dimensions (\cite{K1999,K2001,S:ieee,SBK}).

The one-dimensional $\dtcwt$ employs two trees of the standard discrete orthogonal wavelet transform. The first tree uses the real-valued orthogonal wavelet filter bank $\{a^0; b^0\}$ for the first level and uses the real-valued orthogonal wavelet filter bank $\{a^1; b^1\}$ for the rest of the levels (that is, for the second and higher levels). The second tree uses the real-valued orthogonal wavelet filter bank $\{a^0(\cdot-1); b^0(\cdot-1)\}$ for the first level and uses the real-valued orthogonal wavelet filter bank $\{a^2; b^2\}$ for the rest of the levels. Then the corresponding high-pass wavelet coefficients from these two trees are mixed together pairwise by forming complex wavelet coefficients through averages and differences.
For an excellent detailed explanation on dual tree complex wavelet transform, see the tutorial article \cite{SBK} and many references therein.

We now explain the one-dimensional $\dtcwt$ from the viewpoint of discrete affine systems. Since the first level of $\dtcwt$ uses two orthogonal wavelet filter banks $\{a^0; b^0\}$ and $\{a^0(\cdot-1); b^0(\cdot-1)\}$, putting them together, we have a tight framelet filter bank $2^{-1/2} \{a^0, a^0(\cdot-1); b^0, b^{0}(\cdot-1)\}$, which is exactly the underlying tight framelet filter bank for a one-dimensional undecimated wavelet transform using the orthogonal wavelet filter bank $\{a^0; b^0\}$. The first level of $\dtcwt$ further changes the two high-pass filters $b^0$ and $b^0(\cdot-1)$ by taking averages and differences to form complex-valued high-pass filters as follows:
\be \label{b1pn}
b^{p}_1:=[b^0+i b^0(\cdot-1)]/\sqrt{2} \quad \mbox{and}\quad b^{n}_1:=[b^0-i b^0(\cdot-1)]/\sqrt{2},
\ee
where the superscripts $p$ and $n$ refer to positive and negative in the frequency domain.
It is trivial to directly check that $2^{-1/2} \{a^0, a^0(\cdot-1); b^{p}_1, b^{n}_1\}$ is indeed a tight framelet filter bank. Define
\be \label{a12}
a^1_1:=a^0, \quad a^2_1:=a^0(\cdot-1).
\ee
Then the first level of $\dtcwt$ uses in fact the tight framelet filter bank $2^{-1/2}\{a^1_1, a^2_1; b^{p}_1, b^{n}_1\}$, which has two real-valued low-pass filters $a^1_1, a^2_1$ and two complex-valued high-pass filters $b^{p}_1, b^{n}_1$.
The $1$-level underlying discrete affine system is simply $\DAS_1(2^{-1/2}\{a^1_1, a^2_1; b^{p}_1, b^{n}_1\})$, more explicitly,
\[
\DAS_1(a^0, a^1, a^2 \mid \dtcwt):=\{
a^1_1(\cdot-2k), a^2_1(\cdot-2k) \setsp k\in \Z\} \cup \{ b^{p}_1(\cdot-2k), b^{n}_1(\cdot-2k) \setsp k\in \Z\}.
\]
We now look at $\dtcwt$ for higher levels $J\ge 2$. At level $J$, we use two real-valued orthogonal wavelet filter banks $\{a^1; b^1\}$ and $\{a^2; b^2\}$.
The underlying $J$-level discrete affine system for $J\ge 2$ is
\[
\DAS_J(a^0, a^1, a^2 \mid \dtcwt):=\{a^1_J(\cdot-2^Jk), a^2_J(\cdot-2^J k) \setsp k\in \Z\}
\cup \cup_{j=1}^J \{ b^p_j(\cdot-2^j k), b^n_j(\cdot-2^j k) \setsp k\in \Z\},
\]
where the multilevel filters $a^1_j, a^2_j, b^p_j, b^n_j$ for $j\ge 2$ are defined to be
\begin{align}
&\wh{a^1_j}(\xi):=2^{(j-1)/2}\wh{a^0}(\xi)\wh{a^1}(2\xi) \cdots \wh{a^1}(2^{j-2}\xi) \wh{a^1}(2^{j-1}\xi), \label{aj1}\\
&\wh{a_j^2}(\xi):=2^{(j-1)/2}\wh{a^0(\cdot-1)}(\xi)\wh{a^2}(2\xi) \cdots \wh{a^2}(2^{j-2}\xi) \wh{a^2}(2^{j-1}\xi),\label{aj2}\\
&b_j^p:=[b^1_j+ib^2_j]/\sqrt{2}, \qquad b_j^n:=[b^1_j-i b^2_j]/\sqrt{2}  \label{bjpn}
\end{align}
with
\[
\wh{b^1_j}(\xi):=\wh{a^1_{j-1}}(\xi) \wh{b^1}(2^{j-1}\xi),
\quad
\wh{b^2_j}(\xi):=\wh{a^2_{j-1}}(\xi)\wh{b^2}(2^{j-1}\xi).
\]
In other words,
\be \label{das:dtcwt}
\DAS_J(a^0, a^1, a^2 \mid \dtcwt)=\{a^1_{J; k}, a^2_{J;k} \setsp k\in \Z\}
\cup \cup_{j=1}^J \{ b^p_{j;k}, b^n_{j;k} \setsp k\in \Z\},
\ee
where
\[
a^1_{j;k}:=a^1_j(\cdot-2^jk), \quad a^2_{j;k}:=a^2_j(\cdot-2^jk), \quad b^p_{j;k}:=b^p_j(\cdot-2^jk), \quad b^n_{j;k}:=b^n_j(\cdot-2^jk), \qquad k\in \Z, j\in \N.
\]
Moreover, for every integer $J\in \N$, the $J$-level discrete affine system $\DAS_J(a^0, a^1, a^2 \mid \dtcwt)$ is a (normalized) tight frame for $\lp{2}$, that is,
\[
v=\sum_{k\in \Z} \Big(\la v, a^1_{J;k}\ra a^1_{J;k}+\la v, a^2_{J;k}\ra a^2_{J;k}\Big)
+\sum_{j=1}^J \sum_{k\in \Z} \Big(\la v, b^p_{j;k}\ra b^p_{j;k}+\la v, b^n_{j;k}\ra b^n_{j;k}\Big), \qquad \forall\; v\in\lp{2}
\]
with the series converging unconditionally in $\lp{2}$.
We also have the following cascade structure, on which the fast algorithm of $\dtcwt$ is based:
\[
\sum_{k\in \Z} \Big(\la v, a^1_{j-1;k}\ra a^1_{j-1;k} +\la v, a^2_{j-1; k}\ra a^2_{j-1;k}\Big)
=\sum_{k\in \Z} \Big(\la v, a^1_{j;k}\ra a^1_{j;k}+\la v, a^2_{j;k}\ra a^2_{j;k}
+\la v, b^p_{j;k}\ra b^p_{j;k} +\la v, b^n_{j;k}\ra b^n_{j;k}\Big),
\]
for all $j\in \N$ and $v\in \lp{2}$.

To understand the directionality of the $\dtcwt$, it is important to investigate the frequency separation of all the high-pass filters $b_j^p, b_j^n$ in a $J$-level discrete affine system $\DAS_J(a^0, a^1, a^2 \mid \dtcwt)$.
For every orthogonal low-pass filter $a$ satisfying $\wh{a}(0)=1$ and $\wh{a}(\pi)=0$, it follows from the identity $|\wh{a}(\xi)|^2+|\wh{a}(\xi+\pi)|^2=1$ that $\wh{a}$ largely concentrates on $[-\pi/2,\pi/2]$, in other words, we often have $|\wh{a}(\xi)|^2 \approx \chi_{[-\pi/2, \pi/2]}(\xi)$ for $\xi \in [-\pi,\pi)$.

We first study the frequency separation of $b^p_1$ and $b^n_1$ at level one. Since
$\wh{b^n_1}(\xi)=\ol{\wh{b^p_1}(-\xi)}$ (that is, $b_1^n=\ol{b_1^p}$) by \eqref{b1pn}, it suffices for us to look at the filter $b^p_1$.
Since $\wh{b^p_1}(\xi)=\tfrac{\sqrt{2}}{2}(1+ie^{-i\xi}) \wh{b^0}(\xi)$ and $\wh{b^0}(\xi)=e^{-i\xi} \ol{\wh{a^0}(\xi+\pi)}$, for $\xi\in [-\pi, \pi)$, noting
$|1+ie^{-i\xi}|^2=2+2\sin \xi$, we have
\[
|\wh{b^p_1}(\xi)|^2=(1+\sin \xi) |\wh{a^0}(\xi+\pi)|^2
\approx (1+\sin \xi)\chi_{[-\pi, -\pi/2]}(\xi)+(1+\sin \xi)\chi_{[\pi/2,\pi]}(\xi).
\]
Note that
\be \label{fs:sin}
0\le \sqrt{1+\sin \xi} \le 1 \quad \mbox{(small) on}\; [-\pi, -\pi/2],\qquad
1\le \sqrt{1+\sin \xi}\le \sqrt{2}\quad \mbox{(large) on}\;  [\pi/2, \pi].
\ee
Therefore,
$\wh{b^p_1}$ concentrates more or less on the positive interval $[\pi/2, \pi)\subseteq [0,\pi)$ while $\wh{b^p_1}$ is relatively small on the negative interval $[-\pi, 0]$. Consequently, by $\wh{b^n_1}(\xi)=\ol{\wh{b^p_1}(-\xi)}$, $\wh{b^n_1}$ concentrates more or less on the negative interval $[-\pi,0]$ and $\wh{b^n_1}$ is relatively small on the positive interval $[0,\pi)$.

As noticed in \cite[page 136]{SBK}, the high-pass filters $b^p_1, b^n_1$ for the first level of $\dtcwt$ do not have nearly ideal frequency separation. Ideally, we prefer that $\wh{b^p_1}$ vanishes on $[-\pi, 0]$ so that $\wh{b^p_1}$ concentrates largely on the positive interval $[0,\pi]$, while $\wh{b^n_1}$ vanishes on $[0,\pi]$ so that $\wh{b^n_1}$ concentrates largely on the negative interval $[-\pi,0]$. Hence, a natural quantity to measure frequency separation of $b^p_1$ and $b^n_1$ is $|\wh{b^p_1}(\xi+\pi)|^2+|\wh{b^n_1}(\xi)|^2$ for $\xi\in [0,\pi]$ (the smaller the quantity, the better the frequency separation). However, we always have the following identities
\be \label{directional:initial}
|\wh{b^p_1}(\xi+\pi)|^2+|\wh{b^n_1}(\xi)|^2=1-\sin \xi, \quad \xi\in [-\pi, \pi] \quad \mbox{and}\quad
\int_0^\pi \big[|\wh{b^p_1}(\xi+\pi)|^2+|\wh{b^n_1}(\xi)|^2\big] d\xi=\pi-2.
\ee
Indeed, it is easy to directly check that
\[
|\wh{b^p_1}(\xi+\pi)|^2=(1-\sin \xi) |\wh{a^0}(\xi)|^2, \qquad
|\wh{b^n_1}(\xi)|^2=(1-\sin \xi) |\wh{a^0}(\xi+\pi)|^2.
\]
Now the identity in \eqref{directional:initial} follows directly from the above identities and the fact that $|\wh{a^0}(\xi)|^2+|\wh{a^0}(\xi+\pi)|^2=1$.
Therefore,
%
%
\eqref{directional:initial} implies that it is impossible to achieve $|\wh{b^p_1}(\xi+\pi)|^2+|\wh{b^n_1}(\xi)|^2 \approx 0$ for all $\xi\in [0,\pi]$, regardless of the choice of the initial real-valued orthogonal wavelet filter bank $\{a^0; b^0\}$.

We now study the frequency separation of $b^p_j$ and $b^n_j$ for $j\ge 2$.
By the half-shift condition in \eqref{halfshift} and the definition of $a_j^1$ and $a^2_j$ in \eqref{aj1} and \eqref{aj2}, we have
\be \label{a21}
\wh{a^2_j}(2^{-j}\xi) \approx \wh{a^1_j}(2^{-j}\xi) e^{-i 2^{-j} \xi} e^{i\sum_{\ell=1}^{j-1} \theta(2^{-\ell}\xi)}.
\ee
Using terminating binary representation of a real number $\xi$, we can prove the following identity
\be \label{theta:identity}
\sum_{\ell=1}^\infty \lfloor 2^{-\ell} \xi+\tfrac{1}{2}\rfloor=\lfloor \xi\rfloor+\frac{1-\sgn(\xi)}{2}=
\begin{cases}
\lfloor \xi\rfloor &\text{if $\xi\ge 0$},\\
\lfloor \xi\rfloor+1 &\text{if $\xi<0$}.
\end{cases}
\ee
By the definition of $\theta$ in \eqref{halfshift} and the above identity, we have
\begin{align*}
\sum_{\ell=1}^\infty \theta(2^{-\ell}\xi)-\theta(\xi+\pi)
&=\sum_{\ell=1}^\infty\Big( -2^{-\ell-1}\xi+\pi \lfloor 2^{-\ell}\tfrac{\xi}{2\pi}+\tfrac{1}{2}\rfloor\Big)-\Big( -\tfrac{\xi+\pi}{2}+\pi \lfloor \tfrac{\xi+2\pi}{2\pi}\rfloor\Big)\\
&=-\frac{\pi}{2}+\pi\Big(\sum_{\ell=1}^\infty \lfloor 2^{-\ell}\tfrac{\xi}{2\pi}+\tfrac{1}{2}\rfloor- \lfloor \tfrac{\xi}{2\pi}\rfloor\Big)=-\frac{\pi}{2}\sgn(\xi)=
\begin{cases} -\tfrac{\pi}{2} &\text{if $\xi\ge 0$,}\\
\tfrac{\pi}{2} &\text{if $\xi<0$.}
\end{cases}
\end{align*}
By the above identity and noting that $\sum_{\ell=j}^\infty \theta(2^{-\ell}\xi)=-2^{-j}\xi+\sum_{\ell=j}^\infty \pi \lfloor 2^{-\ell}\tfrac{\xi}{2\pi}+\tfrac{1}{2}\rfloor$, we deduce that
\[
\sum_{\ell=1}^{j-1} \theta(2^{-\ell}\xi)=\sum_{\ell=1}^\infty \theta(2^{-\ell}\xi)-\sum_{\ell=j}^\infty \theta(2^{-\ell}\xi)=
-\tfrac{\xi}{2}+2^{-j}\xi+\pi \Big(\tfrac{1-\sgn(\xi)}{2}+\lfloor \tfrac{\xi}{2\pi}\rfloor
-\sum_{\ell=j}^\infty \lfloor 2^{-\ell}\tfrac{\xi}{2\pi}+\tfrac{1}{2}\rfloor\Big).
\]
When $\xi \in [-2^j \pi, 2^j \pi)$, we have $2^{-\ell}\tfrac{\xi}{2\pi}+\tfrac{1}{2}\in [0,1)$ for all $\ell\ge j$. Hence, it follows from the above identity that
\[
\sum_{\ell=1}^{j-1} \theta(2^{-\ell}\xi)=-\tfrac{\xi}{2}+2^{-j}\xi+\pi \Big(\tfrac{1-\sgn(\xi)}{2}+\lfloor \tfrac{\xi}{2\pi}\rfloor\Big), \qquad \xi\in [-2^j\pi, 2^j\pi).
\]
Therefore, we deduce from \eqref{a21} and the above identity that
\be \label{rel:aj12}
\wh{a^2_j}(\xi) \approx e^{-i 2^{j-1} \xi} \wh{a^1_j}(\xi)  \eta(2^{j}\xi), \qquad \forall\; \xi\in [-\pi, \pi), j\ge 2,
\ee
where
\[
\eta(\xi):=e^{i\pi \big(\tfrac{1-\sgn(\xi)}{2}+\lfloor \tfrac{\xi}{2\pi}\rfloor\big)}=
\sgn(\xi) e^{i\pi \lfloor \tfrac{\xi}{2\pi}\rfloor}.
\]
We see that $\eta(\xi)=(-1)^k$ for all $|\xi|\in [2\pi k, 2\pi (k+1))$ and $k\in \N \cup\{0\}$.
In particular, $\eta(\xi)=1$ for all $\xi\in [-2\pi, 2\pi)$.
Since $|\wh{a^1_j}(\xi)|^2 \approx 2^{j-1} \chi_{2^{-j}[-\pi, \pi)}(\xi)$ for $\xi\in [-\pi, \pi)$ and $\eta(2^j\xi)=1$ for all $\xi \in 2^{1-j}[-\pi, \pi)$, we deduce from \eqref{rel:aj12} that $\wh{a^2_j}(\xi)\approx e^{-i2^{j-1}\xi} \wh{a^1_j}(\xi)$ for all $\xi\in [-\pi, \pi)$, that is, $a^2_j\approx a^1_j(\cdot-2^{j-1})$ for $j\ge 2$.

Note that on the basic frequency interval $[-\pi, \pi)$, $|\wh{a^2_{j-1}}(\xi)|^2 \approx 2^{j-2} \chi_{2^{1-j}[-\pi,\pi]}(\xi)$. Also note that  $\eta(2^{j-1}\xi)=1$ for $\xi\in 2^{2-j}[-\pi, \pi)$. Now by \eqref{rel:b12:basic} and \eqref{rel:aj12}, for $\xi\in [-\pi, \pi)$, we have
\[
\wh{b^2_j}(\xi)=\wh{a^2_{j-1}}(\xi) \wh{b^2}(2^{j-1}\xi)\approx
e^{-i 2^{j-1} \xi}  \wh{a^1_{j-1}}(\xi) \eta(2^{j-1}\xi) (-i)  \sgn(\xi) e^{i 2^{j-1} \xi}\wh{b^1}(2^{j-1}\xi)
=-i \sgn(\xi) \wh{b^1_j}(\xi).
\]
That is, $b^1_j$ and $b^2_j$ are linked to each other through the Hilbert transform. Consequently, we have
\be \label{direction:bpJ}
\wh{b^p_j}(\xi)=[\wh{b^1_j}(\xi)+i\wh{b^2_j}(\xi)]/\sqrt{2}\approx
\wh{b^1_j}(\xi)[1+\sgn(\xi)]/\sqrt{2}=\begin{cases}
0, &\text{if $\xi\in [-\pi, 0)$,}\\
\sqrt{2} \wh{b^1_j}(\xi), &\text{if $\xi\in [0, \pi)$.}
\end{cases}
\ee
By the relation $\wh{b^n_j}(\xi)=\ol{\wh{b^p_j}(-\xi)}$, we see that $\wh{b^n_j}(\xi)\approx 0$ for $\xi \in [0, \pi)$ and $\wh{b^n_j}(\xi)\approx \sqrt{2} \wh{b^1_j}(\xi)$ for $\xi\in [-\pi, 0]$.
Therefore, $b^p_j$ and $b^n_j$ have nearly ideal frequency separation when $j\ge 2$. More precisely, $\wh{b^p_j}$ vanishes nearly on the negative interval $[-\pi, 0)$ and concentrates largely on the positive interval $[0,\pi)$, while $\wh{b^n_j}$ vanishes nearly on the positive interval $[0,\pi)$ and concentrates largely on the negative interval $[-\pi,0)$.

Though algorithmically the two-dimensional $\dtcwt$ can be implemented using tensor product of one-dimensional $\dtcwt$, due to the mixing and pairing of the corresponding high-pass wavelet coefficients to form complex wavelet coefficients after the tensor product wavelet transform, the resulting discrete affine systems for two-dimensional $\dtcwt$ are not tensor product of discrete affine systems for one-dimensional $\dtcwt$, more precisely, they are not obtained by $\{a^1_j, a^2_j; b^p_j, b^n_j\} \otimes \{a^1_j, a^2_j; b^p_j, b^n_j\}$. In fact, to achieve better directionality, there is a further frequency separation for the pair $(a^1_j, a^2_j)$ of low-pass filters by using a similar technique as in \eqref{bjpn} for the high-pass filters $b^1_j$ and $b^2_j$.
Let us explain the details in the following. Define
\be \label{ajpn}
a_j^p:=[a^1_j+ia^2_j]/\sqrt{2}, \qquad a_j^n:=[a^1_j-i a^2_j]/\sqrt{2}, \qquad j\in \N.
\ee
Then it is trivial to see that $\{a^p_{J;k}, a^n_{J;k}\setsp k\in \Z\}\cup\cup_{j=1}^J \{ b^p_{j;k}, b^n_{j,k} \setsp k\in \Z\}$ is still a tight frame for $\lp{2}$ and the following identity holds:
\be \label{twobases}
\sum_{k\in \Z} \Big( \la v, a^1_{J;k}\ra a^1_{J;k}+ \la v, a^2_{J;k}\ra a^2_{J;k}\Big)=
\sum_{k\in \Z} \Big( \la v, a^p_{J;k}\ra a^p_{J;k}+ \la v, a^n_{J;k}\ra a^n_{J;k}\Big), \qquad \forall\; v\in \lp{2}, J\in \N.
\ee
In the $J$-level discrete affine system for two-dimensional $\dtcwt$,
its low-pass part has $4$ real-valued low-pass filters and is obtained from the low-pass part in the tensor product $\{a^1_J, a^2_J; b^p_J, b^n_J\}\otimes \{a^1_J, a^2_J; b^p_J, b^n_J\}$, that is,
\[
\mbox{LP}_J:=\{a^1_J, a^2_J\}\otimes \{a^1_J, a^2_J\}=\{a^1_J\otimes a^1_J, a^1_J\otimes a^2_J, a^2_J\otimes a^1_J, a^2_J\otimes a^2_J\}
\]
and its high-pass part has $12$ complex-valued high-pass filters in total and is taken from the high-pass part in the tensor product $\{a^p_j, a^n_j; b^p_j, b^n_j\}\otimes \{a^p_j, a^n_j; b^p_j, b^n_j\}$, that is,
\[
\mbox{HP}_j:=\{a^p_j\otimes b^p_j, a^p_j\otimes b^n_j, a^n_j\otimes b^p_j, a^n_j\otimes b^n_j,
b^p_j\otimes a^p_j, b^p_j\otimes a^n_j, b^p_j\otimes b^p_j, b^p_j\otimes b^n_j,
b^n_j\otimes a^p_j, b^n_j\otimes a^n_j, b^n_j\otimes b^p_j, b^n_j\otimes b^n_j\}.
\]
Now the $J$-level discrete affine system for two-dimensional $\dtcwt$ with complex-valued high-pass filters is
\[
\DAS_J(a^0, a^1, a^2\, |\, 2\D\; \dtcwt):=\{ u(\cdot-2^J k) \setsp u\in \mbox{LP}_J, k\in \Z^2\}
\cup \cup_{j=1}^J \{ v(\cdot-2^j k) \setsp v\in \mbox{HP}_j, k\in \Z^2\}.
\]

At the level one, on $[-\pi,\pi)$, we have
\be \label{a1p}
\wh{a^p_1}(\xi)=[\wh{a^1_1}(\xi)+i \wh{a^2_1}(\xi)]/\sqrt{2}=
\wh{a^0}(\xi)(1+ie^{-i\xi})/\sqrt{2} \quad \mbox{and}\quad
|\wh{a^p_1}(\xi)|=|\wh{a^0}(\xi)|\sqrt{1+\sin \xi}.
\ee
By \eqref{fs:sin}, we see that $\wh{a^p_1}$ concentrates more or less on the positive interval $[\pi/2, \pi)\subseteq [0,\pi)$ while $\wh{a^p_1}$ is relatively small on the negative interval $[-\pi,0]$. By $\wh{a^n_1}(\xi)=\ol{\wh{a^p_1}(-\xi)}$, we see that $\wh{a^n_1}$ concentrates more or less on the negative interval $[\pi,0]$ while $\wh{a^n_1}$ is relatively small on the positive interval $[0,\pi)$.

By the relation in \eqref{rel:aj12} for $j\ge 2$, we have
\[
\wh{a^p_j}(\xi)=[\wh{a^1_j}(\xi)+i\wh{a^2_j}(\xi)]/\sqrt{2}\approx \wh{a^1_j}(\xi)(1+i e^{-i 2^{j-1}\xi})\eta(2^j\xi)/\sqrt{2}.
\]
Since $|\wh{a^1_j}(\xi)|^2\approx 2^{j-1}\chi_{2^{-j}[-\pi,\pi)}(\xi)$ for $\xi\in [-\pi, \pi)$ and $\eta(2^j\xi)=1$ for $\xi\in 2^{1-j}[-\pi,\pi)$, we conclude that
\be \label{direction:apj}
|\wh{a^p_j}(\xi)|=\sqrt{1+\sin (2^{j-1}\xi)} |\wh{a^1_j}(\xi)| \approx 2^{(j-1)/2} \sqrt{1+\sin (2^{j-1}\xi)} \chi_{2^{-j}[-\pi,\pi)}(\xi), \qquad \xi\in [-\pi,\pi).
\ee
Note that $0\le \sqrt{1+\sin (2^{j-1}\xi)}\le 1$ (small) for $\xi \in 2^{-j}[-\pi, 0]$ and
$1\le \sqrt{1+\sin (2^{j-1}\xi)}\le \sqrt{2}$ (large) for $\xi \in 2^{-j}[0, \pi)$. Therefore, on the basic frequency interval $[-\pi, \pi)$, $\wh{a^p_j}$ concentrates largely inside $[0, 2^{-j}\pi)\subseteq [0,\pi)$ and $\wh{a^p_j}$ is relatively small on $[-\pi,0)$,
while $\wh{a^n_j}$ concentrates largely inside $[-2^{-j}\pi, 0]\subseteq [-\pi,0]$ and $\wh{a^n_j}$ is relatively small on $[0,\pi)$.

For a sequence $u: \dZ \rightarrow \C$, we can write $u=u^{[r]}+i u^{[i]}$ with both $u^{[r]}$ and $u^{[i]}$ being real-valued filters, that is, $u^{[r]}$ and $u^{[i]}$ are the real and imaginary parts of the filter $u$.
Due to the relation $\wh{a^n_j}(\xi)=\ol{\wh{a^p_j}(-\xi)}$ and $\wh{b^n_j}(\xi)=\ol{\wh{b^p_j}(-\xi)}$, we have $a^n_j=\ol{a^p_j}$ and $b^n_j=\ol{b^p_j}$. Hence,
there are essentially $12$ real-valued high-pass filters in $\mbox{HP}_j$ having the following directions:
\begin{enumerate}
\item The real and imaginary parts of $b^p_j\otimes a^p_j$ (or $b^n_j\otimes a^n_j$) have direction along $15^\circ$;

\item The real and imaginary parts of  $b^p_j\otimes a^n_j$ (or $b^n_j\otimes a^p_j$) have direction along $-15^\circ$;

\item The real and imaginary parts of $a^p_j\otimes b^p_j$ (or $a^n_j\otimes b^n_j$) have direction along $75^\circ$;

\item The real and imaginary parts of  $a^p_j\otimes b^n_j$ (or $a^n_j\otimes b^p_j$) have direction along $-75^\circ$;

\item The real and imaginary parts of  $b^p_j\otimes b^p_j$ (or $b^n_j\otimes b^n_j$) have direction along $45^\circ$;

\item The real and imaginary parts of  $b^p_j\otimes b^n_j$ (or $b^n_j\otimes b^p_j$) have direction along $-45^\circ$.
\end{enumerate}

From the above discussion, we see that for level $j\ge 2$, the two-dimensional $\dtcwt$ has strong  directions along $\pm 45^\circ$ due to the nearly ideal frequency separation in \eqref{direction:bpJ}; while the directions along $\pm 15^\circ$ and $\pm 75^\circ$ are not that strong or ideal, due to the weak frequency separation in \eqref{direction:apj}.
For the initial level $j=1$,  the two-dimensional $\dtcwt$ has weak directions along all $\pm 15^\circ, \pm 45^\circ$ and $\pm 75^\circ$.

\section{Image Denoising by $\dtcwt$ Using Frequency-based Filter Banks}

In this section, we shall look at various filter banks used in $\dtcwt$ and then compare their performance for the problem of image denoising.
On one hand, finitely supported filter banks are of importance and interest in many applications, due to their computational efficiency and good space/time localization. On the other hand, it is easy to design filter banks in the frequency domain to satisfy \eqref{tffb:pr:1} and \eqref{tffb:pr:0} for constructing tight framelet filter banks. Moreover, the frequency separation and frequency localization of the elements in discrete affine systems are two critical ingredients for the impressive performance of a discrete framelet/wavelet transform in many applications.
For application of $\dtcwt$ to image denoising, we shall see in this section that $\dtcwt$, employing a pair of frequency-based correlated orthogonal wavelet filter banks, performs equally well as the original $\dtcwt$ employing a pair of finitely supported correlated
orthogonal wavelet filter banks proposed and commonly used in \cite{K1999,K2001,SBK,SS:bs,SS:bslocal} and references therein.

We first recall the finitely supported filter banks which have been commonly used in $\dtcwt$ and have been designed by Kingsbury and Selesnick in \cite{K1999,SBK,SS:bs,SS:bslocal}.

By $\lp{0}$ we denote all finitely supported sequences on $\Z$. Since $e^{i\theta(\xi)}$ in \eqref{halfshift} is not a $2\pi$-periodic trigonometric polynomial, if both $a^1$ and $a^2$ are finitely supported filters from $\lp{0}$, then the half-shift condition in \eqref{halfshift} can be only approximately satisfied. Many finitely supported pairs $\{a^1; b^1\}$ and $\{a^2; b^2\}$ approximately satisfying \eqref{halfshift} have been constructed in \cite{K1999,K2001,S:ieee,SBK} and references therein. Here we only list one pair which has been implemented and frequently used for the purpose of image denoising by the research groups of Kingsbury and Selesnick. Such filters are posted at {\tt http://eeweb.poly.edu/iselesni/WaveletSoftware/}.
The initial filter $a^0$ is given by
\be \label{dtcwt:a0}
\begin{split}
a^0&=\{-\tfrac{1}{16},\tfrac{1}{16}, \tfrac{4+\sqrt{15}}{16}, \tfrac{4+\sqrt{15}}{16}, \tfrac{1}{16},-\tfrac{1}{16},\tfrac{4-\sqrt{15}}{16},\tfrac{4-\sqrt{15}}{16}\}_{[-3,4]}\\
&\approx 2^{-1/2}\{-0.08838834764832, 0.08838834764832, 0.695879989034, 0.695879989034,\\ &\qquad\qquad 0.08838834764832,
 -0.08838834764832, 0.01122679215254, 0.01122679215254\}_{[-3,4]}.
\end{split}
\ee
The correlated pair $(a^1, a^2)$ of real-valued orthogonal filters has been constructed by Kingsbury \cite{K1999} as follows:
\begin{align}
&a^1=2^{-1/2}\{
0.03516384,0,-0.08832942,0.23389032,0.76027237,0.5875183,0,-0.11430184\}_{[-4,3]}, \label{dtcwt:a1}\\
&\wh{a^2}(\xi):=e^{-i\xi}\ol{\wh{a^1}(\xi)}.\label{dtcwt:a2}
\end{align}

\begin{figure}[ht]
\centering
\subfigure[Filter $a^0$]{
\includegraphics[width=1.5in,height=1.0in]
{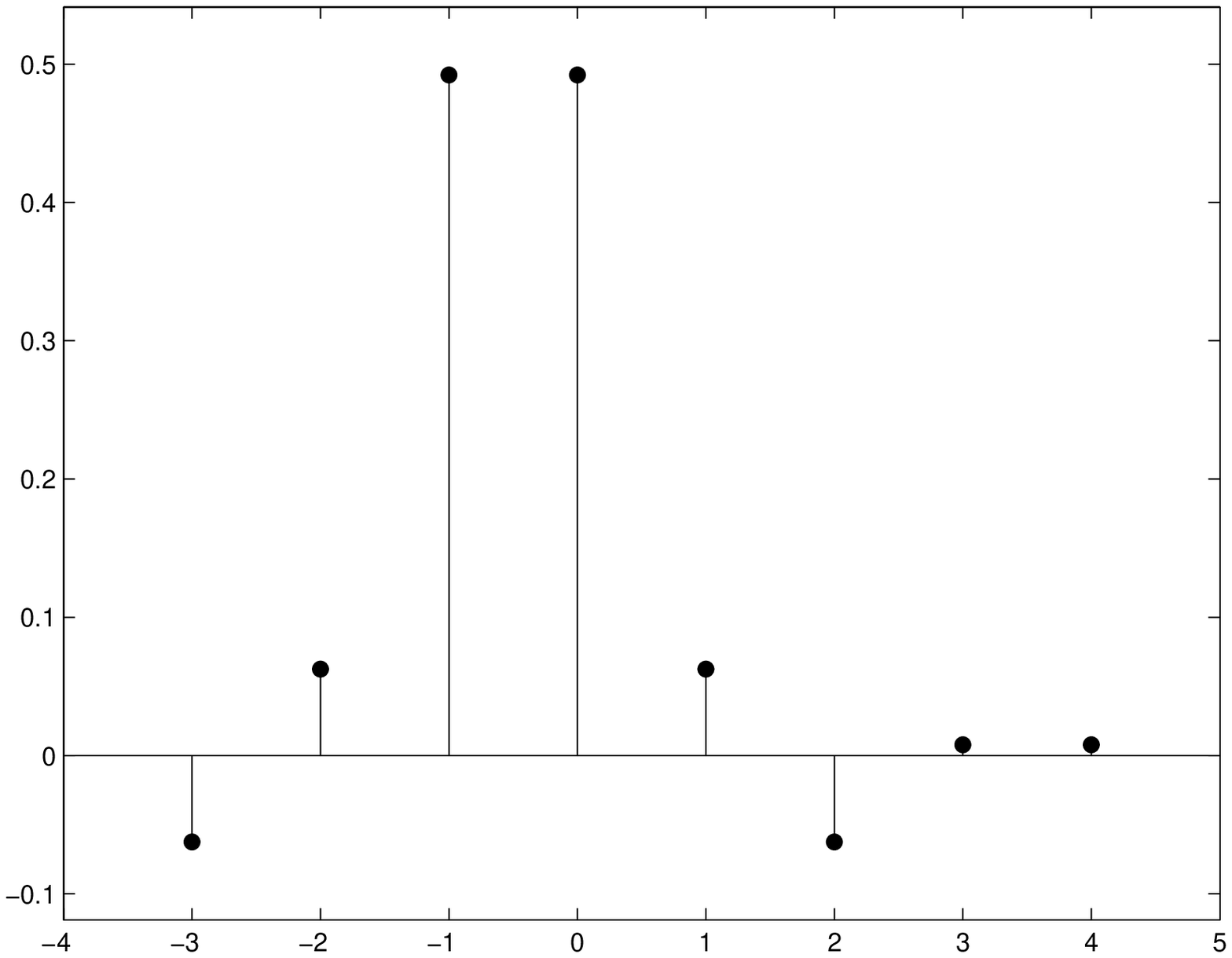}}
\subfigure[$\wh{a^0}$]{
\includegraphics[width=1.5in,height=1.0in]
{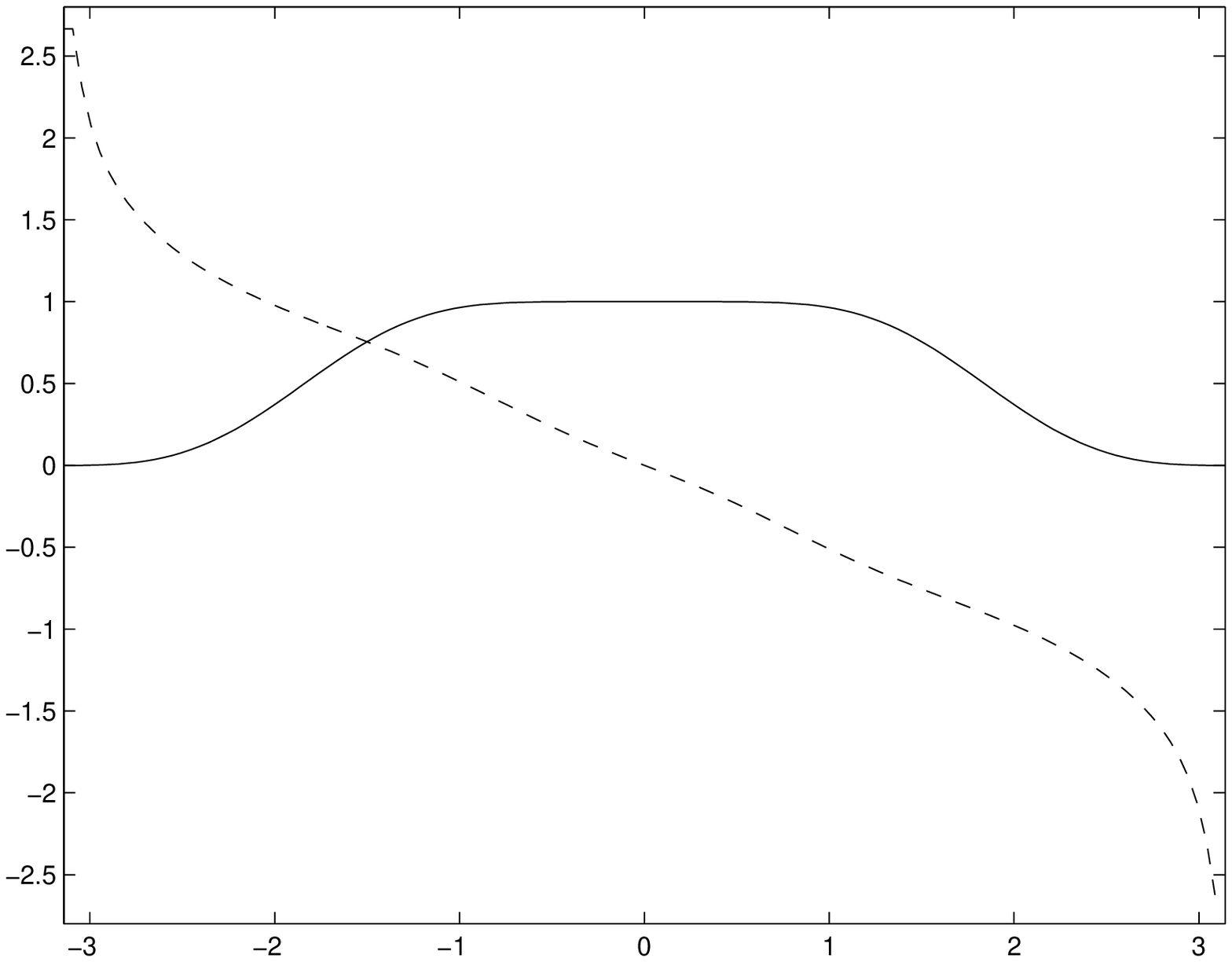}}
\subfigure[$\phi^{a^0}$]{
\includegraphics[width=1.5in,height=1.0in]
{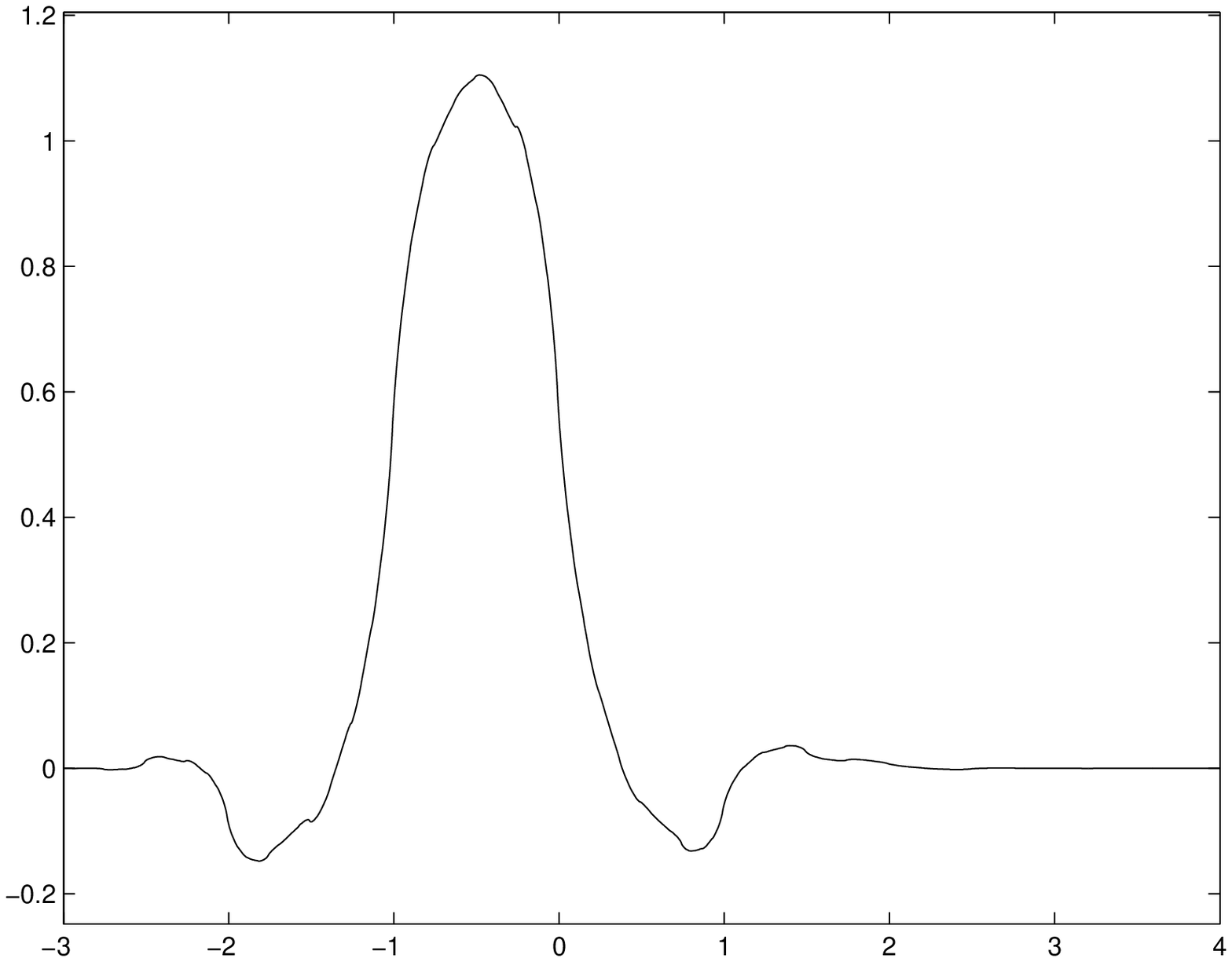}}
\subfigure[$\psi^{a^0}$]{
\includegraphics[width=1.5in,height=1.0in]
{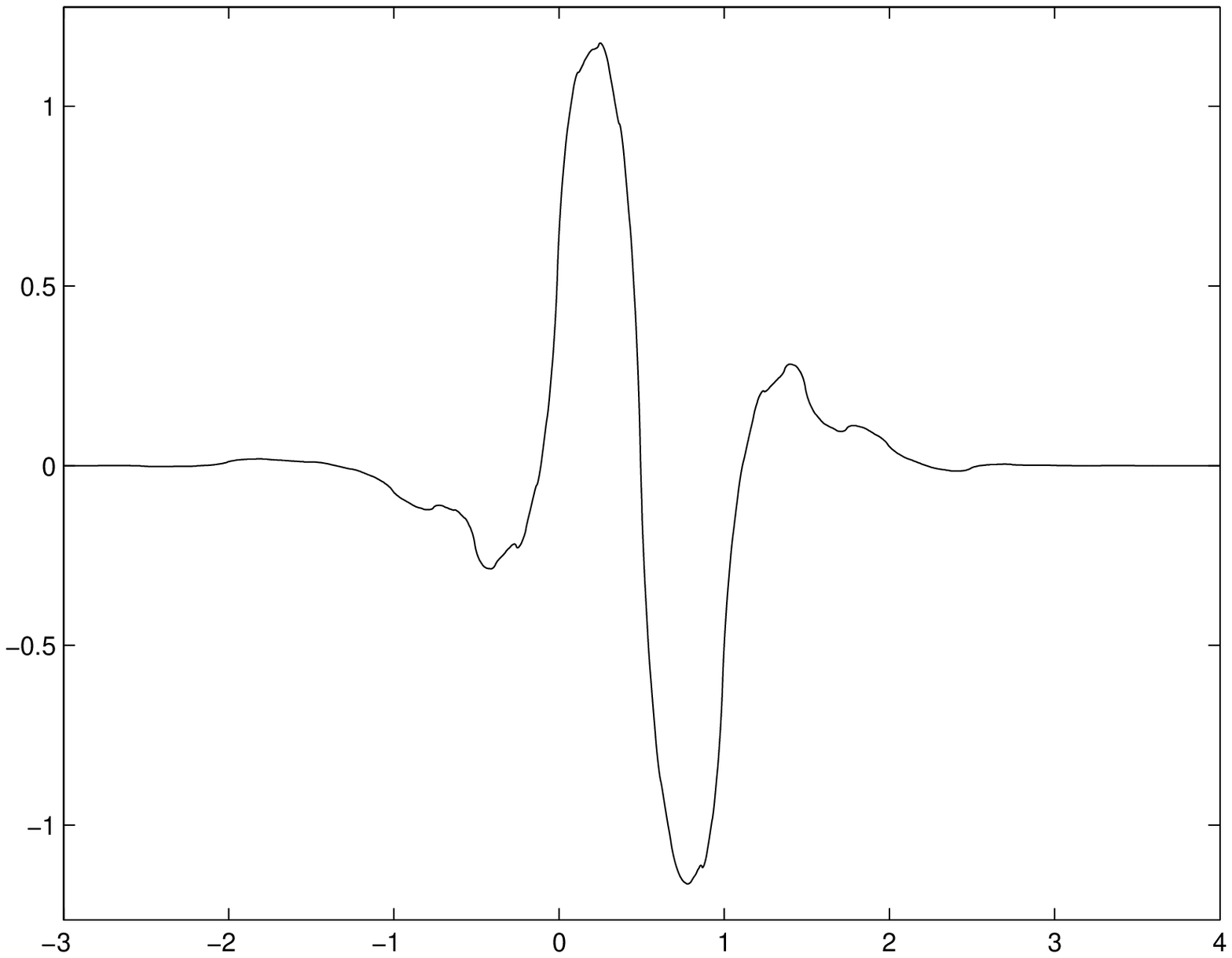}}\\
\subfigure[Filter $a^1$]{
\includegraphics[width=1.5in,height=1.0in]
{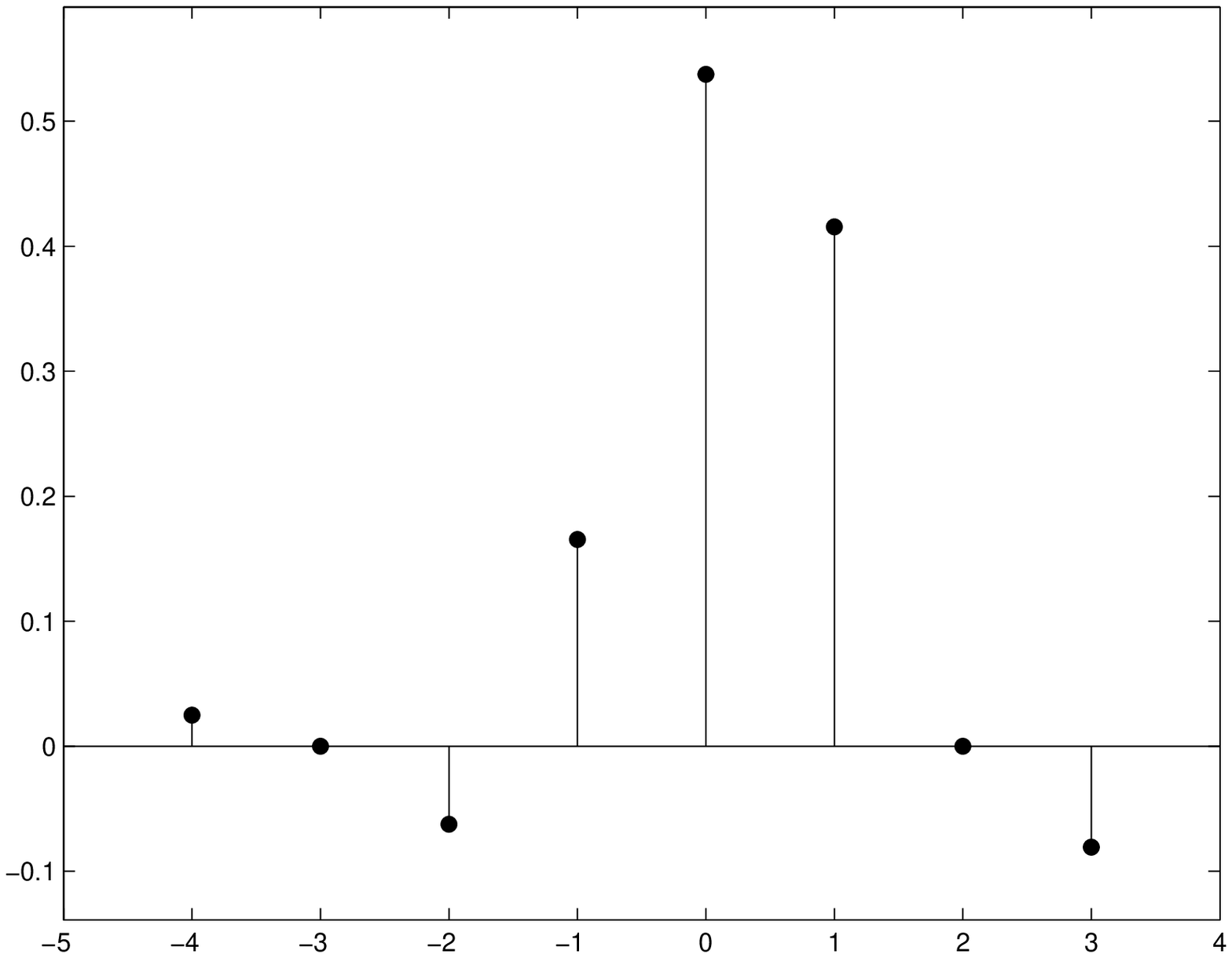}}
\subfigure[$\wh{a^1}$]{
\includegraphics[width=1.5in,height=1.0in]
{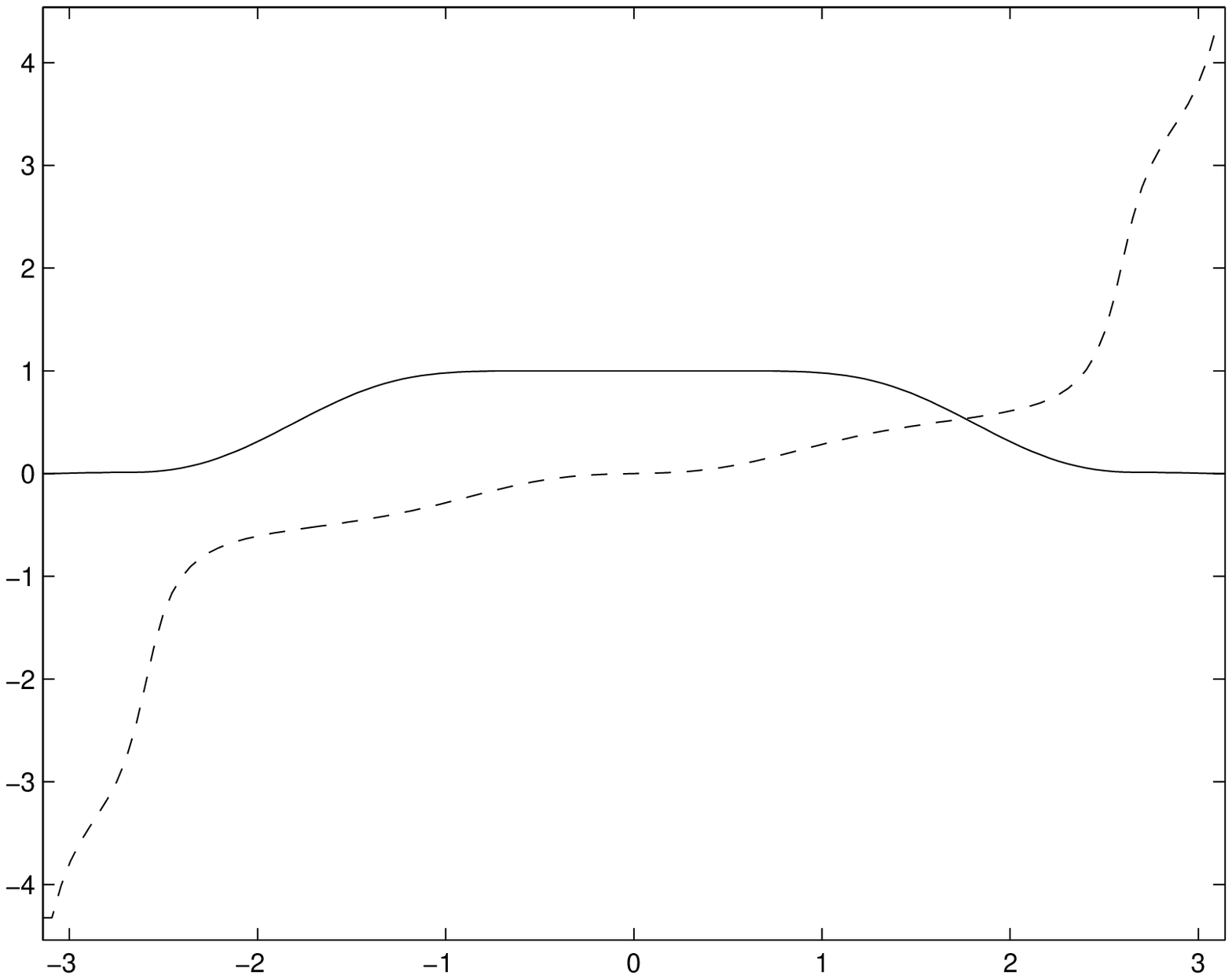}}
\subfigure[$\phi^{a^1}$]{
\includegraphics[width=1.5in,height=1.0in]
{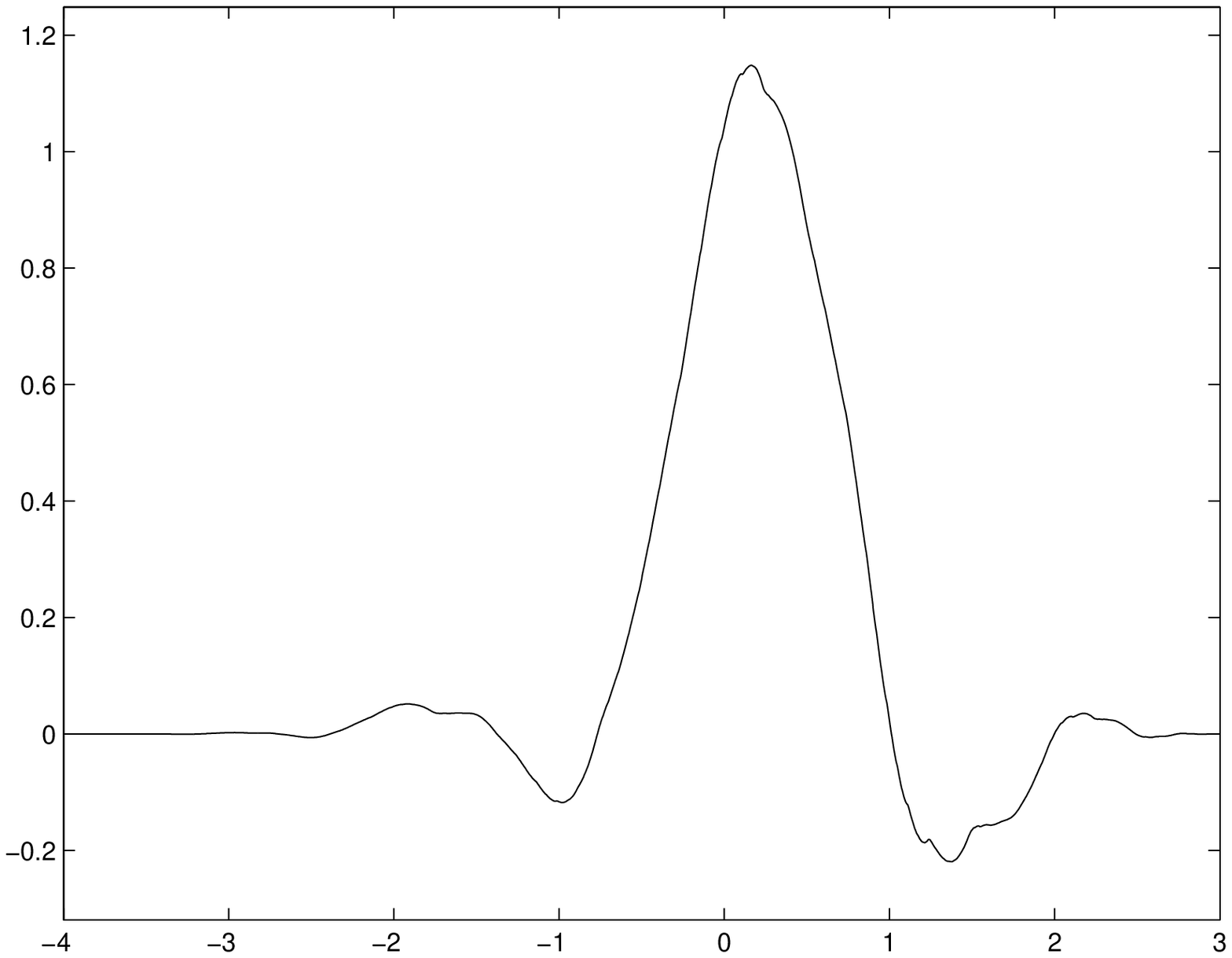}}
\subfigure[$\psi^{a^1}$]{
\includegraphics[width=1.5in,height=1.2in]
{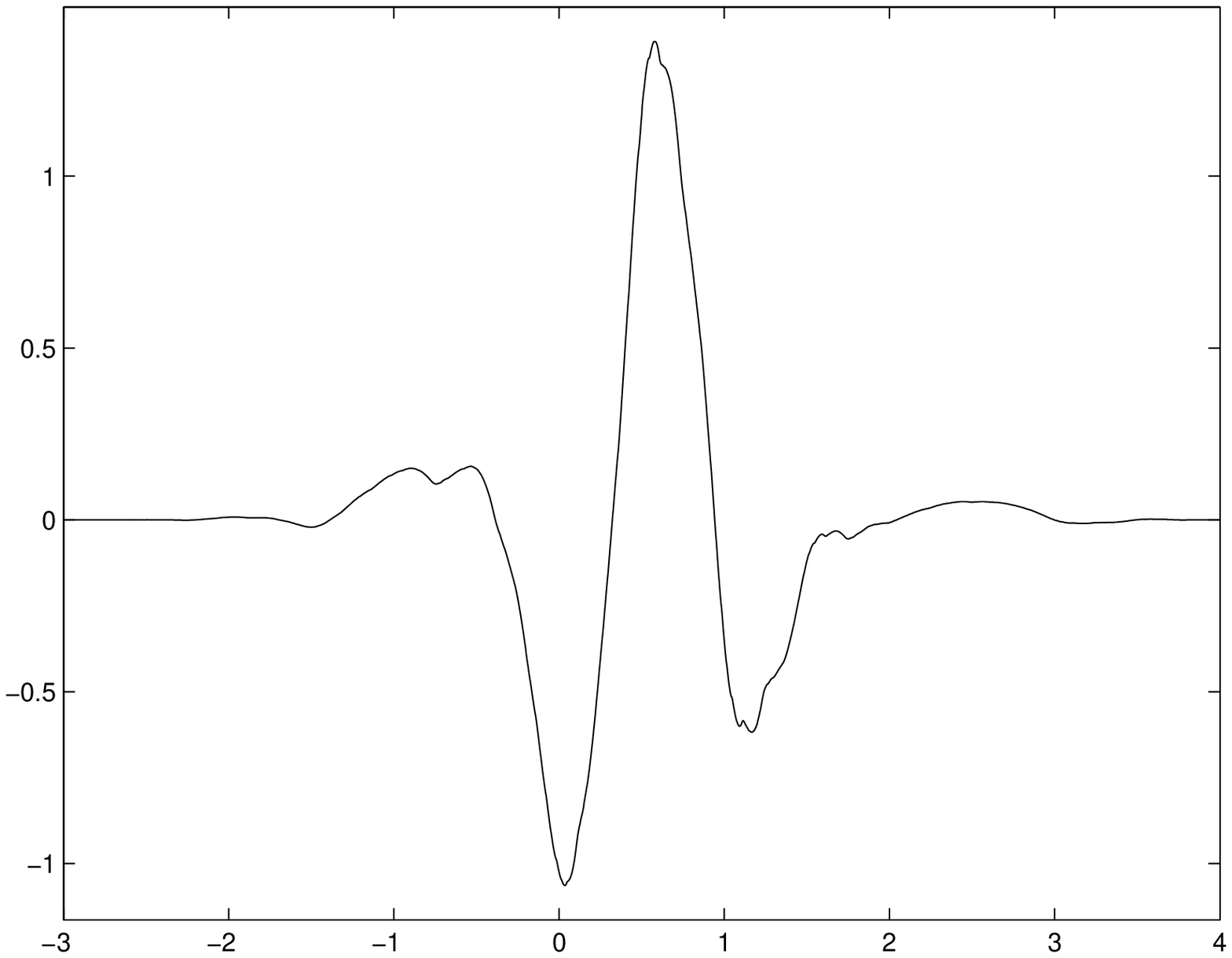}}
\begin{caption}{
Filter, magnitude and phase, refinable function, and wavelet function associated with $a^0$ and $a^1$. The solid lines in (b) and (f) represent the magnitude and the dotted lines in (b) and (f) refer to the phase of the filters in the frequency domain.
} \label{dtcwt}
\end{caption}
\end{figure}

To analyze some properties of the orthogonal low-pass filters $a^0$ and $a^1$, we now recall some definitions and notation. For a finitely supported filter $a: \Z \rightarrow \C$, we define its sum rule order to be $\sr(a):=m$, where $m$ is the largest nonnegative integer such that $\wh{a}(\xi)=(1+e^{-i\xi})^m \wh{u}(\xi)$ for some $2\pi$-periodic trigonometric polynomial $\wh{u}$, in other words, $\wh{a}(\xi+\pi)=\bo(|\xi|^m)$ as $\xi\to 0$. The \emph{smoothness exponent} of the low-pass filter $a$ is defined to be
\be \label{sm:a}
\sm(a):=-1/2-\log_2 \sqrt{\rho(u)},
\ee
where $\rho(u)$ denotes the spectral radius--the largest of the modulus of all the eigenvalues--of the square matrix $(v(2j-k))_{-K\le j,k\le K}$, where $v$ is determined by $\sum_{k=-K}^K v(k) e^{-ik\xi}:=|\wh{u}(\xi)|^2$. The larger the quantity $\sm(a)$, the smoother its associated refinable function $\phi^a$, which is defined to be $\wh{\phi^a}(\xi):=\prod_{j=1}^\infty \wh{a}(2^{-j}\xi)$.
For a finitely supported high-pass filter $b$, we define its vanishing moment order to be $\vmo(b):=n$, where $n$ is the largest integer such that $\wh{b}(\xi)=(1-e^{-i\xi})^n\wh{v}(\xi)$ for some $2\pi$-periodic trigonometric polynomial, that is, $\wh{b}(\xi)=\bo(|\xi|^n)$ as $\xi\to 0$.

The filter $a^0$ in \eqref{dtcwt:a0} is a real-valued orthogonal low-pass filter with $\sr(a)=2$ and $\sm(a^0)\approx 1.509402$.
Hence, its associated high-pass filter $b^0$ has two vanishing moments by $\vmo(b^0)=2$.
Note that $a^0$ is almost symmetric about the point $-1/2$. Therefore, $\wh{a^0}(\xi)\approx |\wh{a^0}(\xi)|e^{i\xi/2}$. The matlab program, which is posted in Selesnick's web page and implements $\dtcwt$, uses the filter bank $\{(a^0)^\star; (b^0)^\star\}$ instead of $\{a^0(\cdot-1); b^0(\cdot-1)\}$
for the first level in the second tree, where $(a^0)^\star$ is the adjoint filter of $a^0$ which is defined by $\wh{(a^0)^\star}(\xi):=\ol{\wh{a^0}(\xi)}$.
This yields the same effect as using $\{a^0(\cdot-1); b^0(\cdot-1)\}$ since we still have the one-shift condition as follows:
\[
\wh{(a^0)^\star}(\xi)=\ol{\wh{a^0}(\xi)}\approx |\wh{a^0}(\xi)| e^{-i\xi/2}=
e^{-i\xi} |\wh{a^0}(\xi)| e^{i\xi/2}\approx e^{-i\xi} \wh{a^0}(\xi).
\]

The filter $a^1$ in \eqref{dtcwt:a1} is a real-valued orthogonal low-pass filter with $\sr(a^1)=1$ and $\sm(a^1)\approx 0.997590$.
Hence, the high-pass filters $b^1$ and $b^2$ have only one vanishing moment by $\vmo(b^1)=\vmo(b^2)=1$.
The filter $a^1$ is designed in such a way that it satisfies the quarter-shift condition
\be \label{quartershift}
\wh{a^1}(\xi)\approx |\wh{a^1}(\xi)|e^{-i\xi/4}, \qquad \xi\in [-\pi,\pi)
\ee
so that we have the half-shift condition in \eqref{halfshift} as follows:
\[
\wh{a^2}(\xi)=e^{-i\xi} \ol{\wh{a^1}(\xi)}\approx
e^{-i\xi} |\wh{a^1}(\xi)| e^{i\xi/4}=
e^{-i\xi/2} |\wh{a^1}(\xi)| e^{-i\xi/4} \approx e^{-i\xi/2} \wh{a^1}(\xi).
\]
See Figure~\ref{dtcwt} for several graphs associated with the orthogonal filters $a^0$ and $a^1$.

As mentioned before, the half-shift condition in \eqref{halfshift} can only be approximately satisfied if we restrict all filters to be finitely supported sequences from $\lp{0}$. However, if we are allowed to use infinitely supported filters, then we indeed can easily satisfy the half-shift condition in \eqref{halfshift} exactly. We now provide a pair of correlated orthogonal wavelet filter banks constructed in the frequency domain.
Let $\pP_{m}(x):=(1-x)^m \sum_{j=0}^{m-1} \binom{m+j-1}{j} x^j$.
Then $\pP_{m}$ satisfies
the identity $\pP_{m}(x)+\pP_{m}(1-x)=1$ (see \cite{Daub:book}).
For $c_L<c_R$ and two positive numbers $\gep_L, \gep_R$ satisfying $\gep_L+\gep_R\le c_R-c_L$, we now define a bump function $\chi_{[c_L, c_R]; \gep_L, \gep_R}$ on $\R$ by
\be \label{bump:func}
\chi_{[c_L, c_R]; \gep_L, \gep_R}(\xi):=
\begin{cases} 0, \quad &\xi\le c_L-\gep_L \; \mbox{or}\; \xi \ge c_R+\gep_R, \\
\sin\big(\tfrac{\pi}{2}\pP_{m}(\tfrac{c_L+\gep_L-\xi}{2\gep_L})\big), \quad &c_L-\gep_L<\xi<c_L+\gep_L,\\
1, \quad &c_L+\gep_L\le \xi\le c_R-\gep_R,\\
\sin\big(\tfrac{\pi}{2}\pP_{m}(\tfrac{\xi-c_R+\gep_R}{2\gep_R})\big),
\quad &c_R-\gep_R<\xi<c_R+\gep_R.
\end{cases}
\ee
%

Let $0<\gep\le \tfrac{\pi}{2}$. Define filters $a^0, a^1, a^2\in \lp{1}$ by
\be \label{a0a1a2:meyer}
\wh{a^0}(\xi):=\wh{a^1}(\xi):=\chi_{[-\tfrac{\pi}{2}, \tfrac{\pi}{2}]; \gep, \gep}(\xi),
\qquad \wh{a^2}(\xi):=e^{-i\xi/2} \wh{a^1}(\xi),
\qquad  \xi \in [-\pi, \pi).
\ee
Then all filters $a^0, a^1, a^2$ are real-valued orthogonal low-pass filters and the half-shift condition in \eqref{halfshift} is satisfied exactly. If $\gep=\frac{\pi}{6}$, then the filter $a^0$ is simply the Meyer orthogonal low-pass filter. Since all $\wh{a^0}, \wh{a^1}$ and $\wh{a^2}$ belong to $C^{m-1}(\T)$, the filters $a^0, a^1$ and $a^2$, though have infinite support, have fast decaying coefficients.
Using discrete Fourier transform, the above frequency-based filter banks can be easily implemented with the same computational complexity as the discrete Fourier transform, that is, $\bo(N\log N)$ with $N$ inputs.

Following the standard practice on image denoising,
we assume that the variance $\sigma_n$ of additive i.i.d. Gaussian noise is known in advance and all the numerical PSNR values are an average over five experiments.
The five standard test images are from {\tt http://decsai.ugr.es/$\sim$javier/denoise/test\_images/index.htm}.
All the PSNR values for dual tree complex wavelet transform ($\dtcwt$) in this paper are obtained using the matlab program posted at Selesnick's web page at {\tt http://eeweb.poly.edu/iselesni/WaveletSoftware/}. This matlab program uses the finitely supported orthogonal filters $a^0, a^1, a^2$ in \eqref{dtcwt:a0}, \eqref{dtcwt:a1}, and \eqref{dtcwt:a2}, and we assume that the variance $\sigma_n$ is known in advance. Note that we use the standard definition $\mbox{PSNR}=10 \log_{10} \frac{255^2}{\operatorname{MSE}}$ instead of
$10 \log_{10} \frac{256^2}{\operatorname{MSE}}$ used in \cite{SS:bs,SS:bslocal}, where $\operatorname{MSE}$ is the mean squared error.

As we shall see in Table~\ref{dtcwtstd} the performance on image denoising of $\dtcwt$ using the above frequency-based orthogonal filters $a^0,a^1,a^2$ in \eqref{a0a1a2:meyer} with $\gep=189/256$ and $m=1$ is comparable with $\dtcwt$ using the finitely supported orthogonal filters $a^0, a^1, a^2$ in \eqref{dtcwt:a0}, \eqref{dtcwt:a1}, and \eqref{dtcwt:a2}. Comparison results of $\dtcwt$ with other transform-based methods for image denoising have been well documented in \cite{K1999,SS:bs,SS:bslocal}.

\begin{table}[hb]
\begin{center}
\begin{tabular}{|c||c|c||c|c||c|c||c|c||c|c||}
\hline
&\multicolumn{2}{|c||}{Lena} &\multicolumn{2}{|c||}{Barbara} &\multicolumn{2}{|c||}{Boat} &\multicolumn{2}{|c||}{House} &\multicolumn{2}{|c|}{Pepper} \\  \hline
$\sigma_n$ &$\tcwt$ &$\fcwt$ &$\tcwt$ &$\fcwt$ &$\tcwt$ &$\fcwt$ &$\tcwt$ &$\fcwt$ &$\tcwt$ &$\fcwt$ \\ \hline
$5$&38.25 &38.25 &37.36 &37.44 &36.77 &36.77 &38.45 &38.41 &37.18 &37.13\\ \hline
$10$&35.19 &35.20  &33.52 &33.60 &33.21 &33.19 &34.78 &34.73 &33.40 &33.31\\ \hline
$15$&33.47 &33.46  &31.38 &31.45 &31.33 &31.29 &32.90 &32.85 &31.29 &31.19\\ \hline
$20$&32.23 &32.22  &29.87 &29.94 &30.01 &29.96 &31.63 &31.58 &29.83 &29.71\\ \hline
$25$&31.26 &31.24  &28.70 &28.78 &28.99 &28.95 &30.65 &30.59 &28.71 &28.57\\ \hline
$30$&30.47 &30.44  &27.77 &27.84 &28.18 &28.14 &29.84 &29.78 &27.80 &27.66\\ \hline
$50$&28.21 &28.18  &25.26 &25.31 &26.01 &25.98 &27.57 &27.52 &25.30 &25.18\\ \hline
\end{tabular}
\medskip
\begin{caption}{
Columns of $\tcwt$ are for PSNR values (an average over five experiments) using bivariate shrinkage in \cite{SS:bslocal} and $\dtcwt$ using finitely supported orthogonal wavelet filter banks in \eqref{dtcwt:a0}--\eqref{dtcwt:a2}. Columns of $\fcwt$ are for PSNR values using the same bivariate shrinkage and $\dtcwt$ using frequency-based orthogonal wavelet filter banks in \eqref{a0a1a2:meyer}.
}\label{dtcwtstd}
\end{caption}
\end{center}
\end{table}

\section{Image Denoising Using Directional Complex Tight Framelets}

In this section we shall first construct one-dimensional complex tight framelet filter banks with good frequency separation property. Then we shall discuss their discrete affine systems and the tensor product complex tight framelet filter banks in dimension two. Finally, we shall address the application of such directional tensor product complex tight framelets for the problem of image denoising. Detailed comparison with $\dtcwt$ for image denoising will be provided in this section.

Directional complex tight framelets have been initially introduced in \cite[Section~7]{Han:MMNP:2013}. It is the purpose of this section to fully and further develop the idea in \cite{Han:MMNP:2013} by providing a systematic study and construction of such complex tight framelets using discrete affine systems, and then compare their performance in image denoising with $\dtcwt$.

The construction of one-dimensional tensor product complex tight framelets $\tpctf_n$ (or more precisely, tensor product complex tight framelet filter banks) is divided into two parts according to the parity of $n$, which is the number of filters in the one-dimensional tight framelet filter bank $\ctf_n$.

\textbf{$\tpctf_n$ with $n=2s+1$ being an odd positive integer.}
Let $0<c_1<c_2<\cdots < c_s<c_{s+1}:=\pi$. Let $\gep_1, \ldots, \gep_s$ be positive numbers satisfying
\be \label{tpctf:condition}
0<\gep_1\le \min(c_1, \tfrac{\pi}{2}-c_1) \quad \mbox{and}\quad
(c_{\ell+1}-c_{\ell})+\gep_{\ell+1}+\gep_\ell \le \pi, \qquad \forall\; \ell=1, \ldots, s.
\ee
Define a real-valued symmetric low-pass filter $a$ by
\be \label{tpctf:odd:a}
\wh{a}:=\chi_{[-c_1, c_1]; \gep_1, \gep_1}
\ee
and define $2s$ number of complex-valued high-pass filters $b^{1,p}, \ldots, b^{s,p}, b^{1,n}, \ldots, b^{s,n}$ by
\be \label{tpctf:odd:b}
\wh{b^{\ell,p}}:=\chi_{[c_\ell, c_{\ell+1}]; \gep_\ell, \gep_{\ell+1}},\qquad
\wh{b^{\ell,n}}:=\ol{\wh{b^{\ell,p}}(-\cdot)},
\qquad \ell=1, \ldots, s.
\ee
Then it is easy to check that $\ctf_n:=\{a; b^{1,p}, \ldots, b^{s,p}, b^{1,n}, \ldots, b^{s,n}\}$ is a one-dimensional tight framelet filter bank such that $a$ is a real-valued low-pass filter, is symmetric about the origin, and satisfies $\wh{a}(0)=1$. Moreover, the high-pass filters $b^{\ell,p}$ and $b^{\ell,n}$ are complex-valued in the time domain and satisfy $b^{\ell,n}=\ol{b^{\ell,p}}$ for all $\ell=1, \ldots, s$.
For simplicity, we often choose $c_1$ and $\gep_1$ as free parameters and take
\be \label{special:c}
c_\ell:=c_1+\tfrac{\pi-c_1}{s}(\ell-1), \qquad \gep_\ell=\gep_1, \qquad \ell=1, \ldots, s.
\ee
For the above particular choice in \eqref{special:c}, the positive parameters $c_1$ and $\gep_1$ must satisfy
\be \label{c1gep1}
0<\gep_1 \le \min(c_1, \tfrac{\pi}{2}-c_1, \tfrac{c_1+(s-1)\pi}{2s}).
\ee
The $J$-level discrete affine system for dimension one is simply $\DAS_J(\{a; b^{1,p}, \ldots, b^{s,p}, b^{1,n}, \ldots, b^{s,n}\})$ which is defined at the beginning of Section~2.
The tensor product complex tight framelet filter bank $\tpctf_n$ for dimension two is simply
\[
\tpctf_n:=\ctf_n\otimes \ctf_n=\{a; b^{1,p}, \ldots, b^{s,p}, b^{1,n}, \ldots, b^{s,n}\}\otimes \{a; b^{1,p}, \ldots, b^{s,p}, b^{1,n}, \ldots, b^{s,n}\}
\]
with $a\otimes a$ being the only low-pass filter and all other $4s(s+1)$ filters being high-pass filters.
The $J$-level discrete affine system for dimension two is simply
\[
\DAS_J(\tpctf_n)=\DAS_J(\{a; b^{1,p}, \ldots, b^{s,p}, b^{1,n}, \ldots, b^{s,n}\}\otimes \{a; b^{1,p}, \ldots, b^{s,p}, b^{1,n}, \ldots, b^{s,n}\}).
\]
$\tpctf_n$ for dimension $d$ can be defined similarly by taking $d$ times tensor product of $\ctf_n$. For simplicity, we also use $\tpctf_n$ to stand for $\ctf_n$ for dimension one.
It is also not very difficult to deduce that the tensor product complex tight framelet $\tpctf_n$ with $n=2s+1$ for dimension two offers $\frac{1}{2}(n-1)(n-3)+4$ directions, that is, $2s(s-1)+4$ directions. For example, $\tpctf_3$ has $4$ directions along $0^\circ, \pm 45^\circ$ and $90^\circ$; $\tpctf_5$ has $8$ directions along $0^\circ, \pm 22.5^\circ, \pm 45^\circ, \pm 67.5^\circ$ and $90^\circ$.

\textbf{$\tpctf_n$ with $n=2s+2$ being an even positive integer.}
This case is almost the same as $\tpctf_{2s+1}$, except that we further split the low-pass filter $a$ into two low-pass filters $a^p, a^n$ in the frequency domain. Let $0<c_1<c_2<\cdots < c_s<c_{s+1}:=\pi$ and let $\gep_0, \gep_1, \ldots, \gep_s$ be positive numbers satisfying \eqref{tpctf:condition} with the additional condition
\be \label{c1gep0}
0<\gep_0<c_1-\gep_1.
\ee
Define three low-pass filters by
\be \label{tpctf:even}
\wh{a^{p}}:=\chi_{[0, c_1]; \gep_0, \gep_1}, \qquad
\wh{a^{n}}:=\ol{\wh{a^p}(-\cdot)},\qquad
\wh{a}:=\chi_{[-c_1, c_1]; \gep_1, \gep_1}.
\ee
The high-pass filters $b^{1,p}, \ldots, b^{s,p}, b^{1,n}, \ldots, b^{s,n}$ are defined as in \eqref{tpctf:odd:b}. Since
\[
|\wh{a}(\xi)|^2=|\wh{a^p}(\xi)|^2+|\wh{a^n}(\xi)|^2 \quad \mbox{and} \quad
\wh{a}(\xi)\ol{\wh{a}(\xi+\pi)}=0=\wh{a^p}(\xi)\ol{\wh{a^p}(\xi+\pi)}
+\wh{a^n}(\xi)\ol{\wh{a^n}(\xi+\pi)},
\]
we can check that both $\ctf_{n-1}:=\ctf_{2s+1}=\{a; b^{1,p}, \ldots, b^{s,p}, b^{1,n}, \ldots, b^{s,n}\}$
and
\[
\ctf_n:=\ctf_{2s+2}=\{a^p, a^n; b^{1,p}, \ldots, b^{s,p}, b^{1,n}, \ldots, b^{s,n}\}
\]
are one-dimensional tight framelet filter banks. Note that the filter
$a$ is a real-valued low-pass filter, is symmetric about the origin, and satisfies $\wh{a}(0)=1$. However, the low-pass filters $a^p$ and $a^n$ are complex-valued and may not have any symmetry but they satisfy $a^n=\ol{a^p}$. In other words, the symmetric real-valued low-pass filter $a$ is split into two complex-valued low-pass filters $a^p$ and $a^n$ satisfying $a^n=\ol{a^p}$.
For simplicity, we often choose $c_1, \gep_0$ and $\gep_1$ as free parameters and take the special choice in \eqref{special:c}.
For this particular case, both \eqref{c1gep1} and \eqref{c1gep0} must be satisfied.

The $J$-level discrete affine system for dimension one is simply $\DAS_J(\{a; b^{1,p}, \ldots, b^{s,p}, b^{1,n}, \ldots, b^{s,n}\})$ which is defined at the beginning of Section~2.
However, the tensor product complex tight framelet filter bank for dimension two is a little bit more complicated by defining the high-pass parts $\tpctf\mbox{-HP}_n$ through deleting all the
low-pass parts $\{a^p, a^n\}\otimes \{a^p, a^n\}$
from the tensor product filter bank
\[
\{a^p, a^n; b^{1,p}, \ldots, b^{s,p}, b^{1,n}, \ldots, b^{s,n}\}\otimes
\{a^p, a^n; b^{1,p}, \ldots, b^{s,p}, b^{1,n}, \ldots, b^{s,n}\}.
\]
More explicitly, $\tpctf\mbox{-HP}_n$ consists of total
$4s(s+1)$ high-pass filters:
\[
a^p\otimes b^{\ell,p}, a^p\otimes b^{\ell,n}, a^n\otimes b^{\ell,p}, a^n\otimes b^{\ell,n},
b^{\ell,p}\otimes b^{m,p}, b^{\ell,p}\otimes b^{m,n},
b^{\ell,n}\otimes b^{m,p}, b^{\ell,n}\otimes b^{m,n},\qquad
\ell,m=1, \ldots, s.
\]
Now it is not difficult to see that the tensor product complex tight framelet filter bank and the $J$-level discrete affine system for dimension two are given by
\[
\tpctf_n:=\{a\otimes a; \tpctf\mbox{-HP}_n\} \qquad \mbox{and}\qquad
\DAS_J(\tpctf_n)
\]
with $a\otimes a$ being the only low-pass filter in the two-dimensional tight framelet filter bank $\tpctf_n$.
$\tpctf_n$ for dimension $d$ can be defined similarly. For simplicity, we also use $\tpctf_n$ to stand for $\ctf_n$ for dimension one.
It is also not very difficult to check that the tensor product complex tight framelet $\tpctf_n$ with $n=2s+2$ offers $\frac{1}{2}(n-4)(n+2)+6$ directions, that is, $2(s-1)(s+2)+6$ directions. For example, $\tpctf_4$ has $6$ directions along $\pm 15^\circ, \pm 45^\circ$ and $\pm 75^\circ$, and $\tpctf_6$ has $14$ directions.

\begin{figure}[ht]
\begin{center}
\includegraphics[width=2.2in,height=1.2in]{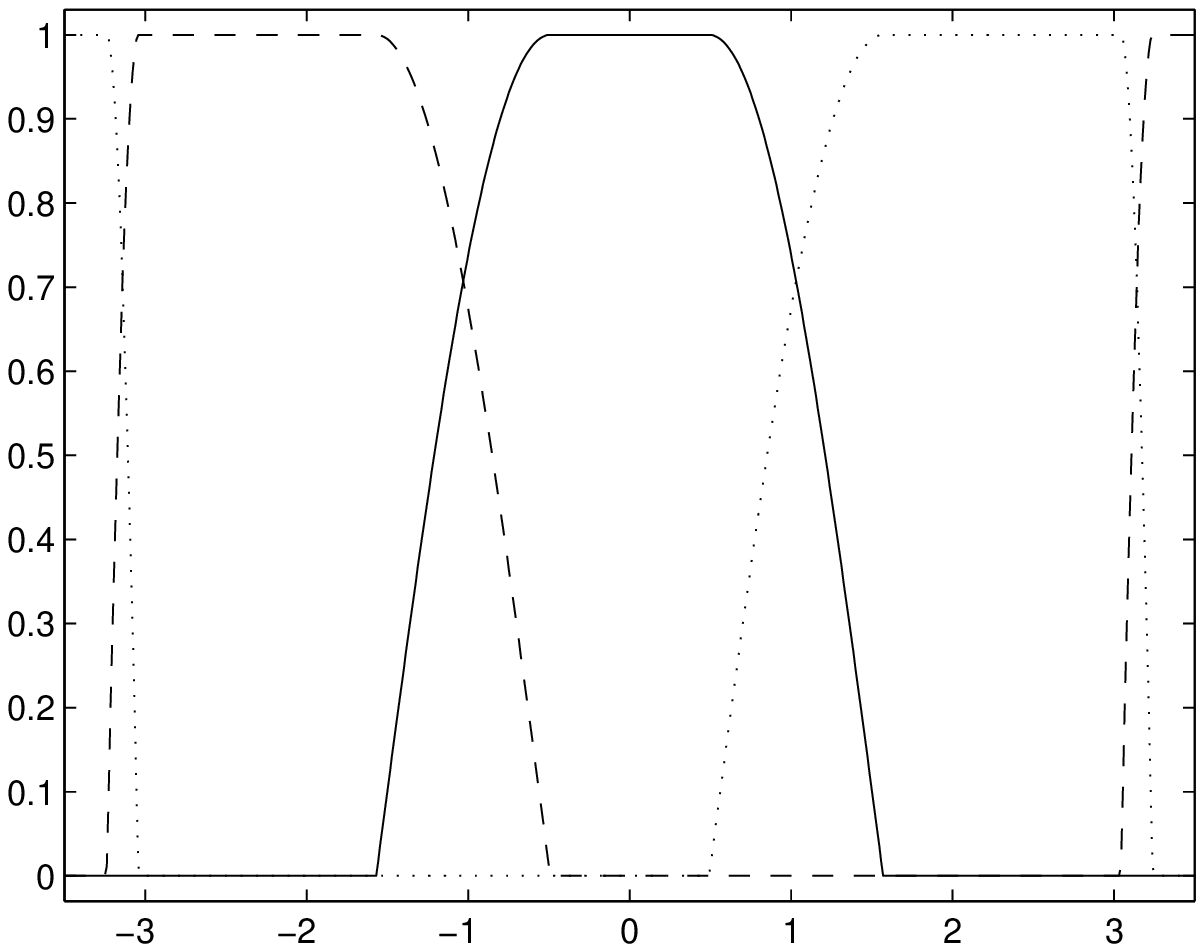}
\includegraphics[width=2.2in,height=1.2in]{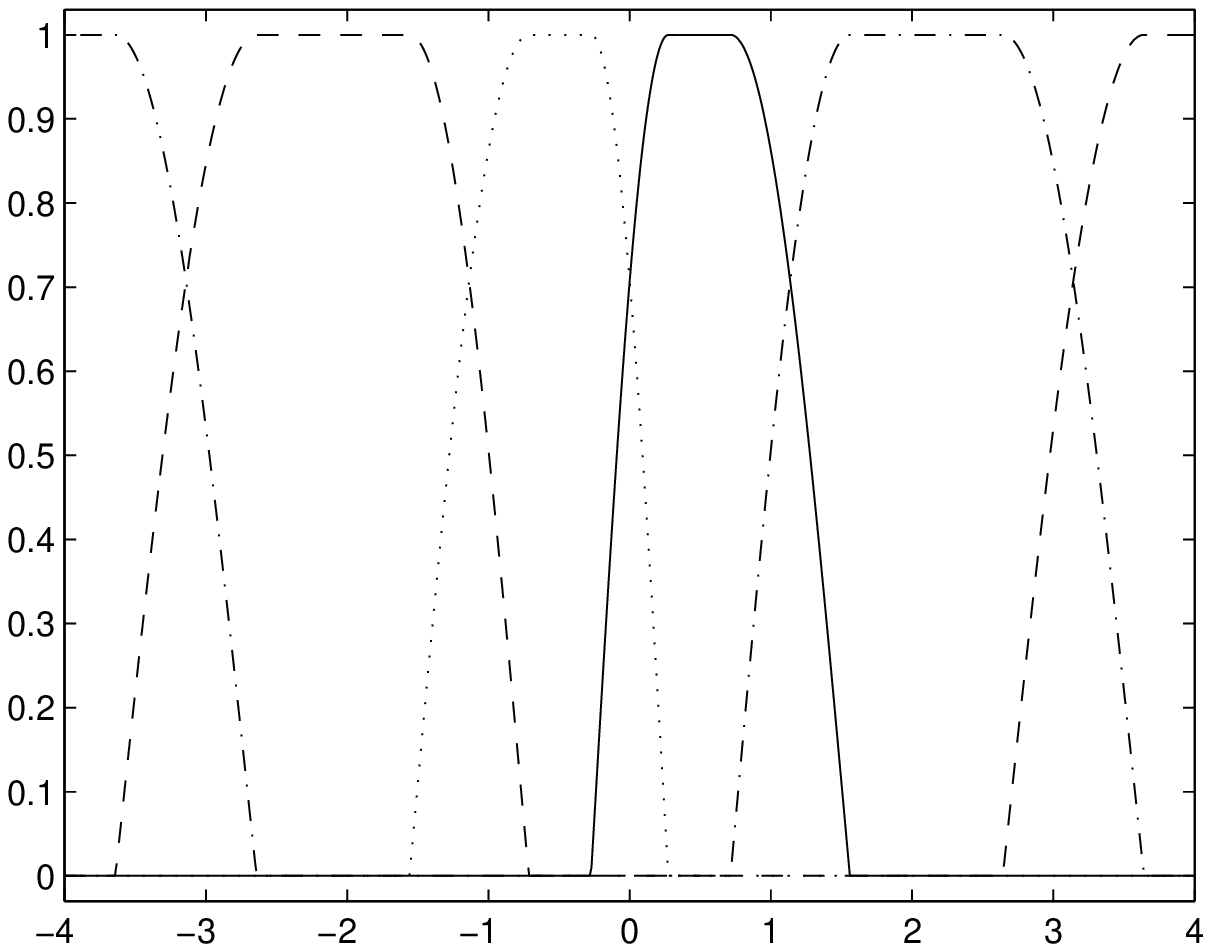}
\includegraphics[width=2.2in,height=1.2in]{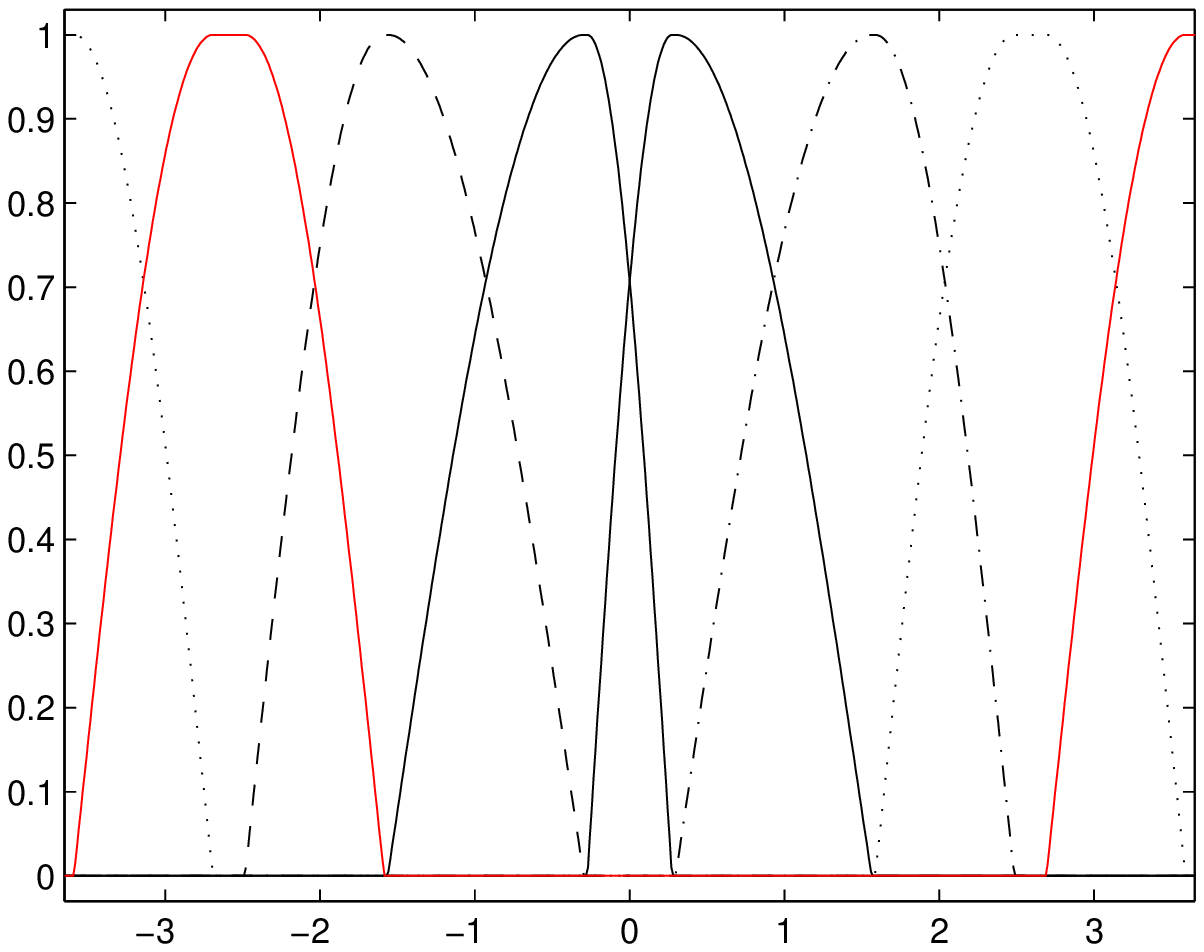}
\begin{caption}{
Left: one-dimensional
$\ctf_3=\{a; b^{1,p}, b^{1,n}\}$ in the frequency domain. Solid line is for low-pass filter $\wh{a}$.
Dotted line is for the high-pass filter $\wh{b^{1,p}}$.
Dashed line is for the high-pass filter $\wh{b^{1,n}}$.
Middle: one-dimensional
$\ctf_4=\{a^p, a^n; b^{1,p}, b^{1,n}\}$ in the frequency domain. Solid line is for the low-pass filter $\wh{a^p}$.
Dotted line is for the low-pass filter $\wh{a^n}$.
Dotted-dashed line is for the high-pass filter $\wh{b^{1,p}}$.
Dashed line is for the high-pass filter $\wh{b^{1,n}}$.
Right: one dimensional
$\ctf_6=\{a^p, a^n; b^{1,p}, b^{2,p}, b^{1,n}, b^{2,n}\}$ in the frequency domain. Right solid line is for $\wh{a^p}$ and left solid line is for $\wh{a^n}$.
Dotted-dashed line is for $\wh{b^{1,p}}$ and dotted line is for $\wh{b^{2,p}}$.
Dashed line is for $\wh{b^{1,n}}$ and red line is for $\wh{b^{2,n}}$.
} \label{ctf:filters}
\end{caption}
\end{center}
\end{figure}

\begin{figure}[ht]
\centerline{
\hbox{\epsfig{file=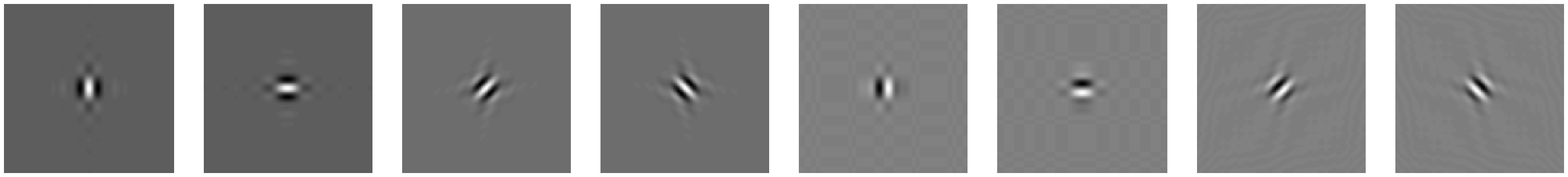,width=5.6in,height=0.68in} }}
\begin{caption}{
The real part (the first four) and the imaginary part (the last four) of the generators at level 5 in $\DAS_6(\tpctf_3)$.
} \label{tpct3graph}
\end{caption}
\end{figure}

\begin{figure}[ht]
\centerline{
\hbox{\epsfig{file=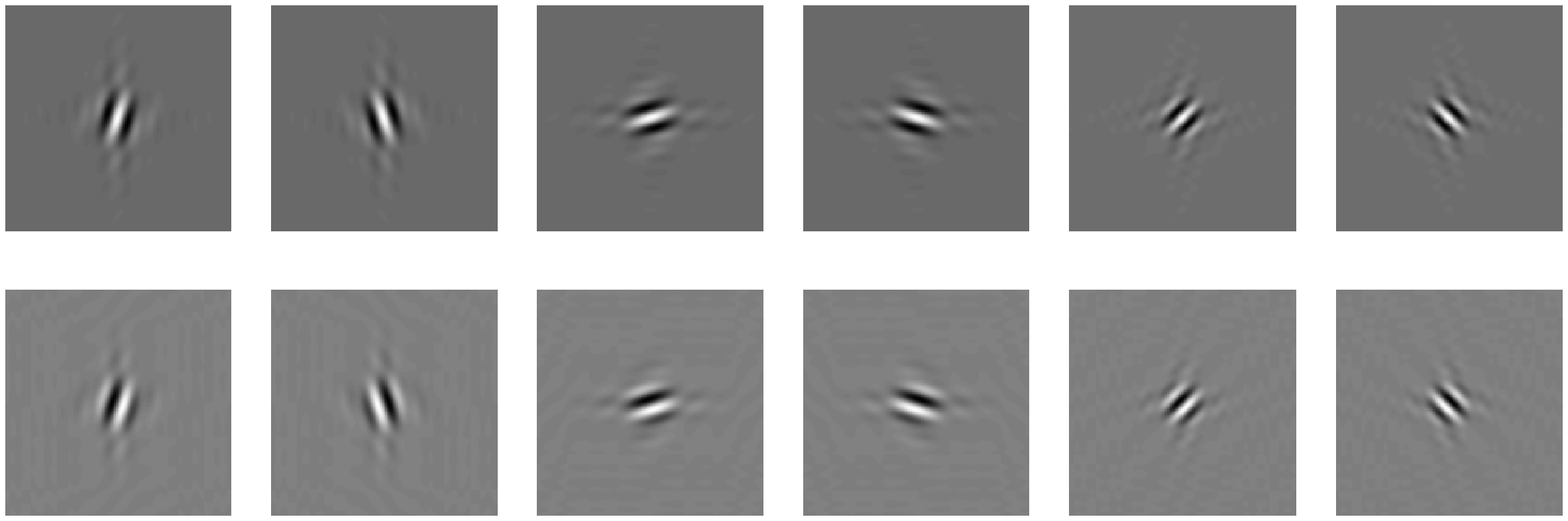,width=4.2in,height=1.3in} }}
\begin{caption}{
The first row shows the real part and the second row shows the imaginary part of the generators at level 5 in $\DAS_6(\tpctf_4)$.
} \label{tpct4graph}
\end{caption}
\end{figure}

\begin{figure}[ht]
\centerline{
\hbox{\epsfig{file=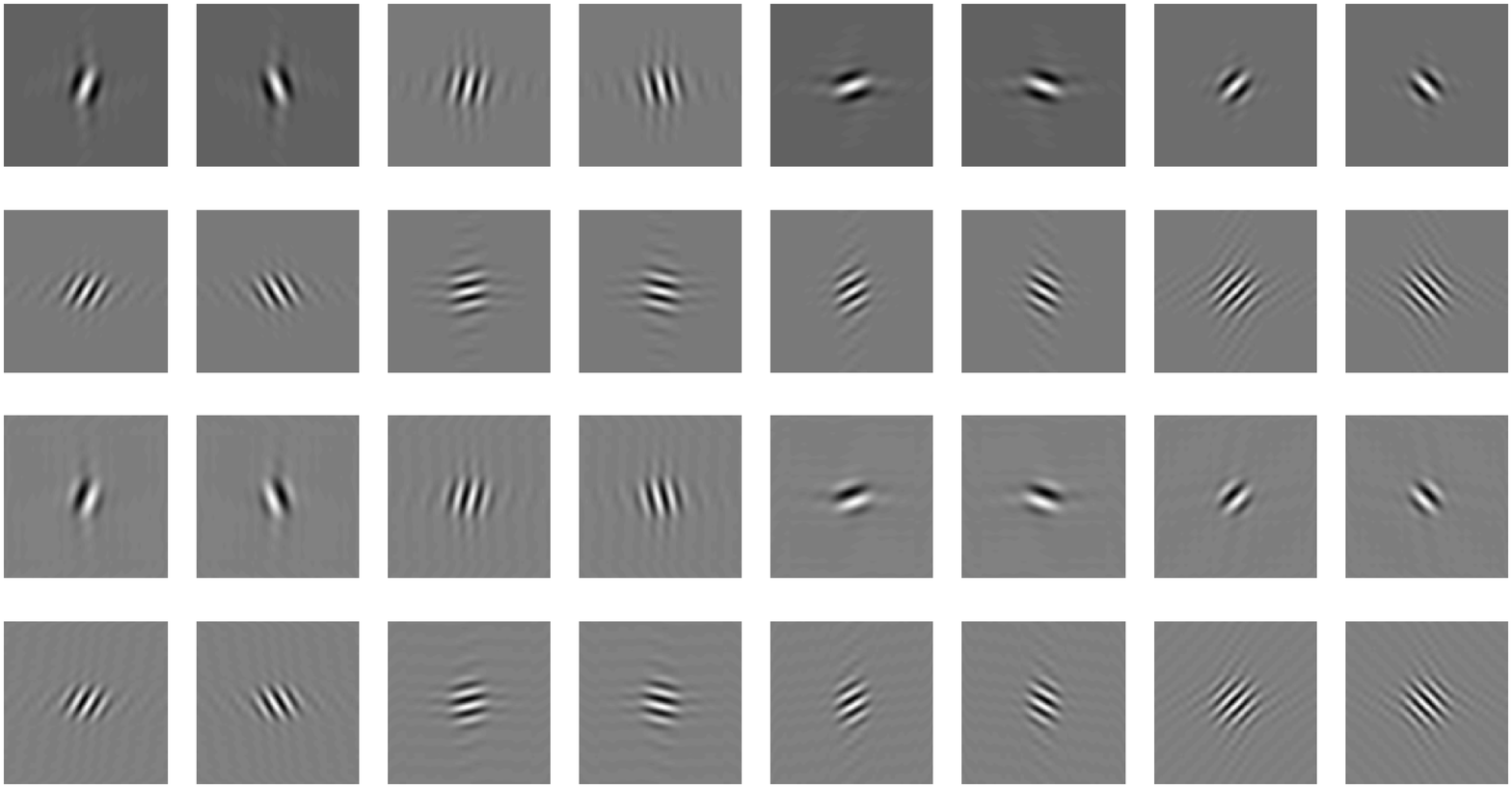,width=5.6in,height=2.8in} }}
\begin{caption}{
The first two rows show the real part and the last two rows show the imaginary part of the generators at level 5 in $\DAS_6(\tpctf_6)$.
Among these 16 graphs, the directions along $\pm 45^\circ$ are repeated twice. Hence, there are totally 14 directions in the discrete affine system $\DAS_J(\tpctf_6)$.
} \label{tpct6graph}
\end{caption}
\end{figure}

Throughout the paper, $\tpctf_3$ uses \eqref{special:c} with $c_1 = \frac{33}{32}$ and $\gep_1 = \frac{69}{128}$, while $\tpctf_4$ uses \eqref{special:c} with $c_1 = \frac{291}{128}$, $\gep_0 = \frac{35}{128}$, and $\gep_1 = \frac{27}{64}$; $\tpctf_6$ uses \eqref{special:c} with $c_1 = \frac{119}{128}$,
$\gep_0 = \frac{35}{128}$, and $\varepsilon_1 = \frac{81}{128}$.
See Figure~\ref{ctf:filters} for graphs of the one-dimensional complex tight framelet filter banks $\ctf_3$, $\ctf_4$, and $\ctf_6$ in the frequency domain. See Figure~\ref{tpct3graph} for the directionality of the two-dimensional tensor product complex tight framelet $\tpctf_3$ (more precisely, the generators in $\DAS_J(\tpctf_3)$),
Figure~\ref{tpct4graph} for the directionality of the two-dimensional $\tpctf_4$, and Figure~\ref{tpct6graph} for the directionality of the two-dimensional $\tpctf_6$.

Now we provide numerical experiments on image denoising using $\tpctf_n$. In all the experiments, bivariate shrinkage proposed in \cite{SS:bs,SS:bslocal} is applied to framelet coefficients. As shown in the following tables, we can see clearly the improved performance in terms of PSNR due to improved directionality. The performance of $\tpctf_4$ is also comparable with that of standard $\dtcwt$ using the finitely supported orthogonal filters in \eqref{dtcwt:a0}--\eqref{dtcwt:a2}.
In addition, the denoising results can be further improved by applying more complicated shrinkages such as the Gaussian scale mixture model in \cite{PSWS} to the real and imaginary parts of the undecimated $\tpctf_n$ coefficients.

\bigskip

\begin{table}[ht]
\begin{center}
\begin{tabular}{|c||c|c|c|c||c|c||}
\hline
&\multicolumn{6}{|c||}{Lena ($512\times 512$)} \\  \hline
$\sigma_n$ &$\dtcwt$ &$\tpctf_3$ &$\tpctf_4$ &$\tpctf_6$ (Gain) &$\ctf_4$-GSM &$\ctf_6$-GSM (Gain)
\\ \hline
5 &38.25 &37.96 &38.10 &\textbf{38.35} (0.10)
&38.43 &\textbf{38.53} (0.28)
\\ \hline
10 &35.19 &34.91 &35.14 &\textbf{35.45} (0.26)  &35.59 &\textbf{35.70} (0.51)
\\ \hline
15 &33.47 &33.25 &33.50 &\textbf{33.77} (0.30) &33.88 &\textbf{34.01} (0.54)
\\ \hline
20 &32.23 &32.07 &32.31 &\textbf{32.55} (0.22) &32.63 &\textbf{32.77} (0.54)
\\ \hline
25 &31.26 &31.15 &31.38 &\textbf{31.58} (0.32) &31.62 &\textbf{31.78} (0.56)
\\ \hline
30 &30.47 &30.40 &30.61 &\textbf{30.78} (0.31) &30.79 &\textbf{30.96} (0.49)
\\ \hline
50 &28.21 &28.29 &28.41 &\textbf{28.50} (0.29) &28.49 &\textbf{28.64} (0.43)
\\ \hline
\end{tabular}
\medskip \caption{
Denoising results for $512\times 512$ Lena image.
Each numerical PSNR value is an average over five experiments.
$\sigma_n$ is the variance of additive i.i.d. Gaussian noise and is assumed to be known in advance.
Column $\dtcwt$ uses $\dtcwt$ with a pair of correlated finitely supported orthogonal wavelet filter banks in \cite{K1999,SBK} (see \eqref{dtcwt:a0}--\eqref{dtcwt:a2}). Column $\tpctf_3$ uses tensor product complex tight framelet $\tpctf_3$. Columns $\tpctf_4$ and $\ctf_4$-GSM use tensor product complex tight framelet $\tpctf_4$. Columns $\tpctf_6$ and $\ctf_6$-GSM use tensor product tight framelet $\tpctf_6$. All the first four columns use the same bivariate shrinkage developed in \cite{SS:bs}. The last two columns use the Gaussian scale mixture in \cite{PSWS}. Gain refers to the PSNR gain of the current column over $\dtcwt$ in column 2.
}\label{table:lena}
\end{center}
\end{table}

\begin{table}[ht]
\begin{center}
\begin{tabular}{|c||c|c|c|c||c|c||}
\hline
&\multicolumn{6}{|c||}{Barbara ($512\times 512$)} \\  \hline
$\sigma_n$ &$\dtcwt$ &$\tpctf_3$ &$\tpctf_4$ &$\tpctf_6$ (Gain) &$\ctf_4$-GSM &$\ctf_6$-GSM (Gain)
\\ \hline
5 &37.36 & 37.16 &37.41 &\textbf{37.82} (0.46) &37.75 &\textbf{38.10} (0.74)
\\ \hline
10 &33.52 &33.17 &33.62 &\textbf{34.14} (0.48) &34.10 &\textbf{34.47} (0.95)
\\ \hline
15 &31.38 &30.89 &31.47 &\textbf{32.02} (0.64) &31.97 &\textbf{32.32} (0.94)
\\ \hline
20 &29.87 &29.27 &29.91 &\textbf{30.49} (0.62) &30.43 &\textbf{30.77} (0.90)
\\ \hline
25 &28.70 &28.01 &28.71 &\textbf{29.31} (0.61) &29.26 &\textbf{29.57} (0.87)
\\ \hline
30 &27.77 &27.01 &27.74 &\textbf{28.34} (0.57) &28.32 &\textbf{28.61} (0.84)
\\ \hline
50 &25.26 &24.51 &25.21 &\textbf{25.71} (0.45) &25.69 &\textbf{26.02} (0.76)
\\ \hline
\end{tabular}
\end{center}
\end{table}

\begin{table}[ht]
\begin{center}
\begin{tabular}{|c||c|c|c|c||c|c||}
\hline
&\multicolumn{6}{|c||}{Boat ($512\times 512$)} \\  \hline
$\sigma_n$ &$\dtcwt$ &$\tpctf_3$ &$\tpctf_4$ &$\tpctf_6$ (Gain) &$\ctf_4$-GSM &$\ctf_6$-GSM (Gain)
\\ \hline
5 &36.77 &36.44 &36.52 &\textbf{36.90} (0.13) &36.90 &\textbf{37.07} (0.30)
\\ \hline
10 &33.21 &32.96 &33.08 &\textbf{33.39} (0.17) &33.57 &\textbf{33.69} (0.48)
\\ \hline
15 &31.33 &31.15 &31.28 &\textbf{31.53} (0.20) &31.69 &\textbf{31.81} (0.48)
\\ \hline
20 &30.01 &29.91 &30.01 &\textbf{30.22} (0.21) &30.35 &\textbf{30.48} (0.47)
\\ \hline
25 &28.99 &28.94 &29.03 &\textbf{29.22} (0.23) &29.33 &\textbf{29.46} (0.47)
\\ \hline
30 &28.18 &28.16 &28.22 &\textbf{28.41} (0.23) &28.51 &\textbf{28.63} (0.45)
\\ \hline
50 &26.01 &26.00 &26.04 &\textbf{26.19} (0.18) &26.27 &\textbf{26.39} (0.38)
\\ \hline
\end{tabular}
\end{center}
\end{table}

\begin{table}[ht]
\begin{center}
\begin{tabular}{|c||c|c|c|c||c|c||}
\hline
&\multicolumn{6}{|c||}{House ($256\times 256$)} \\  \hline
$\sigma_n$ &$\dtcwt$ &$\tpctf_3$ &$\tpctf_4$ &$\tpctf_6$ (Gain) &$\ctf_4$-GSM &$\ctf_6$-GSM (Gain)
\\ \hline
5 &38.45 &38.40 &38.54 &\textbf{38.91} (0.46) &38.82 &\textbf{39.13} (0.68)
\\ \hline
10 &34.78 &34.76 &34.94 &\textbf{35.43} (0.65) &35.42 &\textbf{35.77} (0.99)
\\ \hline
15 &32.90 &32.97 &33.13 &\textbf{33.57} (0.67) &33.65 &\textbf{33.99} (1.09)
\\ \hline
20 &31.63 &31.76 &31.90 &\textbf{32.32} (0.69) &32.36 &\textbf{32.72} (1.09)
\\ \hline
25 &30.65 &30.81 &30.94 &\textbf{31.34} (0.69) &31.34 &\textbf{31.69} (1.04)
\\ \hline
30 &29.84 &30.04 &30.15 &\textbf{30.52} (0.68) &30.49 &\textbf{30.82} (0.98)
\\ \hline
50 &27.57 &27.89 &27.90 &\textbf{28.14} (0.57) &28.11 &\textbf{28.36} (0.79)
\\ \hline
\end{tabular}
\end{center}
\end{table}

\begin{table}[ht]
\begin{center}
\begin{tabular}{|c||c|c|c|c||c|c||}
\hline
&\multicolumn{6}{|c||}{Pepper ($256\times 256$)} \\  \hline
$\sigma_n$ &$\dtcwt$ &$\tpctf_3$ &$\tpctf_4$ &$\tpctf_6$ (Gain) &$\ctf_4$-GSM &$\ctf_6$-GSM (Gain)
\\ \hline
5 &37.18 &36.98 &37.07 &\textbf{37.25} (0.07) &37.55 &\textbf{37.61} (0.43)
\\ \hline
10 &33.40 &33.29 &33.41 &\textbf{33.61} (0.21) &34.02 &\textbf{34.07} (0.67)
\\ \hline
15 &31.29 &31.28 &31.38 &\textbf{31.60} (0.31) &31.93 &\textbf{32.01} (0.72)
\\ \hline
20 &29.83 &29.90 &29.96 &\textbf{30.20} (0.37) &30.42 &\textbf{30.54} (0.71)
\\ \hline
25 &28.71 &28.82 &28.86 &\textbf{29.11} (0.40) &29.27 &\textbf{29.43} (0.72)
\\ \hline
30 &27.80 &27.95 &27.97 &\textbf{28.23} (0.43) &28.34 &\textbf{28.53} (0.73)
\\ \hline
50 &25.30 &25.52 &25.51 &\textbf{25.77} (0.47) &25.80 &\textbf{26.02} (0.72)
\\ \hline
\end{tabular}
\end{center}
\end{table}

\begin{figure}[ht]
\centering
\subfigure[Original image]{
\includegraphics[width=1.35in,height=1.35in]
{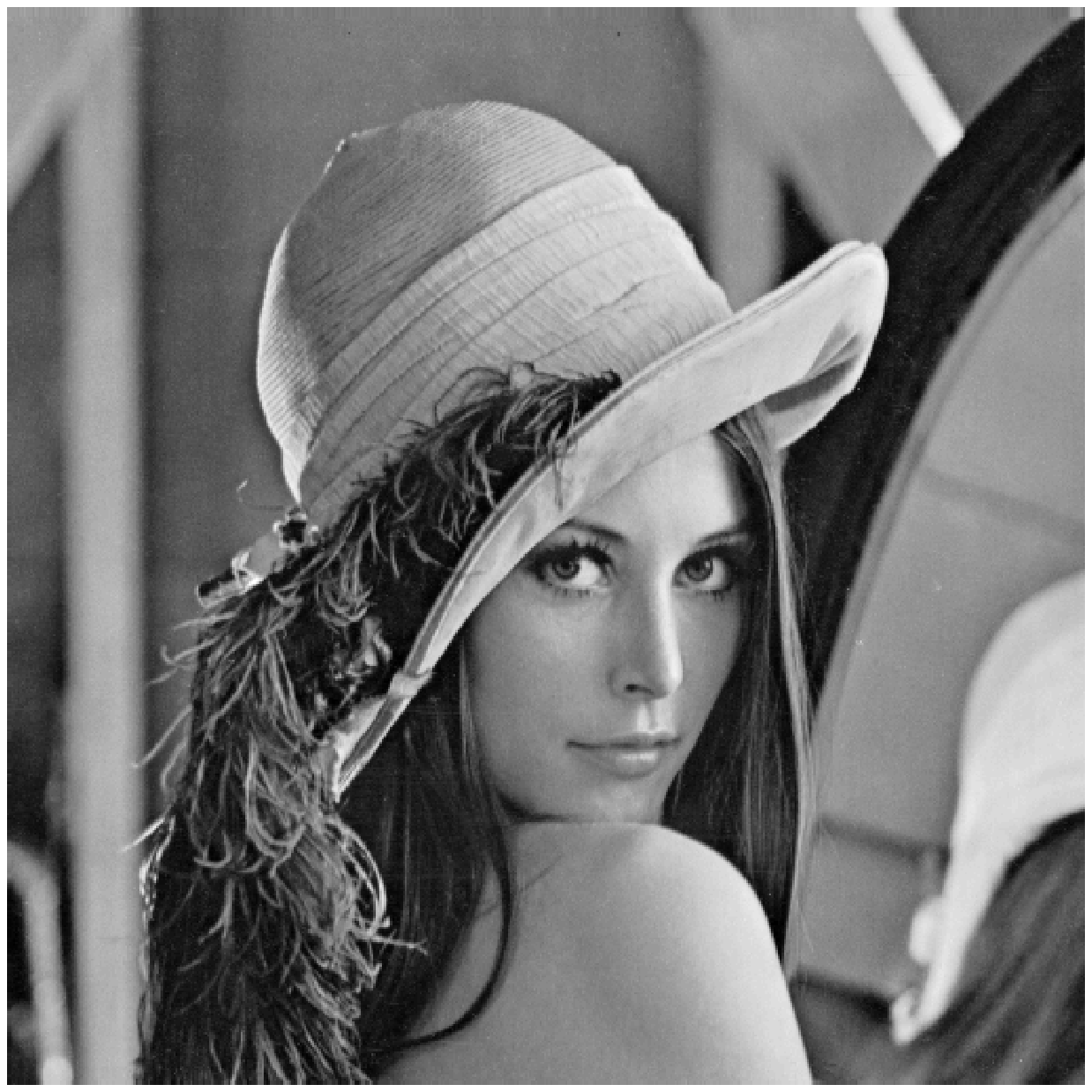}}
\subfigure[Noisy image]{
\includegraphics[width=1.35in,height=1.35in]
{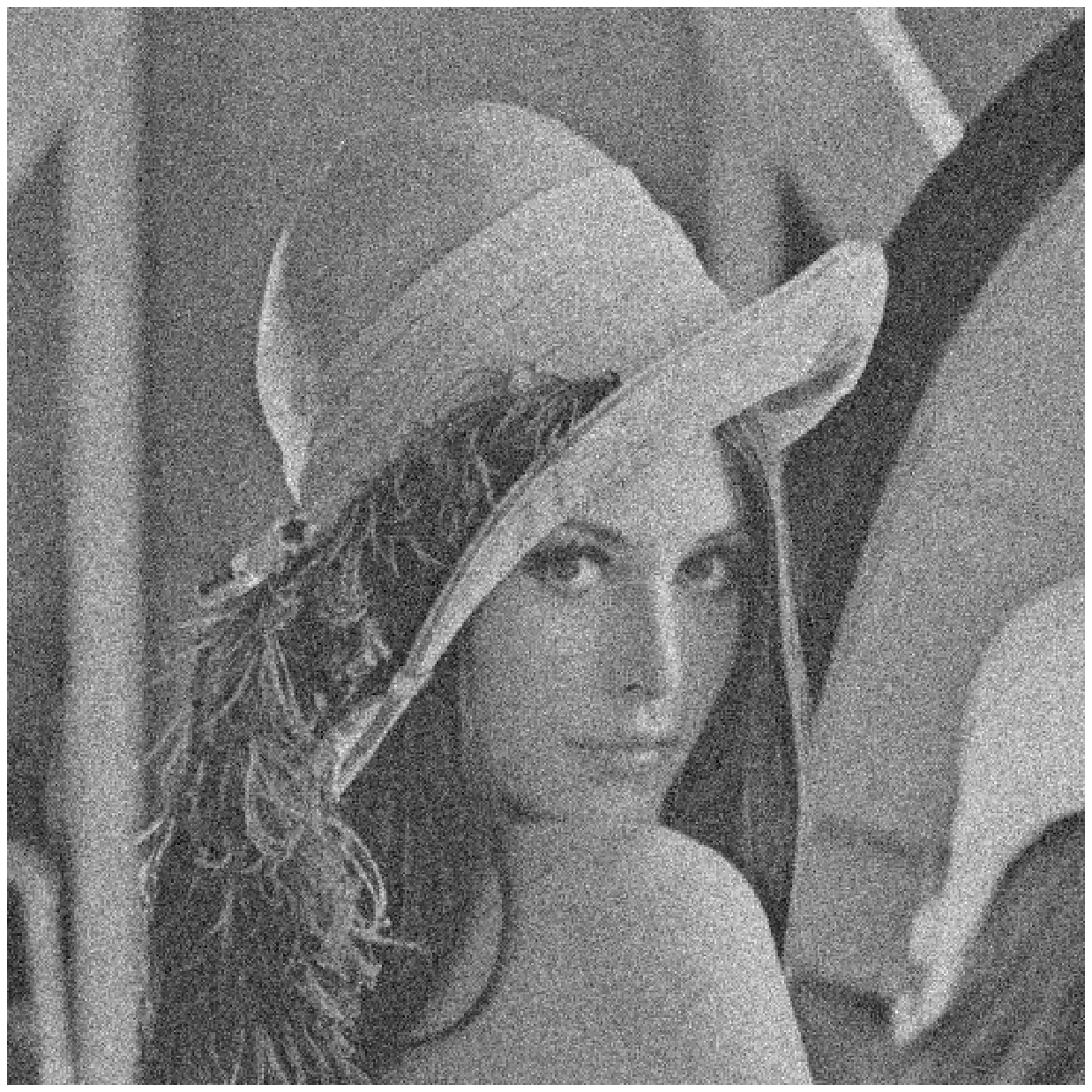}}
\subfigure[By $\dtcwt$]{
\includegraphics[width=1.35in,height=1.35in]
{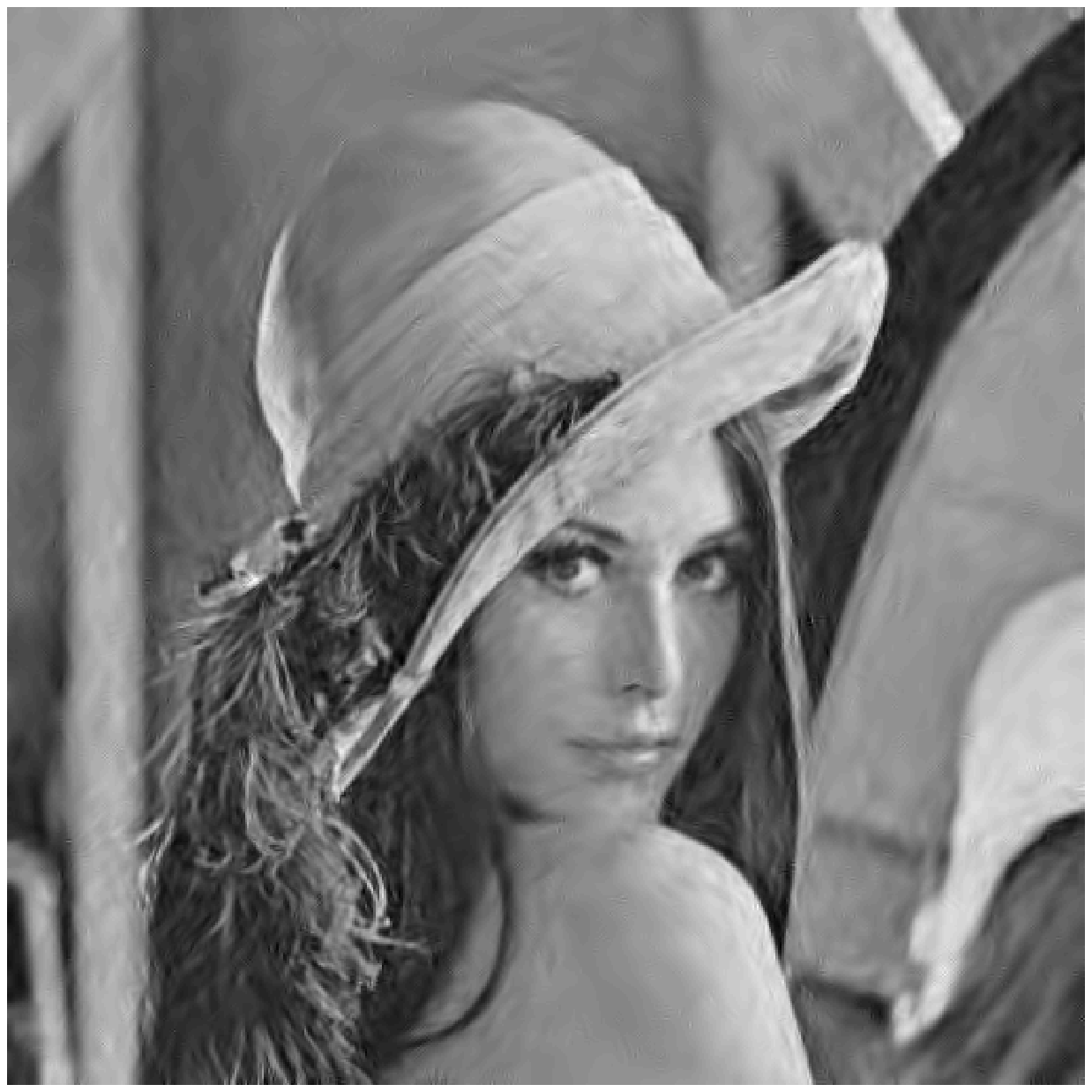}}
\subfigure[By $\tpctf_6$]{
\includegraphics[width=1.35in,height=1.35in]
{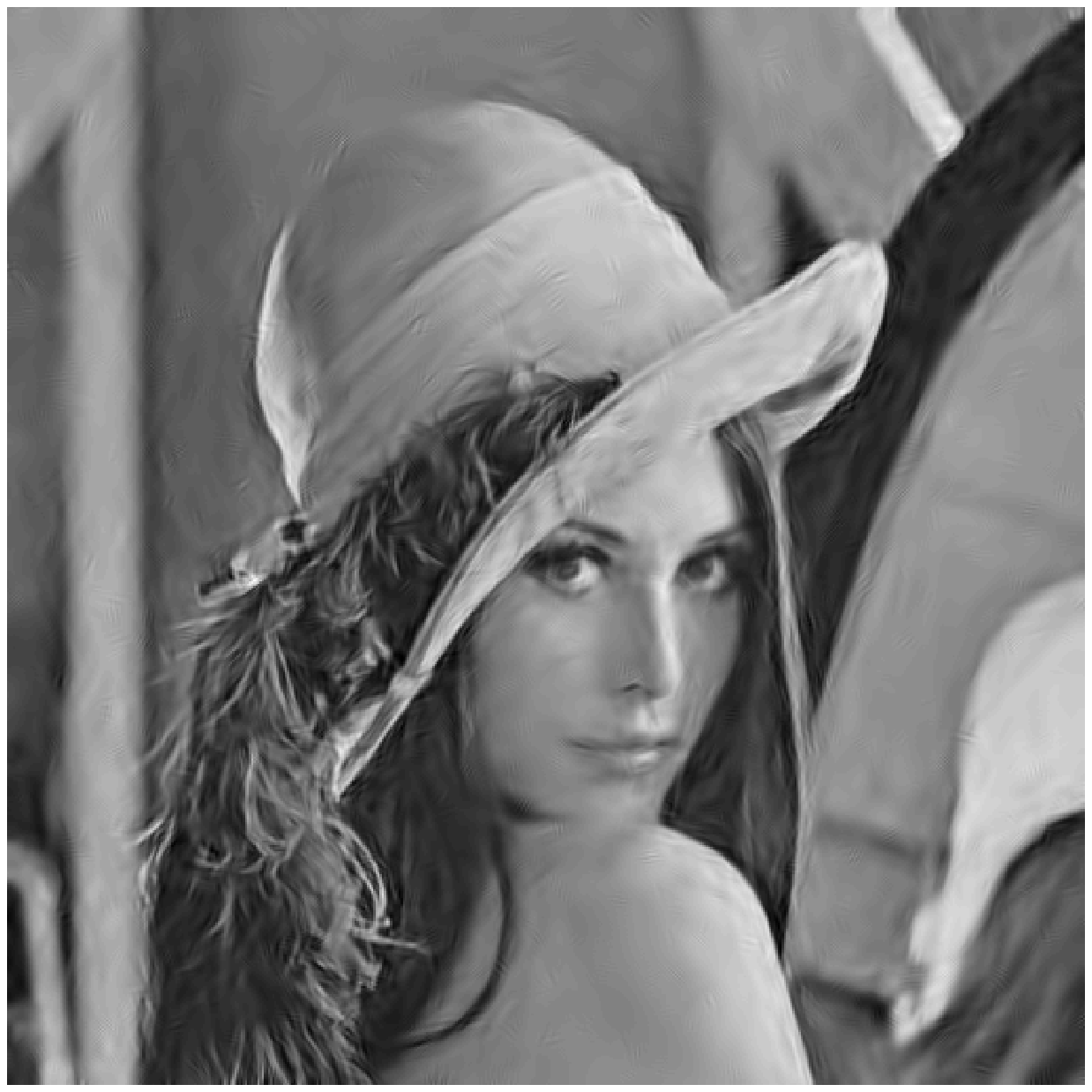}}
\subfigure[By $\tpctf_6$ \& GSM]{
\includegraphics[width=1.35in,height=1.35in]
{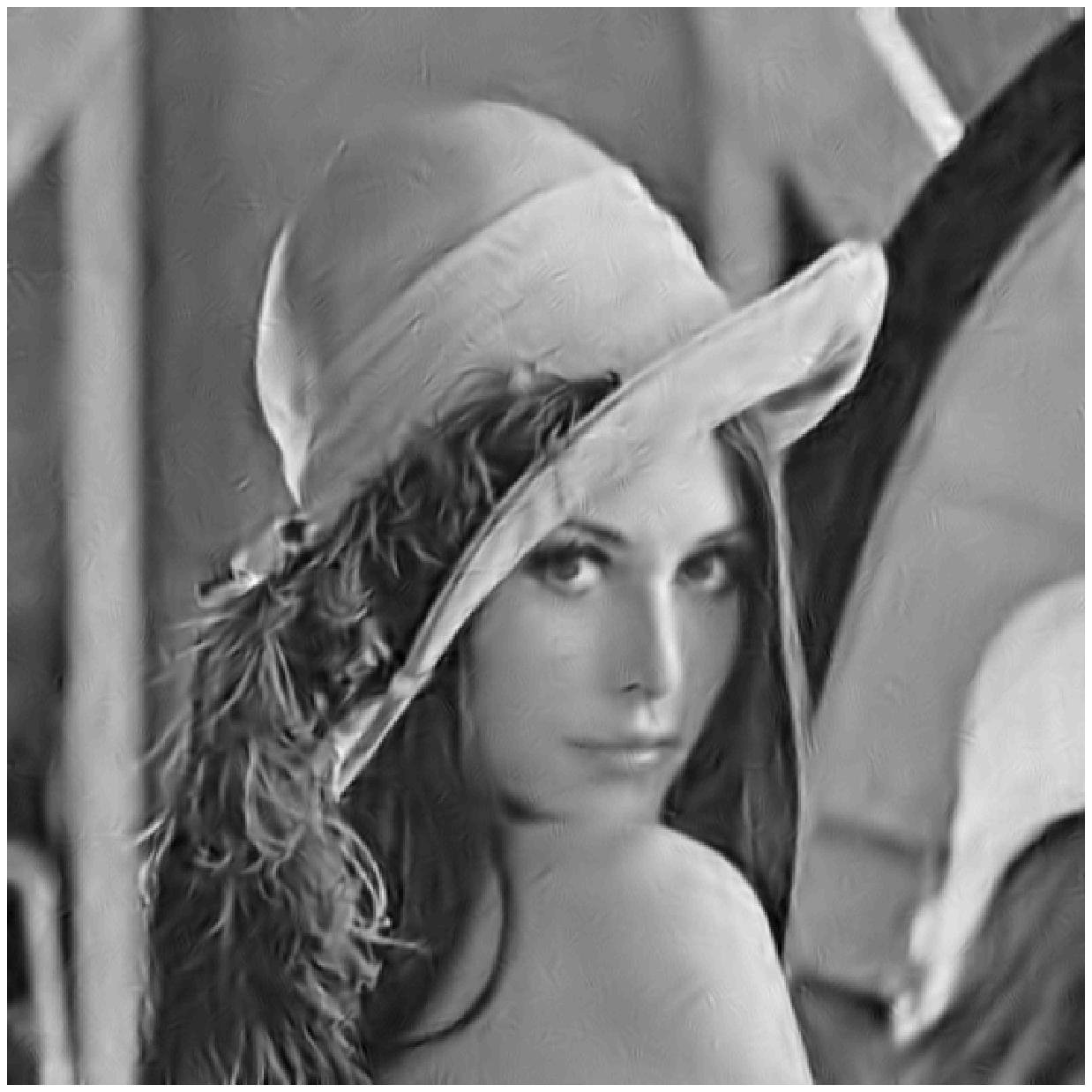}}
\begin{caption}{(a) Original $512\times 512$ image of Lena. (b) Noisy image with $\sigma_n=30$ (PSNR=$18.60$). (c) Denoised image by $\dtcwt$ (PSNR=$30.47$). (d) Denoised image by $\tpctf_6$ using bivariate shrinkage (PSNR=$30.76$). (e) Denoised image by $\tpctf_6$ using Gaussian scale mixture (PSNR=$30.94$).
} \label{barbara}
\end{caption}
\end{figure}

\begin{figure}[ht]
\centering
\subfigure[Original image]{
\includegraphics[width=1.35in,height=1.35in]
{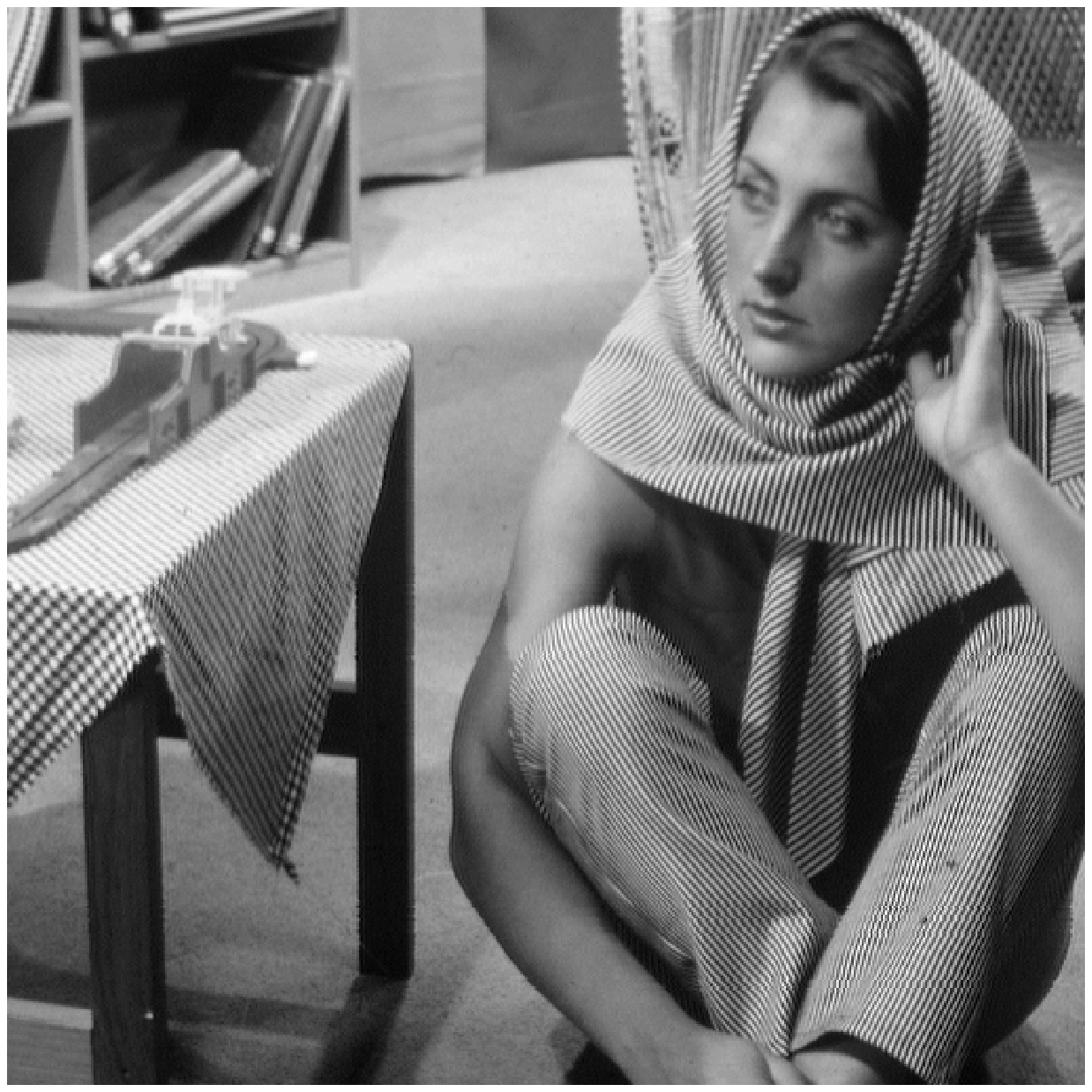}}
\subfigure[Noisy image]{
\includegraphics[width=1.35in,height=1.35in]
{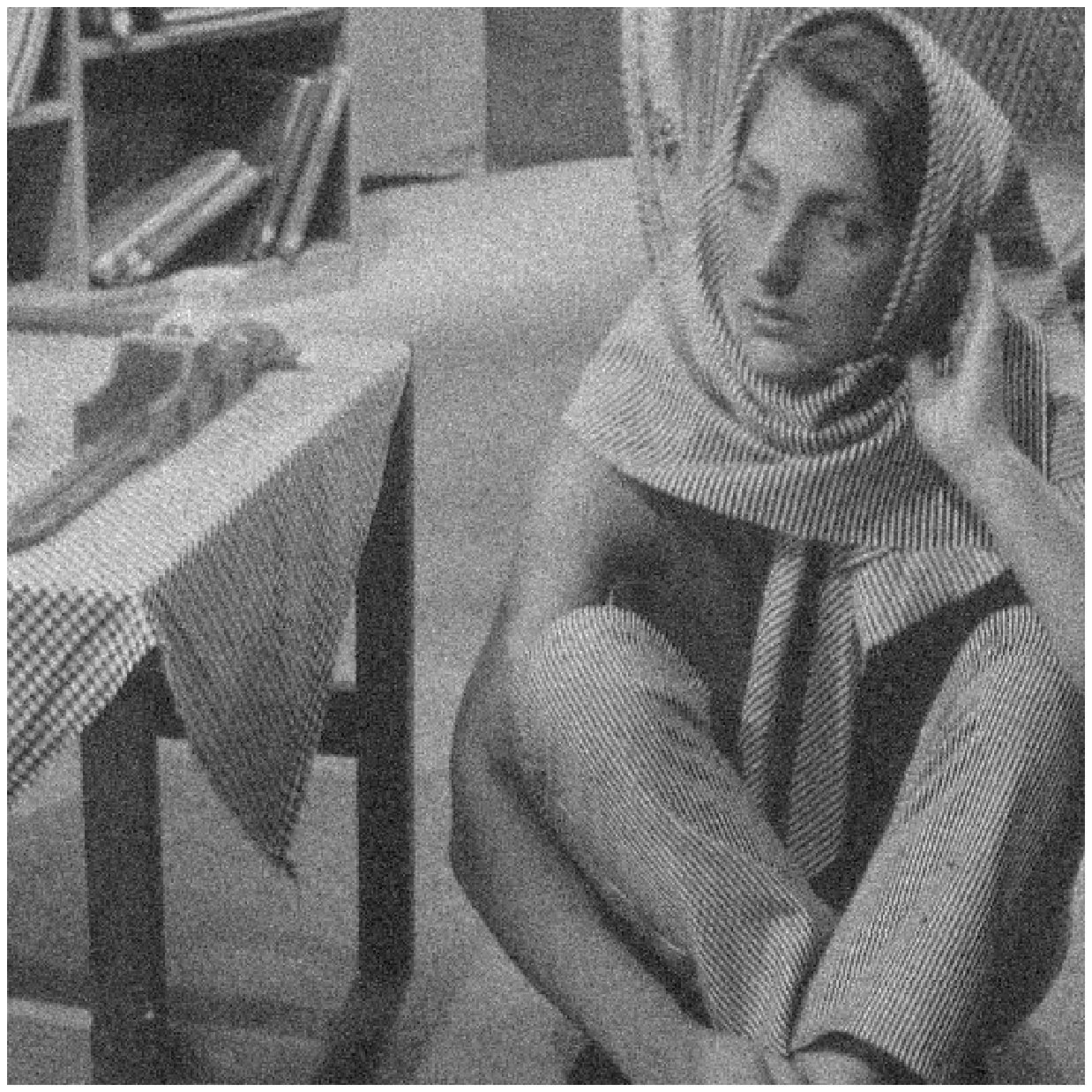}}
\subfigure[By $\dtcwt$]{
\includegraphics[width=1.35in,height=1.35in]
{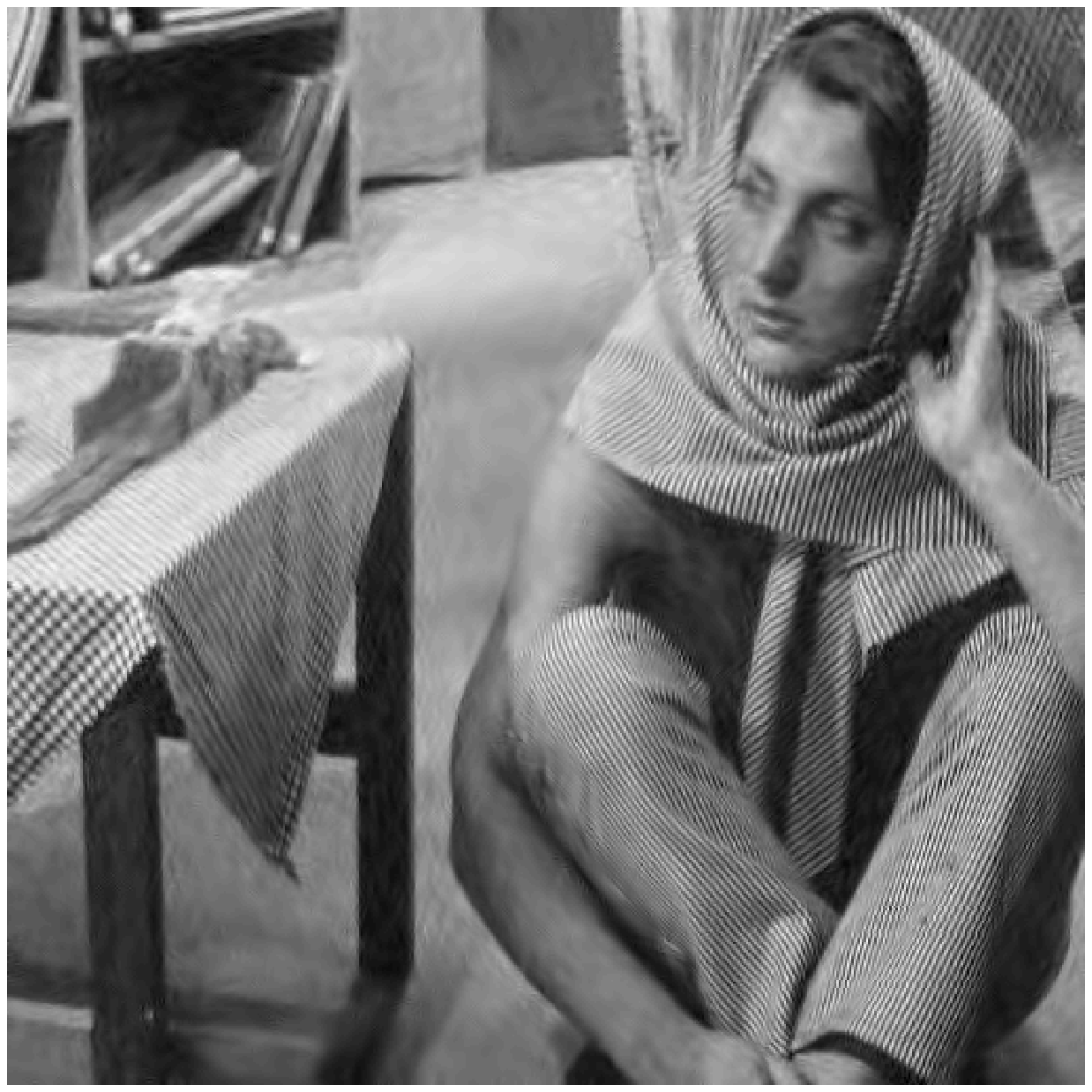}}
\subfigure[By $\tpctf_6$]{
\includegraphics[width=1.35in,height=1.35in]
{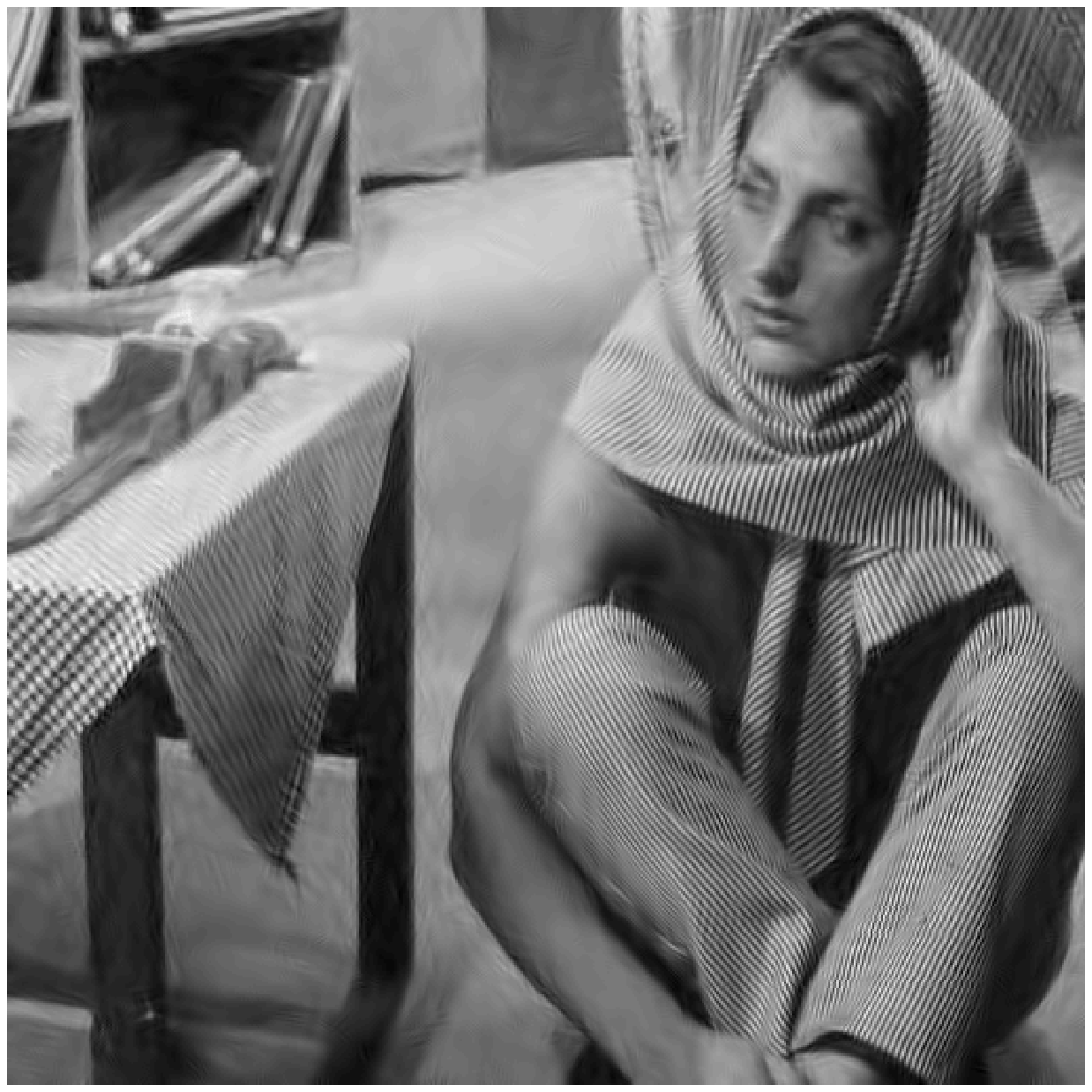}}
\subfigure[By $\tpctf_6$ \& GSM]{
\includegraphics[width=1.35in,height=1.35in]
{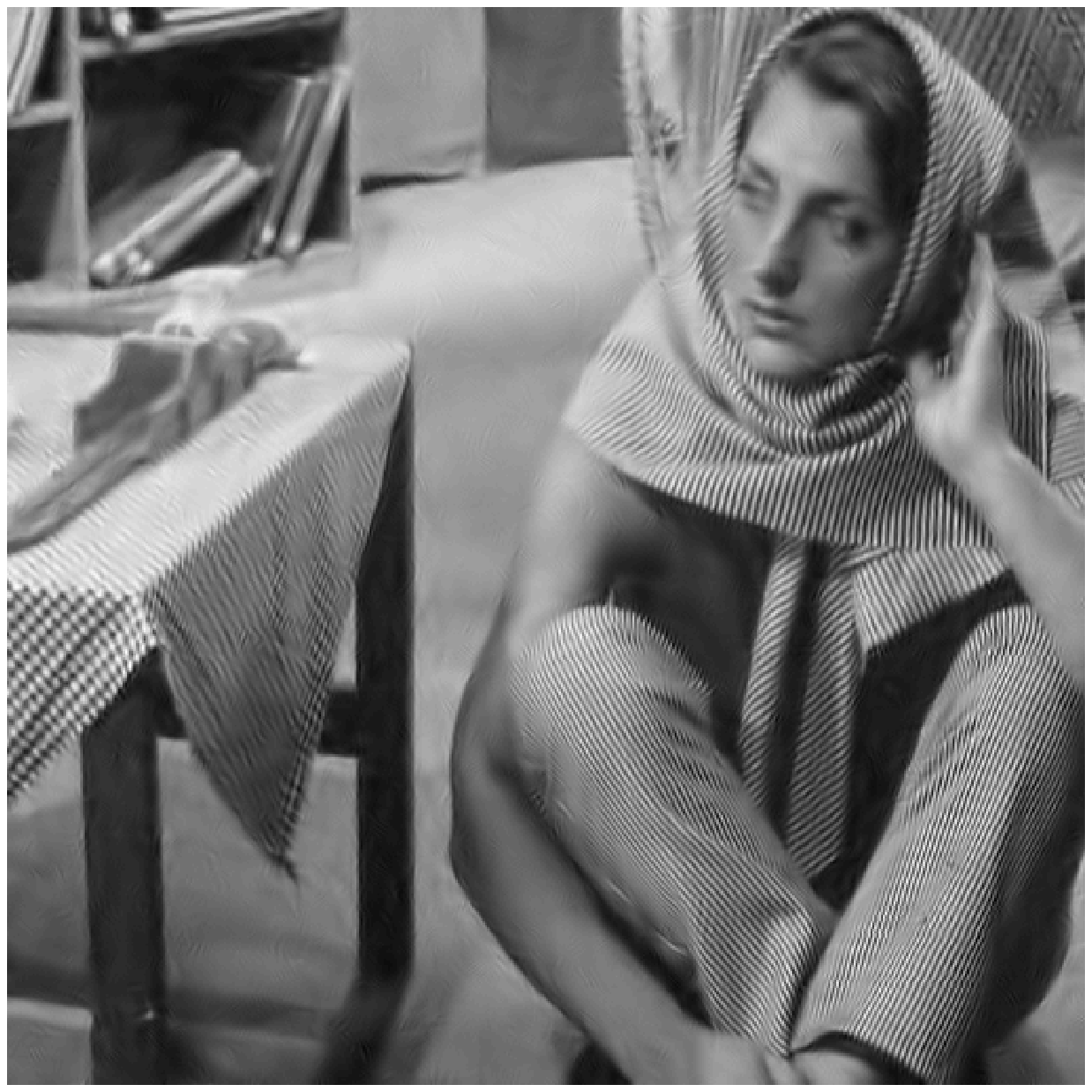}}
\begin{caption}{(a) Original $512\times 512$ image of Barbara. (b) Noisy image with $\sigma_n=20$ (PSNR=$22.12$). (c) Denoised image by $\dtcwt$ (PSNR=$29.85$). (d) Denoised image by $\tpctf_6$ using bivariate shrinkage (PSNR=$30.48$). (e) Denoised image by $\tpctf_6$ using Gaussian scale mixture (PSNR=$30.75$).
} \label{barbara}
\end{caption}
\end{figure}

\begin{figure}[ht]
\centering
\subfigure[Original image]{
\includegraphics[width=1.35in,height=1.35in]
{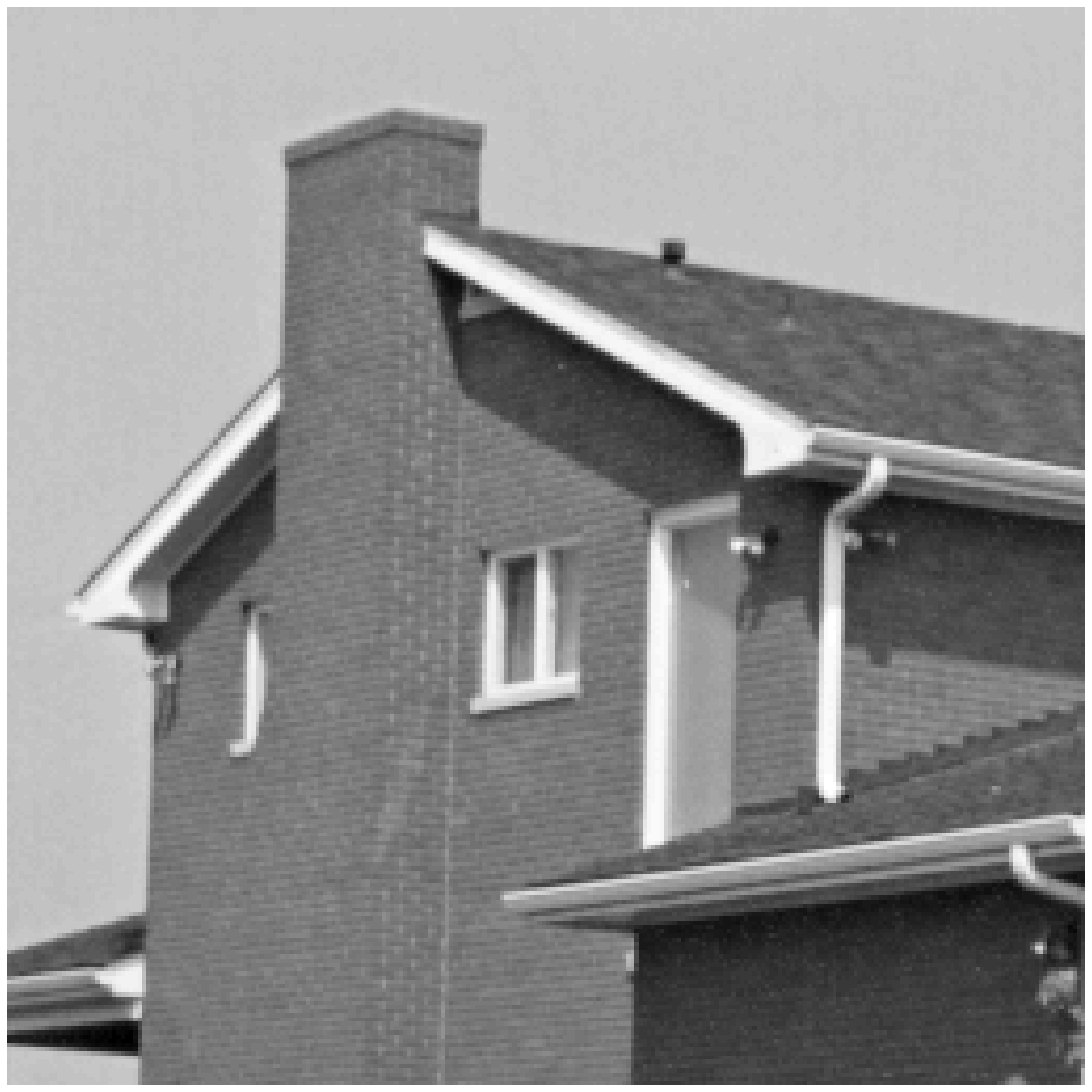}}
\subfigure[Noisy image]{
\includegraphics[width=1.35in,height=1.35in]
{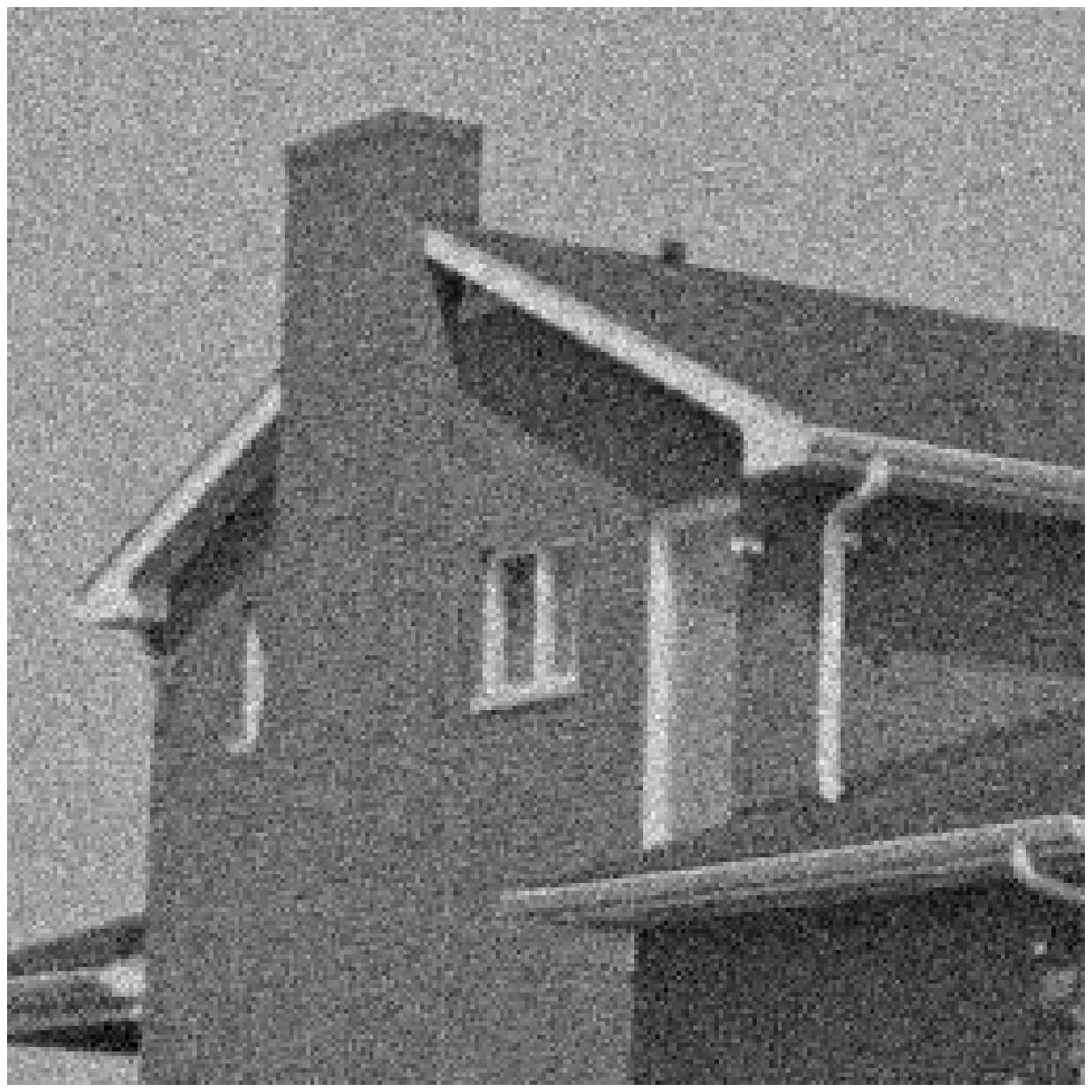}}
\subfigure[By $\dtcwt$]{
\includegraphics[width=1.35in,height=1.35in]
{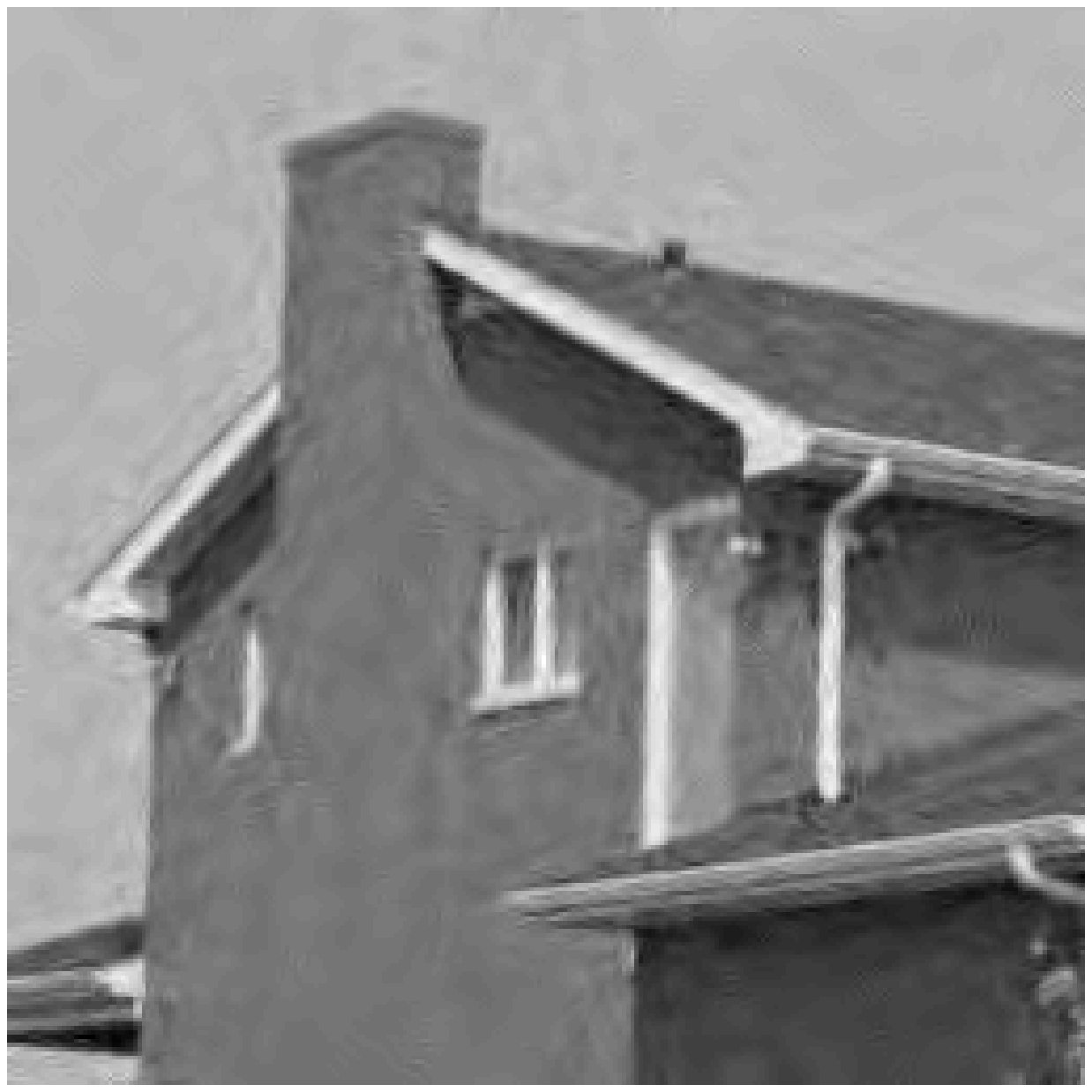}}
\subfigure[By $\tpctf_6$]{
\includegraphics[width=1.35in,height=1.35in]
{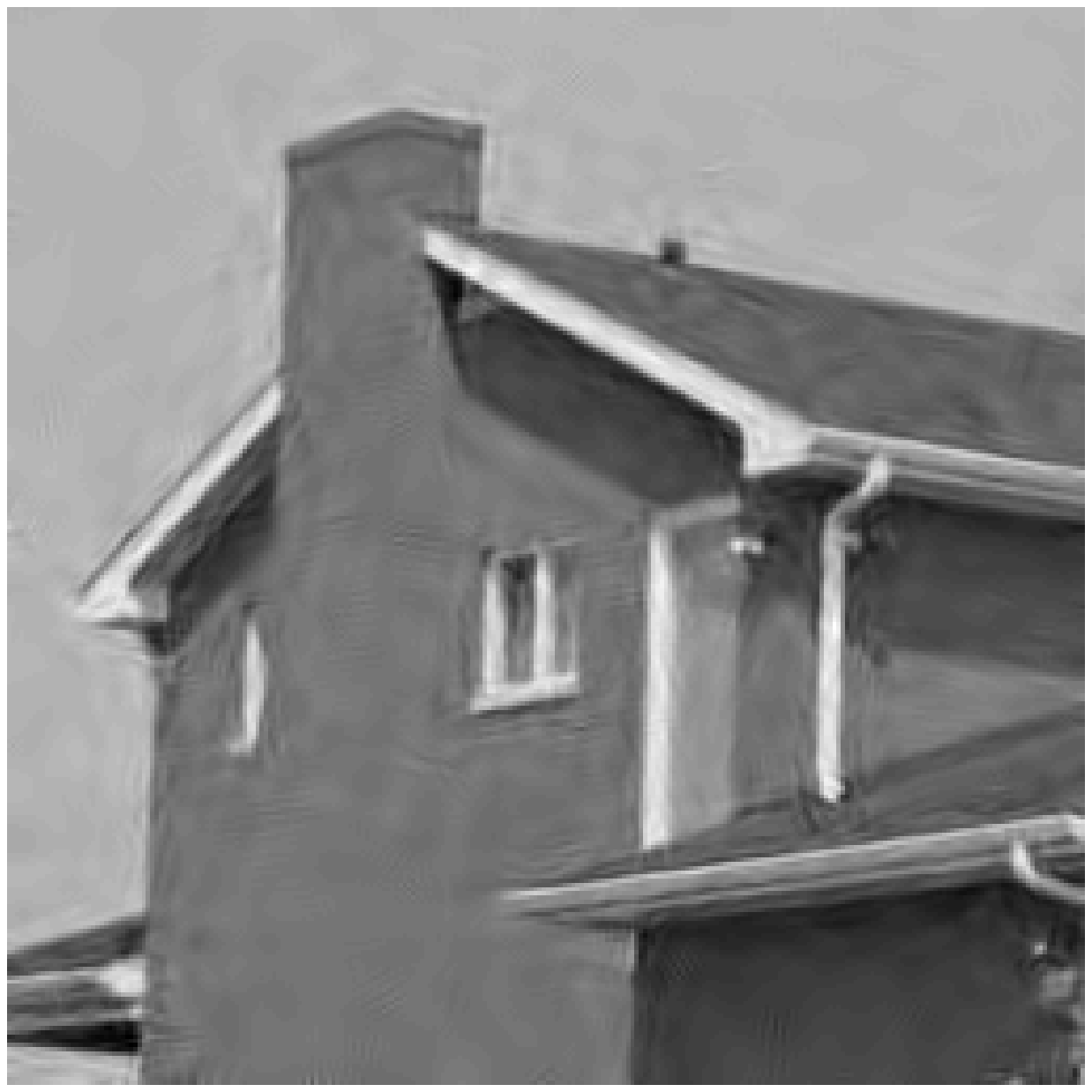}}
\subfigure[By $\tpctf_6$ \& GSM]{
\includegraphics[width=1.35in,height=1.35in]
{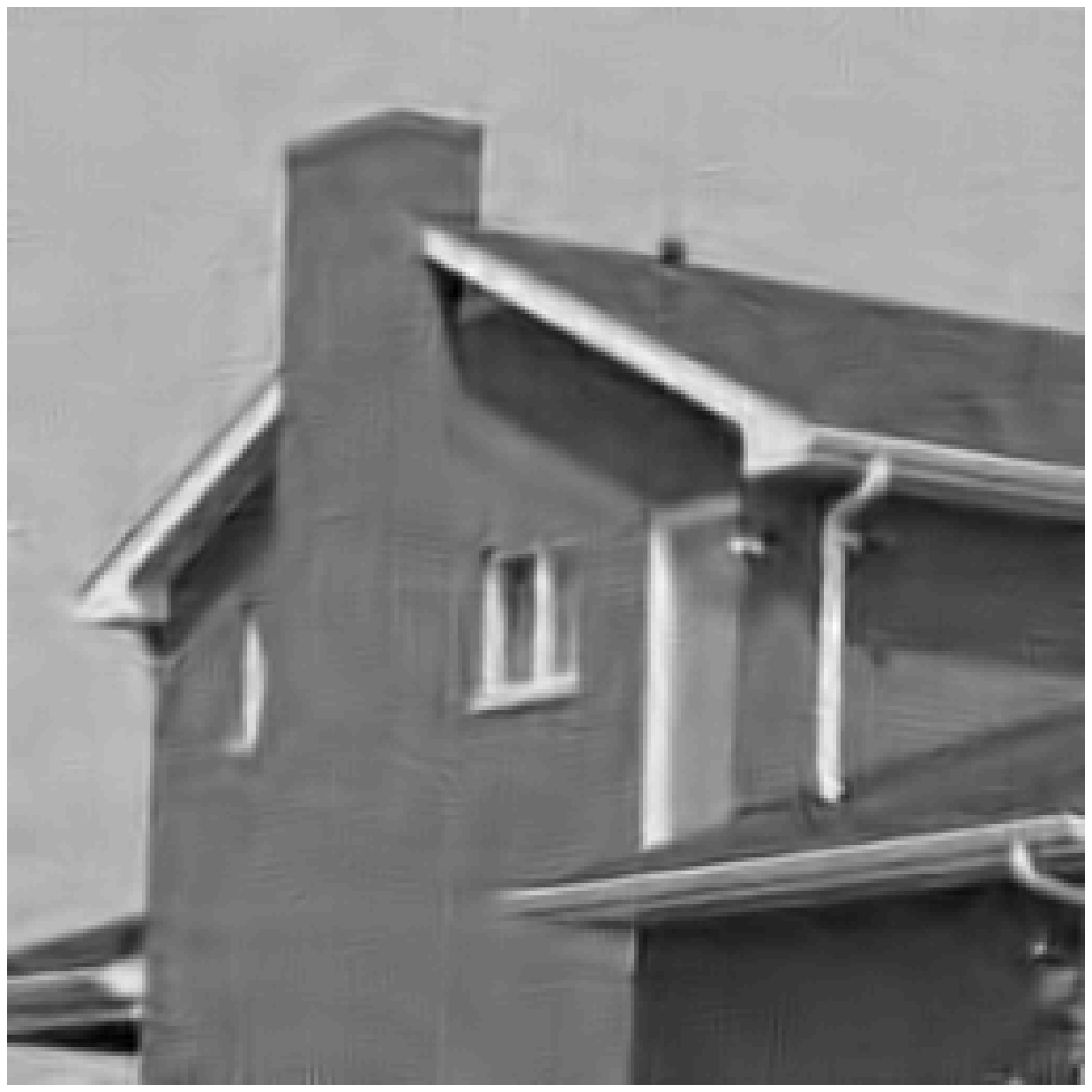}}
\begin{caption}{(a) Original $256\times 256$ image of House. (b) Noisy image with $\sigma_n=20$ (PSNR=$22.15$). (c) Denoised image by $\dtcwt$ (PSNR=$31.64$). (d) Denoised image by $\tpctf_6$ using bivariate shrinkage (PSNR=$32.35$). (e) Denoised image by $\tpctf_6$ using Gaussian scale mixture (PSNR=$32.73$).
} \label{house}
\end{caption}
\end{figure}

\section{Initial Filter Banks of $\dtcwt$ and Some Remarks on $\tpctf_n$}

In this section we shall discuss the choice of the initial filter banks for the level one of $\dtcwt$. We shall see that tensor product complex tight framelets $\tpctf_n$ can be used as the initial filter banks in $\dtcwt$ to further improve the directionality of the first level of $\dtcwt$.
Performance of $\dtcwt$ using $\tpctf_n$ as the initial filter bank for image denoising will be provided in this section.
Finally, we shall also make some remarks on $\tpctf_n$ for possible further improvements.

The original $\dtcwt$ proposed by Kingsbury \cite{K1999,K2001} uses the initial filter banks $\{a^1; b^1\}$ for tree one and $\{a^2; b^2\}$ for tree two, instead of $\{a^0; b^0\}$ for tree one and $\{a^0(\cdot-1); b^0(\cdot-1)\}$ for tree two as proposed in \cite{SBK} (see Section~2). Note that for the approach in \cite{K1999,K2001} we have $\wh{a^2}(\xi)/\wh{a^1}(\xi)\approx e^{-i\xi/2}$ (that is, half-shift difference between tree two and tree one) while for the approach in \cite{SBK} $\wh{a^0(\cdot-1)}(\xi)/\wh{a^0}(\xi)=e^{-i\xi}$ (full-shift difference).

In the function setting, there is no difference between these two approaches. Indeed, the refinable functions for the approach in \cite{SBK} (that is, Section~2) are
\[
\wh{\phi^1}(\xi):=\lim_{J\to \infty} 2^{(1-J)/2} \wh{a^1_J}(2^{-J}\xi)=\lim_{J\to \infty} \wh{a^1}(2^{-1}\xi)\cdots \wh{a^1}(2^{1-J}\xi) \wh{a^0}(2^{-J}\xi)=\prod_{j=1}^\infty \wh{a^1}(2^{-j}\xi)=:\wh{\phi^{a^1}}(\xi)
\]
and similarly,
\[
\wh{\phi^2}(\xi):=\lim_{J\to \infty} 2^{(1-J)/2} \wh{a^2_J}(2^{-J}\xi)=\lim_{J\to \infty} \wh{a^2}(2^{-1}\xi)\cdots \wh{a^2}(2^{1-J}\xi) \wh{a^0(\cdot-1)}(2^{-J}\xi)=\prod_{j=1}^\infty \wh{a^2}(2^{-j}\xi)=:\wh{\phi^{a^2}}(\xi).
\]
That is, $\phi^1=\phi^{a^1}$ and $\phi^2=\phi^{a^2}$.
Consequently, the half-shift condition in \eqref{halfshift} implies
\[
\wh{\phi^{a^2}}(\xi)\approx e^{-i\xi/2} \wh{\phi^{a^1}}(\xi)
\quad \mbox{and}\quad
\wh{\phi^2}(\xi)\approx e^{-i\xi/2} \wh{\phi^1}(\xi).
\]
Define $\psi^1$ and $\psi^2$ by $\wh{\psi^1}(\xi):=\wh{b^1}(\xi/2)\wh{\phi^{a^1}}(\xi/2)$ and
$\wh{\psi^2}(\xi):=\wh{b^2}(\xi/2)\wh{\phi^{a^2}}(\xi/2)$. Now it follows from the relation in \eqref{rel:b12:basic} that
\[
\wh{\psi^2}(2\xi)=\wh{b^2}(\xi)\wh{\phi^{a^2}}(\xi)
\approx -i \sgn(\xi) e^{i\xi/2} \wh{b^1}(\xi) e^{-i\xi/2} \wh{\phi^{a^1}}(\xi)=-i\sgn(\xi) \wh{\psi^1}(2\xi).
\]
Therefore, the Hilbert transform relation in the function setting still holds for both approaches. However, in terms of discrete affine systems, which faithfully reflect the dual tree complex wavelet transform, these two approaches have nontrivial differences.

Let us first consider level $j\ge 2$ by replacing $\{a^0; b^0\}$ and $\{a^0(\cdot-1); b^0(\cdot-1)\}$ with $\{a^1; b^1\}$ and $\{a^2; b^2\}$, respectively. In this case, \eqref{aj1} and \eqref{aj2} becomes
\[
\wh{\mathring{a}^1_j}(\xi):=2^{(j-1)/2} \wh{a^1}(\xi)\wh{a^1}(2\xi)\cdots \wh{a^1}(2^{j-1}\xi), \quad
\wh{\mathring{a}^2_j}(\xi):=2^{(j-1)/2} \wh{a^2}(\xi)\wh{a^2}(2\xi)\cdots \wh{a^2}(2^{j-1}\xi).
\]
To distinguish the two approaches, here we add a small circle over the multilevel filters for the approach in \cite{K1999,K2001}. Then by the half-shift condition in \eqref{halfshift}, we have
\[
\wh{\mathring{a}^2_j}(2^{-j}\xi)\approx \wh{\mathring{a}^1_j}(2^{-j}\xi) e^{-i 2^{-j-1}\xi} e^{i\sum_{\ell=1}^{j-1} \theta(2^{-\ell}\xi)}.
\]
That is, we lost a factor $e^{-i2^{-j-1}\xi}$ in \eqref{a21}, or equivalently, the above is obtained by multiplying $e^{i2^{-j-1}\xi}$ to \eqref{a21}. Consequently, the same analysis as in Section~2 shows that \eqref{rel:aj12} now becomes
\[
\wh{\mathring{a}^2_j}(\xi)\approx e^{-i2^{j-1}\xi} e^{i\xi/2} \wh{\mathring{a}^1_j}(\xi) \eta(2^j\xi)\approx e^{-i 2^{j-1}\xi} e^{i\xi/2} \wh{\mathring{a}^1_j}(\xi), \qquad \xi\in [-\pi, \pi), j\ge 2,
\]
which implies $\mathring{a}^2_j\approx \mathring{a}^1_j(\cdot-2^{j-1}+1/2)$ in the time domain.
For the low-pass filters in \eqref{ajpn}, we have
\[
\wh{\mathring{a}_j^p}(\xi)=[\wh{\mathring{a}_j^1}(\xi)+i \wh{\mathring{a}^2_j}(\xi)]/\sqrt{2}\approx \wh{\mathring{a}_j^1}(\xi)[1+i e^{-i 2^{j-1}\xi+i\xi/2}]/\sqrt{2}, \qquad \xi\in [-\pi, \pi).
\]
Since $|\wh{\mathring{a}_j^1}(\xi)|\approx 2^{j-1} \chi_{2^{-j}[-\pi, \pi)}$ for $\xi\in [-\pi, \pi)$ and $|1+i e^{-i 2^{j-1}\xi+i\xi/2}|^2=2+2\sin (2^{j-1}\xi-2^{-1}\xi)$, we have
\[
|\wh{\mathring{a}_j^p}(\xi)| \approx \sqrt{1+\sin (2^{j-1}\xi-2^{-1}\xi)} |\wh{\mathring{a}_j^1}(\xi)|
\approx \sqrt{1+\sin (2^{j-1}\xi-2^{-1}\xi)} 2^{(j-1)/2} \chi_{2^{-j}[-\pi, \pi)}(\xi), \qquad \xi\in [-\pi,\pi).
\]
When $j$ is large, the frequency separation factor $\sqrt{1+\sin (2^{j-1}\xi-2^{-1}\xi)}  \approx \sqrt{1+\sin (2^{j-1}\xi)}$ is more or less the same as in \eqref{direction:apj} on $\xi\in 2^{-j}[-\pi,\pi)$. But when $j$ is small (in particular, $j=2$ and $3$), by plotting and comparing these functions, we see that the frequency separation factor $\sqrt{1+\sin (2^{j-1}\xi-2^{-1}\xi)}$ is always slightly worse than the frequency separation factor $\sqrt{1+\sin (2^{j-1}\xi)}$ on the interval $2^{-j}[-\pi,\pi)$.

For high-pass filters, \eqref{direction:bpJ} becomes
\[
\wh{\mathring{b}^p_j}(\xi):=[\wh{\mathring{b}^1_j}(\xi)+i \wh{\mathring{b}^2_j}(\xi)]/\sqrt{2}=
\wh{\mathring{b}^1_j}(\xi)[1+e^{i\xi/2}\sgn(\xi)]/\sqrt{2}.
\]
Note that
\be \label{fs:exp}
|[1+e^{i\xi/2}\sgn(\xi)]/\sqrt{2}|=\sqrt{1+\cos(\xi/2)\sgn(\xi)}
=\begin{cases}
\sqrt{1+\cos(\xi/2)} &\text{if $\xi\in [0,\pi)$,}\\
\sqrt{1-\cos(\xi/2)} &\text{if $\xi\in [-\pi,0)$.}
\end{cases}
\ee
For the approach in Section~2 (that is, \cite{SBK}), for $j\ge 2$, we have the ideal frequency separation in \eqref{direction:bpJ}, that is, $\wh{b^p_j}\approx \sqrt{2} \wh{b^1_j} \chi_{[0,\pi)}$ and $\wh{b^n_j}\approx \sqrt{2}\wh{b^1_j} \chi_{[-\pi,0)}$ on the basic frequency interval $[-\pi,\pi)$. That is, we are using the ideal frequency separation factor $\sqrt{2}\chi_{[0,\pi)}$ for the approach in Section~2 when $j\ge 2$. However, the frequency separation factor in \eqref{fs:exp} is not ideal and is much worse than the ideal frequency separation factor when $j$ is small. However, in every application, directionality for small decomposition levels is much more important. When the decomposition level $J$ increases, the resolution of the processed image decreases by a factor of $2$ (the processed image at decomposition level $J$ becomes smoother when $J$ becomes larger). However, it is very natural that when the resolution level is higher, we need more and better directions. In this sense, for level $J\ge 2$, the approach proposed in \cite{SBK} has better directionality than the original approach of Kingsbury \cite{K1999,K2001}.

\begin{figure}[th]
\begin{center}
\includegraphics[width=1.5in,height=1.2in]{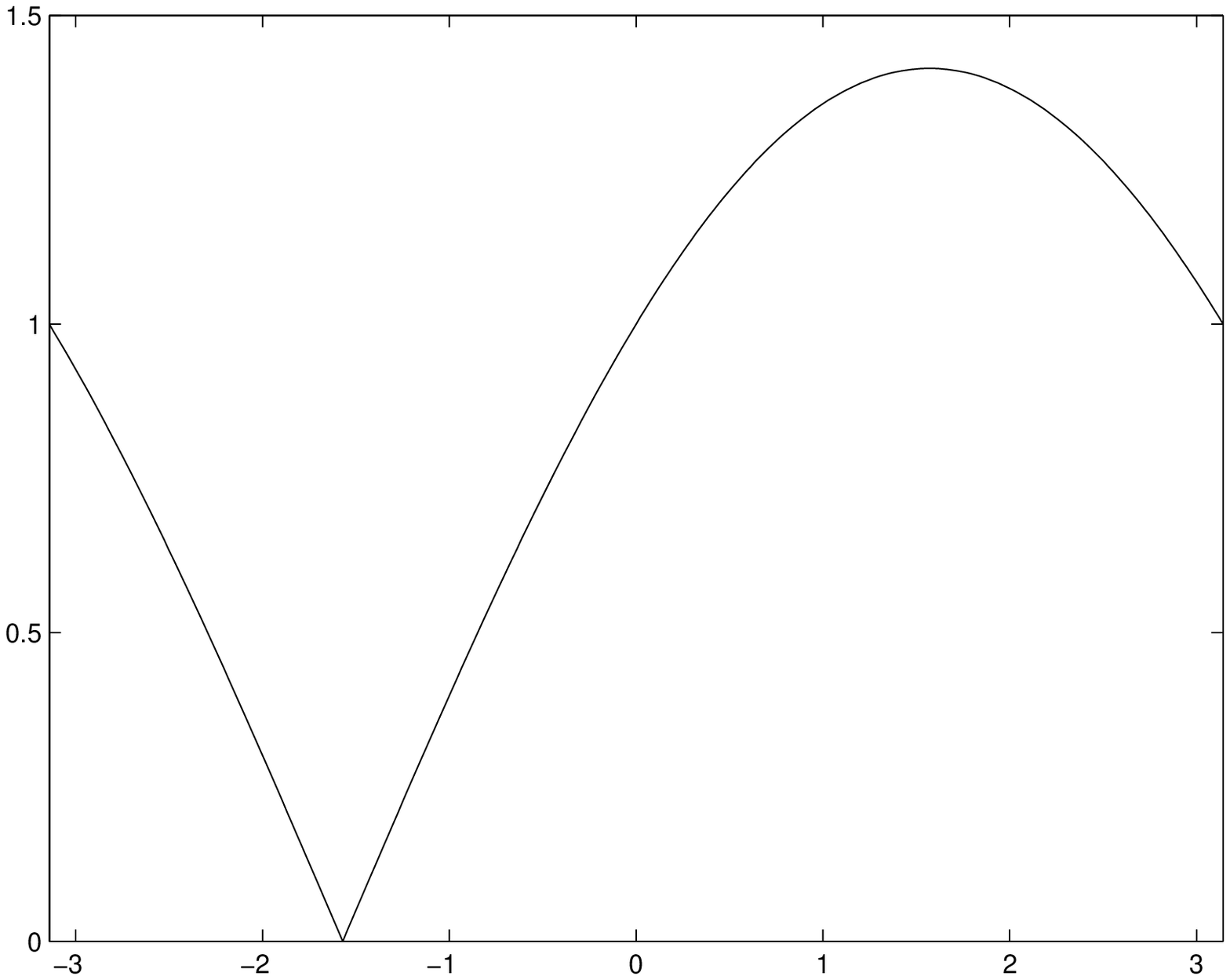}
\includegraphics[width=1.5in,height=1.2in]{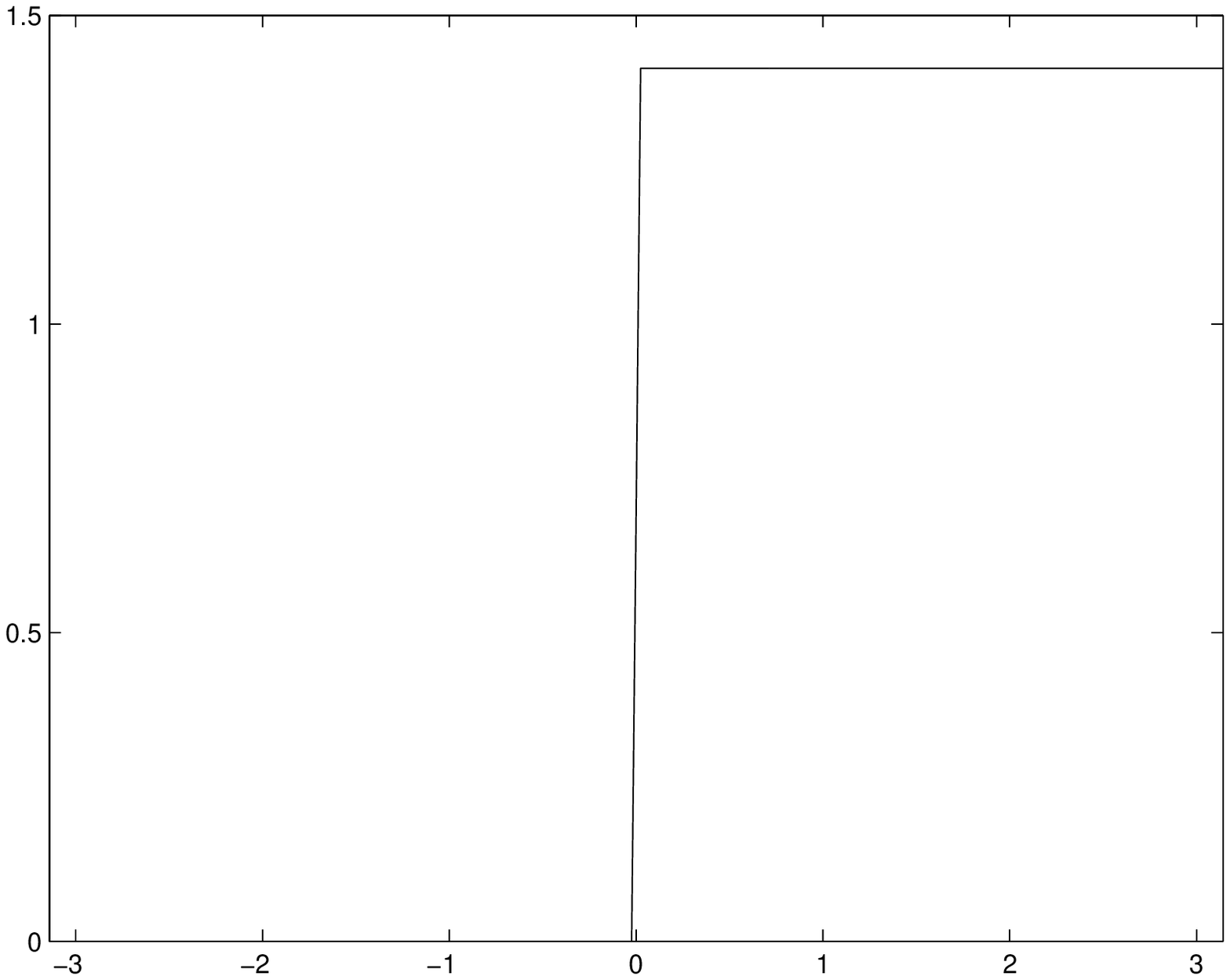}
\includegraphics[width=1.5in,height=1.2in]{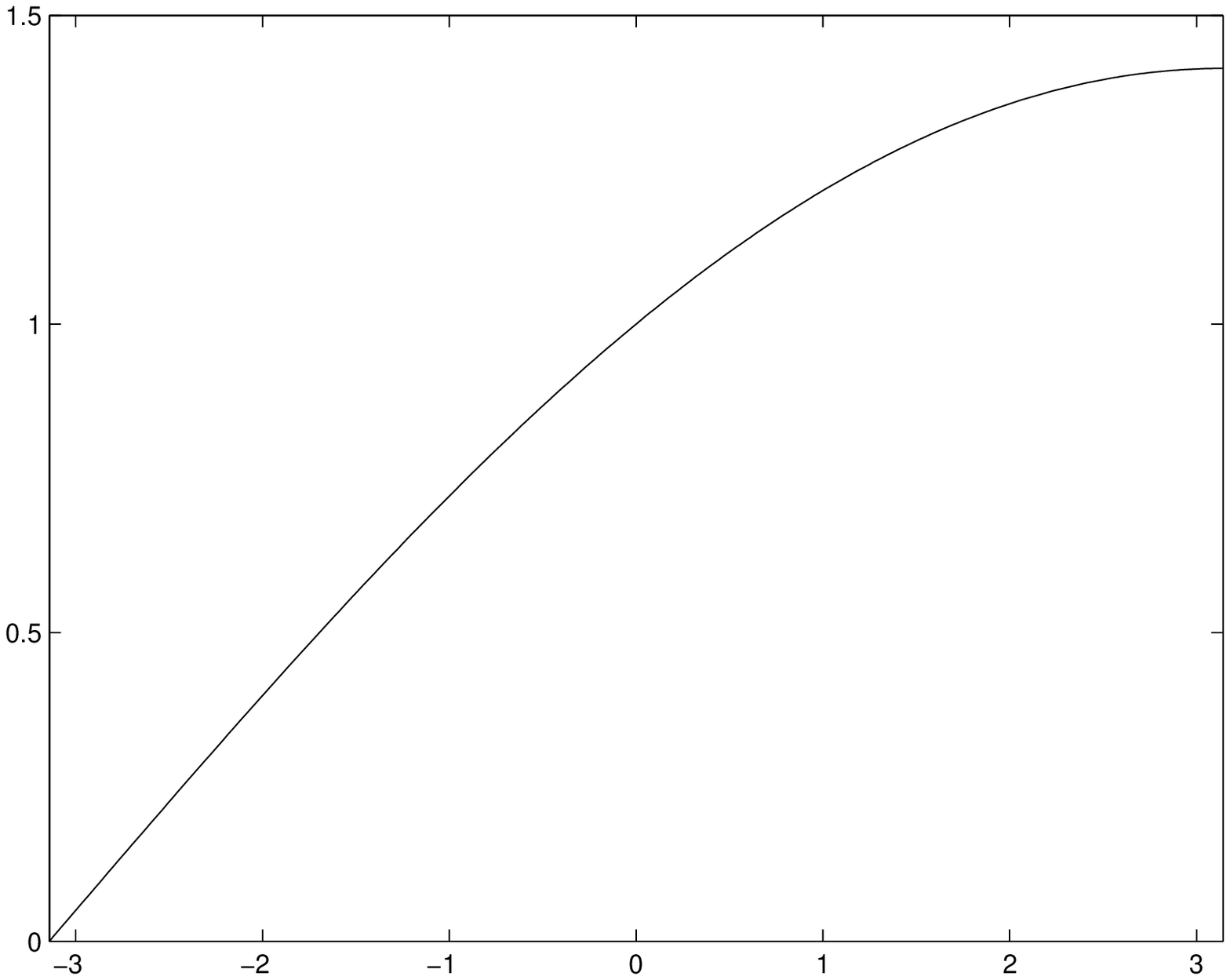}
\includegraphics[width=1.5in,height=1.2in]{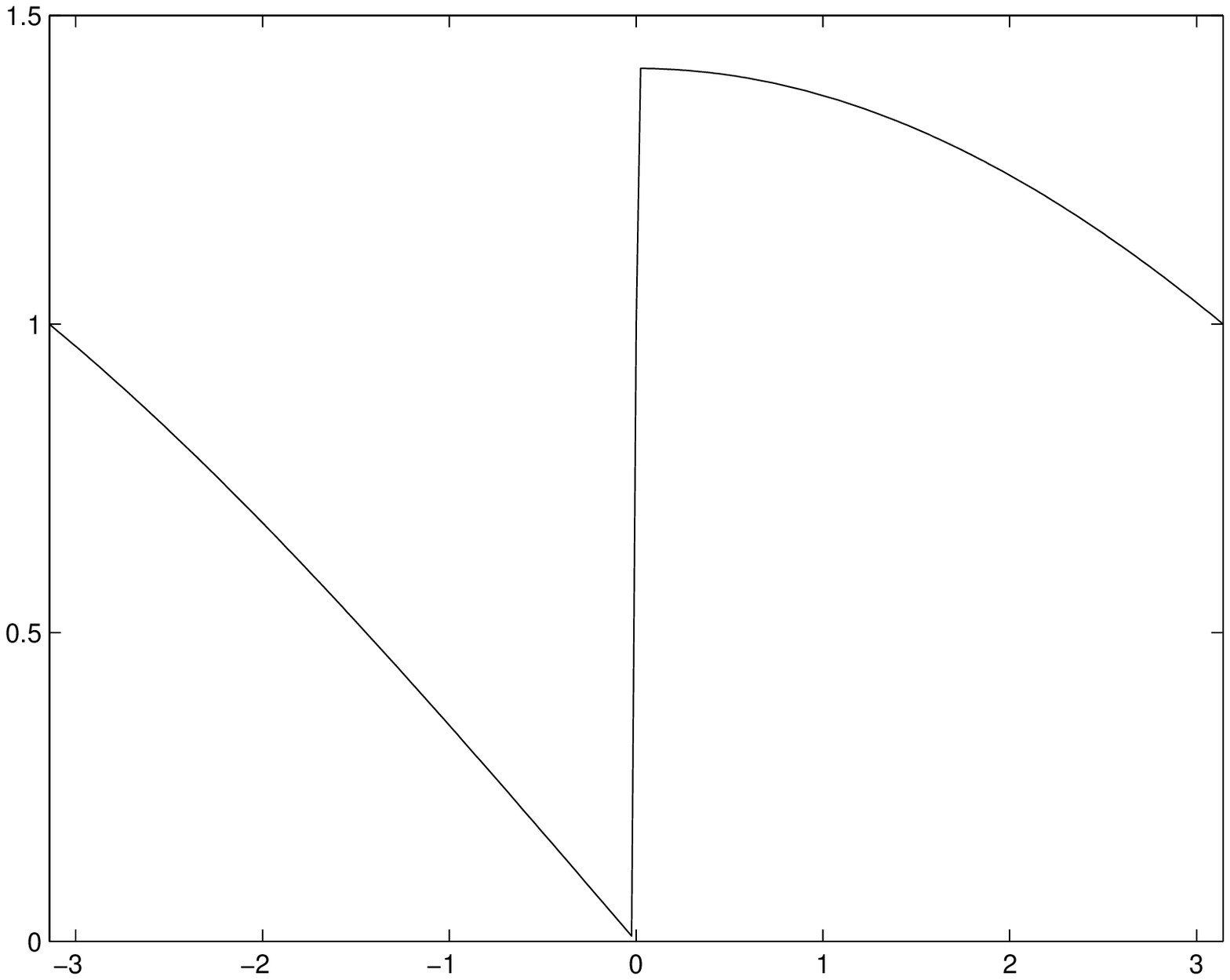}
\begin{caption}{
From the left to the right, the frequency separation factor $\sqrt{1+\sin \xi}$ in \eqref{fs:sin},
the ideal frequency separation factor $(1+\sgn(\xi))/\sqrt{2}$ in \eqref{direction:bpJ},  the frequency separation factor $\sqrt{1+\sin(\xi/2)}$ in \eqref{fs:sin2}, and the frequency separation factor $\sqrt{1+\cos(\xi/2) \sgn(\xi)}$ in \eqref{fs:exp}.
Note that
$\int_{-\pi}^\pi |\sqrt{1+\sin \xi}|^2 d\xi=
\int_{-\pi}^\pi |(1+\sgn(\xi))/\sqrt{2}|^2 d\xi=
\int_{-\pi}^\pi |\sqrt{1+\sin(\xi/2)}|^2 d\xi=
\int_{-\pi}^\pi |\sqrt{1+\cos(\xi/2) \sgn(\xi)}|^2d\xi=2\pi$.
} \label{freqsep}
\end{caption}
\end{center}
\end{figure}

Let us now consider level one. Then we have
\[
\mathring{a}_1^p:=[a^1+ia^2]/\sqrt{2}, \quad \mathring{a}^n_1:=[a^1-ia^2]/\sqrt{2}\quad \mbox{and}\quad
\mathring{b}_1^p:=[b^1+ib^2]/\sqrt{2}, \quad \mathring{b}^n_1:=[b^1-ib^2]/\sqrt{2}.
\]
For low-pass filters, we have
\[
\wh{\mathring{a}_1^p}(\xi)=[\wh{a^1}(\xi)+i\wh{a^2}(\xi)]/\sqrt{2}
\approx \wh{a^1}(\xi) [1+i e^{-i\xi/2}]/\sqrt{2}, \qquad \xi\in [-\pi,\pi).
\]
By $|1+i e^{-i\xi/2}|=2+\sin(\xi/2)$, we have
\be \label{fs:sin2}
|\wh{\mathring{a}_1^p}(\xi)|\approx \sqrt{1+\sin(\xi/2)} |\wh{a^1}(\xi)|, \qquad \xi\in [-\pi, \pi).
\ee
See Figure~\ref{freqsep} for graphs of the several frequency separation factors.
From the graphs in Figure~\ref{freqsep}, since $|\wh{a^1}|^2\approx \chi_{[-\pi/2, \pi/2]}$ for $\xi\in [-\pi, \pi)$, we see that the frequency separation factor $\sqrt{1+\sin(\xi/2)}$ is slightly worse than the frequency separation factor $\sqrt{1+\sin \xi}$ for splitting the low-pass filter $a^1$.

For high-pass filters, we have
\[
\wh{\mathring{b}_1^p}(\xi)=[\wh{b^1}(\xi)+i \wh{b^2}(\xi)]/\sqrt{2}=
[e^{-i\xi} \ol{\wh{a^1}(\xi+\pi)}+i e^{-i\xi} \ol{\wh{a^2}(\xi+\pi)}]/\sqrt{2}
\approx e^{-i\xi} \ol{\wh{a^1}(\xi+\pi)}[1+ie^{-i\theta(\xi+\pi)}]/\sqrt{2}.
\]
Since
\[
e^{-i\theta(\xi+\pi)}=e^{i(\xi+\pi)/2} e^{-i \pi \lfloor \tfrac{\xi+2\pi}{2\pi}\rfloor}
=e^{i\xi/2} e^{i\pi \lfloor \tfrac{\xi}{2\pi}\rfloor}=\sgn(\xi) e^{i\xi/2}
\]
for $\xi \in (-2\pi, 2\pi)$. We conclude that
\[
\wh{\mathring{b}^p_1}(\xi) \approx \wh{b^1}(\xi)[1+e^{i\xi/2} \sgn(\xi)]/\sqrt{2}, \qquad \xi\in [-\pi, \pi).
\]
Now we see that the above frequency separation factor is the same as in \eqref{fs:exp} for $j\ge 2$.
Since $|\wh{b^1}|^2\approx 2^{j-1}\chi_{[-\pi, -\pi/2]\cup[\pi/2, \pi]}$ on $[-\pi, \pi)$, from Figure~\ref{freqsep}, we see that the frequency separator factor in \eqref{fs:exp} for $j=1$ is slightly worse than the frequency separation factor $\sqrt{1+\sin \xi}$ for $\xi\in [-\pi, \pi)$ in \eqref{fs:sin} for splitting the high-pass filter $b^1$. Moreover, by \eqref{fs:exp} and $|\wh{a^1}(\xi)|^2+|\wh{a^1}(\xi+\pi)|^2=1$, we deduce that
\begin{align*}
\int_0^\pi &\big[|\wh{b^p_1}(\xi+\pi)|^2+|\wh{b^n_1}(\xi)|^2\big] d\xi \approx \int_0^\pi \big[|\wh{a^1}(\xi)|^2(1-\sin(\xi/2))+ |\wh{a^1}(\xi+\pi)|^2(1-\cos(\xi/2))\big]
d\xi\\
&\qquad =\pi-\int_{-\pi}^{\pi} |\wh{a^1}(\xi)|^2 |\sin(\xi/2)|d\xi
\approx \pi-\int_{-\pi/2}^{\pi/2} |\sin(\xi/2)|d\xi=\pi-(4-2\sqrt{2})>\pi-2.
\end{align*}
Comparing with \eqref{directional:initial}, the directionality for level one using the approach in \cite{K1999,K2001} is slightly worse than the approach in \cite{SBK}.
In conclusion, we see that the approach in \cite{SBK} has slightly better directionality than the approach in \cite{K1999,K2001}.

However, as we pointed out in \eqref{directional:initial} of Section~2, the directionality for the first level of $\dtcwt$ is not very strong.
To remedy this shortcoming, we can replace the initial tight framelet filter bank $2^{-1/2}\{a^0, a^0(\cdot-1); b^1, b^2\}$ by the undecimated version of tensor product complex tight framelet filter bank $\tpctf_n$ with $n$ relatively large, more precisely, we use the following tight framelet filter bank as the initial filter bank in $\dtcwt$:
\[
2^{-1/2}\{a, a(\cdot-1); b^{1,p}, \ldots, b^{s,p}, b^{1,n}, \ldots, b^{s,n}, b^{1,p}(\cdot-1), \ldots, b^{s,p}(\cdot-1), b^{1,n}(\cdot-1), \ldots, b^{s,n}(\cdot-1)\}.
\]

Their performance on image denoising is reported in Table~\ref{dtcwtfirstlevel}.
We can see the improvement due to more directions in the first stage filter bank, especially when the image contains many details such as Barbara. Since the directions are not the same between the first level and the second level, the coefficients along approximately the same direction share a common parent. When the noise variance is high, the parent coefficients can only provide information to predict the threshold value on the same direction as the child coefficients. This is the main reason for the limited or no improvement of the PSNR values for high noise variance.

\begin{table}[ht]
\begin{center}
\begin{tabular}{|c||c|c||c|c||c|c||c|c||c|c||}
\hline
&\multicolumn{2}{|c||}{Lena} &\multicolumn{2}{|c||}{Barbara} &\multicolumn{2}{|c||}{Boat} &\multicolumn{2}{|c||}{House} &\multicolumn{2}{|c|}{Pepper} \\  \hline
$\sigma_n$ &$\tcwt$ &$\ncwt$ &$\tcwt$ &$\ncwt$ &$\tcwt$ &$\ncwt$ &$\tcwt$ &$\ncwt$ &$\tcwt$ &$\ncwt$ \\ \hline
$5$ &38.25 &38.27  &37.36 &37.73 &36.77 &36.70 &38.45 &38.67 &37.18 &37.19\\ \hline
$10$ &35.19 &35.35  &33.52 &33.96 &33.21 &33.28 &34.78 &35.05 &33.40 &33.50\\ \hline
$15$&33.47 &33.61  &31.38 &31.73 &31.33 &31.36 &32.90 &33.16 &31.29 &31.36\\ \hline
$20$&32.23 &32.34  &29.87 &30.13 &30.01 &30.00 &31.63 &31.80 &29.83 &29.85\\ \hline
$25$&31.26 &31.34  &28.70 &28.90 &28.99 &28.96 &30.65 &30.74 &28.71 &28.71\\ \hline
$30$&30.47 &30.52  &27.77 &27.90 &28.18 &28.14 &29.84 &29.88 &27.80 &27.78\\ \hline
$50$&28.21 &28.24  &25.26 &25.27 &26.01 &25.97 &27.57 &27.53 &25.30 &25.27\\ \hline
\end{tabular}
\medskip
\begin{caption}{
Columns of $\tcwt$ are for PSNR values (an average over five experiments) using bivariate shrinkage in \cite{SS:bslocal} and $\dtcwt$ using finitely supported orthogonal wavelet filter banks in \eqref{dtcwt:a0}--\eqref{dtcwt:a2}. Columns of $\ncwt$ are for PSNR values using the same bivariate shrinkage and $\dtcwt$ except that the first level transform uses the tight framelet filter bank $\tpctf_6$ (14 directions) instead of the filter bank in \eqref{dtcwt:a0}.
}\label{dtcwtfirstlevel}
\end{caption}
\end{center}
\end{table}

We complete this paper by some remarks on $\tpctf_n$. If one insists on using tensor product filter banks for high dimensional problems, then the approach of $\tpctf_n$ is probably the most natural choice. For the convenience of the reader, we list some possible advantages of $\tpctf_n$ as follows:
\begin{enumerate}
\item $\tpctf_n$ has more directions than $\dtcwt$ when $n$ increases.

\item $\tpctf_n$ enjoys the same simple tensor product structure as $\dtcwt$. Therefore, its algorithm is essentially the same as a standard discrete wavelet transform using tensor product.

\item $\tpctf_4$ offers an alternative to $\dtcwt$ while enjoying less redundancy than $\dtcwt$, since $\tpctf_4$ uses only one low-pass filter and $\dtcwt$ uses four low-pass filters for dimension two.

\item The low-pass filters used in $\tpctf_n$ are not only real-valued but also have symmetry, which is one of the desired properties of a filter bank in applications.
    The finitely supported low-pass filters used in $\dtcwt$ in \cite{K1999,SBK} do not have symmetry, due to the quarter-shift condition in \eqref{quartershift}. In fact, except the variants of the Haar orthogonal low-pass filter, any  finitely supported real-valued orthogonal low-pass filter cannot have symmetry (\cite{Daub:book}).
\end{enumerate}

One possible shortcoming of the tensor product complex tight framelets $\tpctf_n$ considered in this paper is that the complex tight framelet filter banks do not have compact support in the space/time domain, which is one of the most desirable properties in wavelet analysis. One may suspect that it may be difficult to have $\tpctf_n$ with finitely supported filters. Fortunately, due to recent developments on one-dimensional complex tight framelet filter banks, finitely supported complex tight framelet filter banks with or without symmetry have been well studied in \cite{Han:acha:2013} and references therein. It turns out that it is quite flexible to construct finitely supported tensor product complex tight framelet filter banks with directionality from any low-pass filter $a$ satisfying $|\wh{a}(\xi)|^2+|\wh{a}(\xi+\pi)|^2\le 1$, which is a necessary condition for constructing tight framelet filter banks.
For construction of compactly supported tensor product complex tight framelets $\tpctf_3$, this has been fully developed in \cite{HMZ}. We shall report elsewhere the detailed construction of compactly supported tensor product complex tight framelets $\tpctf_n$ for any integer $n\ge 3$ and their performance for certain applications.

\end{document}